\documentclass[aps,prd,showpacs,superscriptaddress,groupedaddress,nofootinbib]{revtex4-2}  % for review and submission
\usepackage[T1]{fontenc}
\usepackage{lmodern} 
\usepackage{graphicx}  % needed for figures
\usepackage{dcolumn}   % needed for some tables
\usepackage{bm}        % for math
\usepackage{amssymb}   % for math
\usepackage{amsmath}
\usepackage{bbold}
\usepackage{array}
\usepackage{makecell}
\usepackage{braket}
\usepackage{longtable}
\usepackage{supertabular,booktabs}

\usepackage{titlesec} %for hyperref to detect names of section etc
\usepackage[colorlinks,citecolor=blue,urlcolor=blue,hypertexnames=true]{hyperref}
\usepackage{subfigure}
\usepackage[usenames,dvipsnames,svgnames,table]{xcolor} % use color 

\newcommand{\be}{\begin{equation}}
\newcommand{\ee}{\end{equation}}
\newcommand{\bea}{\begin{eqnarray}}
\newcommand{\eea}{\end{eqnarray}}
\newcommand{\bml}{\begin{subequations}}
\newcommand{\eml}{\end{subequations}}
\newcommand{\bfig}{\begin{figure}}
\newcommand{\efig}{\end{figure}}

\newcommand{\slashed}{\hspace{-1.1ex}/}

\newcommand{\bmat}{\begin{pmatrix}}
\newcommand{\emat}{\end{pmatrix}}
\usepackage{graphicx,slashed,booktabs,xcolor,multirow,float,
amsfonts,bbold,mathtools,sidecap,tikz,bm,enumitem}
\usepackage{multirow}
\usepackage{titlesec}
\usepackage{hyperref}
\usepackage{amssymb}% http://ctan.org/pkg/amssymb
\usepackage{pifont}% http://ctan.org/pkg/pifont

\titleclass{\subsubsubsection}{straight}[\subsection]

\newcounter{subsubsubsection}[subsubsection]
\renewcommand\thesubsubsubsection{\thesubsubsection.\arabic{subsubsubsection}}
\titleformat{\subsubsubsection}
  {\normalfont\normalsize\bfseries}{\thesubsubsubsection}{1em}{}
\titlespacing*{\subsubsubsection}
{0pt}{3.25ex plus 1ex minus .2ex}{1.5ex plus .2ex}

\makeatletter
\renewcommand\paragraph{\@startsection{paragraph}{5}{\z@}%
  {3.25ex \@plus1ex \@minus.2ex}%
  {-1em}%
  {\normalfont\normalsize\bfseries}}
\renewcommand\subparagraph{\@startsection{subparagraph}{6}{\parindent}%
  {3.25ex \@plus1ex \@minus .2ex}%
  {-1em}%
  {\normalfont\normalsize\bfseries}}
\def\toclevel@subsubsubsection{4}
\def\toclevel@paragraph{5}
\def\toclevel@paragraph{6}
\def\l@subsubsubsection{\@dottedtocline{4}{7em}{4em}}
\def\l@paragraph{\@dottedtocline{5}{10em}{5em}}
\def\l@subparagraph{\@dottedtocline{6}{14em}{6em}}
\makeatother

\begin{document}

% The following information is for internal review, please remove them for submission
%\widetext

\definecolor{lime}{HTML}{A6CE39}
\DeclareRobustCommand{\orcidicon}{\hspace{-2.1mm}
\begin{tikzpicture}
\draw[lime,fill=lime] (0,0.0) circle [radius=0.13] node[white] {{\fontfamily{qag}\selectfont \tiny \,ID}}; \draw[white, fill=white] (-0.0525,0.095) circle [radius=0.007]; 
\end{tikzpicture} \hspace{-3.7mm} }
\foreach \x in {A, ..., Z}{\expandafter\xdef\csname orcid\x\endcsname{\noexpand\href{https://orcid.org/\csname orcidauthor\x\endcsname} {\noexpand\orcidicon}}}
\newcommand{\orcidauthorA}{0000-0002-0459-3873}
\newcommand{\orcidauthorB}{0000-0002-1466-8525}
\newcommand{\orcidauthorC}{0000-0003-1081-0632}

% the following line is for submission, including submission to the arXiv!!
%\hspace{5.2in} \mbox{Fermilab-Pub-04/xxx-E}

\title{\textcolor{Sepia}{\textbf \huge\Large No-go for the formation of heavy mass Primordial Black Holes in Single Field Inflation}}

%Can superluminality save Primordial Black Hole formation\\ from\\ Effective Field Theory of Single Field Inflation?}}
%\input author_list.tex       % D0 authors (remove the first 3 lines
                             % of this file prior to submission, they
                             % contain a time stamp for the authorlist)
                             % (includes institutions and visitors)
%\date{\today}

%\date{\today}
\author{{\large  Sayantan Choudhury\orcidA{}${}^{1}$}}
\email{sayantan\_ccsp@sgtuniversity.org,  \\ sayanphysicsisi@gmail.com (Corresponding author)}
\author{\large Mayukh~R.~Gangopadhyay\orcidB{}${}^{1}$}
\email{mayukh\_ccsp@sgtuniversity.org,  mayukhraj@gmail.com }
\author{ \large M.~Sami\orcidC{}${}^{1,2,3}$}
\email{ sami\_ccsp@sgtuniversity.org,  samijamia@gmail.com}

\affiliation{ ${}^{1}$Centre For Cosmology and Science Popularization (CCSP),\\
        SGT University, Gurugram, Delhi- NCR, Haryana- 122505, India,}
\affiliation{${}^{2}$Center for Theoretical Physics, Eurasian National University, Astana 010008, Kazakhstan.}
	\affiliation{${}^{3}$Chinese Academy of Sciences,52 Sanlihe Rd, Xicheng District, Beijing.}

\begin{abstract}
%%%%%%%%%%%%%%%%%%%%%%%%%%%%%%%%%%%%%%%%%%%
We examine the possibility of Primordial Black Holes (PBHs) formation in single field models of inflation. \textcolor{black}{Using the adiabatic or wave function renormalization scheme in the short range modes, we show that one-loop correction to the power spectrum is free from quadratic UV divergence.} 
We consider a framework in which PBHs are produced during the transition from Slow Roll (SR) to Ultra Slow Roll (USR) followed by the end of inflation.
We demonstrate that the renormalized power spectrum \textcolor{black}{soften the contribution of the logarithmic IR divergence and} severely restricts the possible mass range of produced PBHs in the said transition, namely, $M_{\rm PBH}\sim 10^{2}{\rm gm}$ $\hat{\rm a}${\it  la} a no-go theorem.  In particular, we find that the produced PBHs are short lived ($t^{\rm evap}_{\rm PBH}\sim 10^{-20}{\rm sec}$) and the corresponding number of e-folds in the USR region is restricted to $\Delta{\cal N}_{\rm USR}\approx 2$.

%%%%%%%%%%%%%%%%%%%%%%%%%%%%%%%%%%%%%%%%%%%
\end{abstract}

\maketitle
\tableofcontents
\newpage

\section{Introduction}
	The exponential expansion of the nascent universe, dubbed the inflationary paradigm along with the hot big bang model, stands as the most promising theoretical epistemology of the Universe. Invoking inflation from the philosophical rationale to overcome the otherwise muddles like horizon and flatness problems, it was quickly realized that the formation of the structures in our Universe can be a manifestation of the fluctuations
    generated quantum mechanically during inflation. It was also discovered that large fluctuations associated with specific scales can cause gravitational collapse as the mode reenters the later radiation-dominated era, resulting in the formation of astrophysical objects behaving like black holes and christened to be Primordial Black Holes (PBHs) \cite{Hawking:1974rv,Carr:1974nx,Carr:1975qj,Chapline:1975ojl,Carr:1993aq,Kawasaki:1997ju,Yokoyama:1998pt,Kawasaki:1998vx,Rubin:2001yw,Khlopov:2002yi,Khlopov:2004sc,Saito:2008em,Khlopov:2008qy,Carr:2009jm,Choudhury:2011jt,Lyth:2011kj,Drees:2011yz,Drees:2011hb,Ezquiaga:2017fvi,Kannike:2017bxn,Hertzberg:2017dkh,Pi:2017gih,Gao:2018pvq,Dalianis:2018frf,Cicoli:2018asa,Ozsoy:2018flq,Byrnes:2018txb,Ballesteros:2018wlw,Belotsky:2018wph,Martin:2019nuw,Ezquiaga:2019ftu,Motohashi:2019rhu,Fu:2019ttf,Ashoorioon:2019xqc,Auclair:2020csm,Vennin:2020kng,Nanopoulos:2020nnh,Gangopadhyay:2021kmf,Inomata:2021uqj,Stamou:2021qdk,Ng:2021hll,Wang:2021kbh,Kawai:2021edk,Solbi:2021rse,Ballesteros:2021fsp,Rigopoulos:2021nhv,Animali:2022otk,Correa:2022ngq,Frolovsky:2022ewg,Escriva:2022duf,Kristiano:2022maq,Kristiano:2023scm,Riotto:2023hoz,Riotto:2023gpm,Karam:2022nym,Ozsoy:2023ryl,Choudhury:2023jlt,Choudhury:2023rks,Choudhury:2023hvf,Choudhury:2023fjs, Bhattacharya:2023ysp,Choudhury:2023kdb,Choudhury:2023hfm,Choudhury:2023fwk,Choudhury:2024one,Choudhury:2024ybk,Choudhury:2024jlz,Choudhury:2024dei,Choudhury:2024dzw,Choudhury:2024aji,Motohashi:2023syh,Firouzjahi:2023ahg,Franciolini:2023lgy,Firouzjahi:2023aum,Cheng:2023ikq,Tasinato:2023ukp} to distinguish them from stellar black holes, which are formed by the death of a star, while keeping Chandrasekhar's limits in mind. In that regard, the mass of PBHs can be as small as the Planck mass ($M_{\rm pl}$) or the associated cut-off of the effective theory under consideration. PBHs can be the riposte to the curve ball thrown to us by nature: the identity of Dark Matter (DM) \cite{Ivanov:1994pa,Afshordi:2003zb,Frampton:2010sw,Carr:2016drx,Kawasaki:2016pql,Inomata:2017okj,Espinosa:2017sgp,Ballesteros:2017fsr,Sasaki:2018dmp,Ballesteros:2019hus,Dalianis:2019asr,Cheong:2019vzl,Green:2020jor,Carr:2020xqk,Ballesteros:2020qam,Carr:2020gox,Ozsoy:2020kat}. Apart from the potential solution of the identity conundrum of dark matter, PBHs have gained a lot of interest in the theoretical physics community due to the recent observations of Gravitational Waves (GWs) \cite{Baumann:2007zm,Saito:2008jc,Saito:2009jt,Choudhury:2013woa,Sasaki:2016jop,Raidal:2017mfl,Ali-Haimoud:2017rtz,Di:2017ndc,Raidal:2018bbj,Cheng:2018yyr,Vaskonen:2019jpv,Drees:2019xpp,Hall:2020daa,Ballesteros:2020qam,Ragavendra:2020sop,Carr:2020gox,Ozsoy:2020kat,Ashoorioon:2020hln,Ragavendra:2020vud,Papanikolaou:2020qtd,Ragavendra:2021qdu,Wu:2021zta,Kimura:2021sqz,Solbi:2021wbo,Teimoori:2021pte,Cicoli:2022sih,Ashoorioon:2022raz,Papanikolaou:2022chm,Wang:2022nml} from merging black holes by Laser Interferometer Gravitational Wave Observatory(LIGO) \cite{LIGOScientific:2016aoc}.The large mass and distance of the BHs indicate their primordial origin. In the past few years, the study of PBHs has emerged as one of the most active fields of research in theoretical cosmology. A nice statistical analysis of interest can be found in ref. \cite{Ozsoy:2023ryl}. 
 
 The formation of PBHs in the early stages of the Universe is mainly attributed to an enhancement of fluctuation at a certain scale due to some mechanism associated with the motion of the inflaton field on the flat potential. There are two main paths explored in a cold inflationary scenario. In one case, the inflaton field encounters a single/multiple tiny, short-lived bump(s) \cite{Mishra:2019pzq,ZhengRuiFeng:2021zoz} on its path along the otherwise flat potential, causing a large enough fluctuation of the field to be imprinted on the associated scale. In another case, there is a transition at an inflection point from the SR to the USR region of the potential, and due to this change in the dynamics of the motion, again there is an enhancement of the fluctuation. On that note, in the alternate dynamical realization of inflation dubbed Warm Inflation (WI), this enhancement of fluctuation is quite natural, and thus the production of PBHs is quite easier to understand \cite{Arya:2019wck,Bastero-Gil:2021fac,Correa:2022ngq}. The cold inflationary dynamics and associated PBHs production are the focus of this work. Although placing a bump on an otherwise flat potential may seem to be phenomenologically viable, it becomes a liability from the point of view of  theoretical motivation. A transition from SR to USR, on the other hand, has the potential to be the cause of the increase in fluctuation responsible for PBH production at later stages \cite{Gangopadhyay:2021kmf,Cicoli:2022sih}. In this framework, quantum effects can be investigated in a model independent fashion without the knowledge of the inflaton potential.
 %Thus, simply from the theoretical perspective, a detailed study of this mechanism from a generalized standpoint is required.
 
 In a recent  Ref. \cite{Kristiano:2022maq}, the authors claim that {\it PBH formation from single-field inflation is ruled out}. Their findings are based on the argument that a one-loop correction to the tree-level power spectrum is extremely large on a large scale. The main reason behind this strong argument is based on the presence of quadratic divergent contributions at the one-loop corrected result of the primordial power spectrum on top of logarithmic divergent effects. 
 
 It was previously argued in Refs.\cite{Sloth:2006az,Seery:2007we,Seery:2007wf,Bartolo:2007ti,Senatore:2009cf,Seery:2010kh,Bartolo:2010bu,Senatore:2012ya,Senatore:2012nq,Pimentel:2012tw,Chen:2016nrs,Markkanen:2017rvi,Higuchi:2017sgj,Syu:2019uwx,Rendell:2019jnn,Cohen:2020php,Green:2022ovz,Premkumar:2022bkm}  that only logarithmic divergences survive in one-loop corrections in the Slow Roll (SR) regime, which is generated on the sub-horizon scale due to quantum fluctuations, and the same behaviour persists up to the super-horizon scales. It was shown with detailed computation in ref. \cite{Chen:2016nrs} that though the quadratic and other power law divergent effects come on the sub-horizon scale computation of the one-loop calculation, they will not survive at the super-horizon scale due to the late time limit in the dimensionally regularized and properly renormalized version of the cosmological one-loop corrected two-point correlation function. Once the late time limit is correctly implemented, one can explicitly show that such quadratic and other power law divergent effects are completely absent in the final form of the two-point correlator, leaving only logarithmic effects.  Though these computations have been done previously in the SR period, a similar argument also applies to the model independent framework where the PBHs formation takes place during the SR-USR transition followed by the end of inflation. However, in Ref. \cite{Kristiano:2022maq} , due to the appearance of quadratic divergence, the authors rule out the formation of PBHs in single field models of inflation.
 %effects can't be absorbed. Hence the strong claim regarding not having PBH formation from any models of single field inflation having one loop quantum effect is not correct. 
 This claim was recently refuted in Ref.\cite{Riotto:2023hoz} using a model independent approach showing that  short scale loop effects do not alter the large-scale primordial power spectrum.\footnote{We thank Antonio Riotto for fruitful communication on the subject. }.

 We re-examined in great detail the one-loop corrections, renormalization and Dynamical Renormlization Group (DRG) resummation method \cite{Boyanovsky:1998aa,Boyanovsky:2001ty,Boyanovsky:2003ui,Burgess:2009bs,Dias:2012qy,Chen:2016nrs,Baumann:2019ghk} to better understand the claim of Ref.\cite{Kristiano:2022maq}.

The paper is organized as follows: In section \ref{basics}, we discuss the basics of the theoretical background of single field inflationary paradigm, which will going to be extremely useful to understand the rest of the computation performed in this paper. Next, in section \ref{tree}, we provide the detailed computation of tree level scalar power spectrum, where we show the explicitly contributions from SR and USR regions in detail. \textcolor{black}{In both cases, we have studied the behaviour of the tree level scalar power spectrum, which can be analyzed in the sub-horizon (quantum), horizon exit point (semi-classical), and super-horizon (classical) regions, respectively.} In section \ref{oneloop}, we compute the one-loop corrected unrenormalized scalar power spectrum both from the SR and USR regions. \textcolor{black}{Such one-loop effects are generated in the sub-horizon region, which has a purely quantum mechanical origin and originates from short-range UV modes. We use the cut-off regularization technique to compute the contributions from one-loop integrals, where we introduce both UV and IR cut-offs in the momentum dependent integrals as appearing in both the SR and USR phases.} Then, in section \ref{ren}, considering both the contributions from the SR and USR regions, we compute the renormalized version of the one-loop corrected scalar power spectrum \textcolor{black}{by making use of the well-known adiabatic or wave function renormalization scheme
\cite{Durrer:2009ii,Wang:2015zfa,Boyanovsky:2005sh,Marozzi:2011da,Finelli:2007fr,Ford:1977in,Parker:1974qw,Fulling:1974pu,Parker:2012at,Ford:1986sy}. By implementing the renormalization condition correctly within the framework of adiabatic or wave function renormalization scheme, we explicitly compute the expression for the counter terms both in the SR and USR periods, which fix the form of the one-loop renormalized spectrum by removing the quadratic UV divergence. Further using this result we perform the renormalization in the power spectrum which in turn softens the effect of long range modes by shifting the logarithmic IR divergence in the next order of perturbation theory.} Next, in section \ref{DRG}, we compute the Dynamical Renormalization Group (DRG) re-summed course grained version of the scalar power spectrum by applying the non-perturbative but convergent exponenciation. \textcolor{black}{Additionally, we discuss the validity of the DRG re-summation method  and provide a numerical estimate which justifies its applicability and significance within the present context where we consider sharp transition from SR to USR phase.} Using these results, in section \ref{num}, we provide a no-go theorem for the mass of the PBH formation in a single field inflationary paradigm. We show that large mass PBHs are not allowed by the present prescription. We also comment on the constraints on the evaporation time scale and on the corresponding number of e-foldings in the USR region using the no-go result. Finally, in section \ref{conclu}, we conclude with some possible promising future prospects. Additionally, we include two appendices \ref{App:1} and \ref{App:B},\textcolor{black}{ where we discuss the details of the computation of the one-loop correction to the primordial power spectrum for the scalar modes using in-in formalism in the USR phase, the technical results of the one-loop momentum integrals in the SR and USR phases, and the appearances of the quadratic UV and logarithmic IR divergent contributions, respectively.}

\section{The single field inflation}
\label{basics}

Let us consider the following representative action for the single field inflation:
\begin{eqnarray}
\label{model} S=\frac{1}{2}\int d^4x \sqrt{-g} \left[M^2_{\rm pl} R - (\partial\phi)^2 - 2 V(\phi) \right],
\end{eqnarray}
where the inflation field $\phi$ is a scalar field which is minimally coupled to the gravity.  In the above action the canonical kinetic term of the scalar field $ (\partial\phi)^2=g^{\mu\nu}\left(\partial_{\mu}\phi\right)\left(\partial_{\nu}\phi\right)$ and  $V(\phi)$ is the effective potential for the scalar field,  $M_{\rm pl}$ is the reduced Planck mass scale,  $R$ is the Ricci scalar. 
In the spatially flat Friedmann-Lemaitre-Robertson-Walker (FLRW) Universe, the metric is 
given by,
\begin{eqnarray}
ds^2=a^2(\tau)\left(-d\tau^2+d{\bf x}^2\right)=a^2(\tau)\left(-d\tau^2+\delta_{ij}dx^{i}dx^{j}\right),
\end{eqnarray}
where $\tau$ is the conformal time coordinate.  The scale factor $a(\tau)$ in the conformal coordinate is given by the following de Sitter solution,$a(\tau)=-1/H\tau$ where $-\infty<\tau<0$. In this expression $`H'$ represents the Hubble parameter 
which is not exactly constant and we, in reality, deal with quasi-de Sitter.  To this effect, let us first write down the evolution equations, 
\begin{eqnarray}
{\cal H}^2=\frac{1}{3M^2_{\rm pl}}\Bigg(\frac{1}{2}\phi^{'2}+a^2V(\phi)\Bigg),\quad\quad{\cal H}^{'}=-\frac{1}{2M^2_{\rm pl}}\phi^{'2},\quad\quad
\phi^{''}+2{\cal H}\phi^{'}+a^2\frac{dV(\phi)}{d\phi}=0.
\end{eqnarray}
where we have introduced a notation $'$ which represents the derivative with respect to the conformal time coordinate $\tau$.  Additionally,  we have used the definition of the Hubble parameter $\displaystyle {\cal H}=a^{'}/a=aH$.

Now, to validate and properly end inflation at a particular point in the field space one needs to introduce the following deviation parameters from the exact de Sitter solution commonly known as the slow-roll parameters,
\begin{eqnarray}
\epsilon &=&\bigg(1-\frac{{\cal H}^{'}}{{\cal H}^2}\bigg)=\frac{1}{2M^2_{\rm pl}}\frac{\phi^{'2}}{a^2{\cal H}^2},\quad\quad
\eta =\frac{\epsilon^{'}}{\epsilon {\cal H}}=\bigg(2\epsilon+\frac{\phi^{''}}{\phi^{'} {\cal H}}-1\bigg).
\end{eqnarray}
To realize inflation in the SR region, one needs to satisfy the following constraints, $\epsilon\ll 1$,$|\eta|\ll 1$ and $\left|\frac{\phi^{''}}{\phi^{'} {\cal H}}-1\right|\ll 1$.
	In SR regime both $\epsilon$ and $\eta$ approximately constant.  We assume that  SR region
 is followed by the USR regime where the inflationary potential becomes extremely flat such that $dV/d\phi\approx 0$,  which further implies the following constraints,
$\frac{\phi^{''}}{\phi^{'} {\cal H}}\approx -2\Longrightarrow\phi^{'} \propto a^{-2}$ which implies $\epsilon \propto a^{-6}$ and $\eta\approx -6$, which will be important for PBH formation.
 \section{Computation of tree level scalar power spectrum}
\label{tree}

   Let us now consider the small  perturbations around the spatially flat FLRW background,   where the linearised version of the field and the metric perturbations are given by,
   \begin{eqnarray}
  && \phi({\bf x},\tau)=\overline{\phi}(\tau)+\delta\phi({\bf x},\tau),\quad\quad
   ds^2=a^2(\tau)\Bigg[-d\tau^2+\bigg\{\bigg(1+2\zeta({\bf x},\tau)\bigg)\delta_{ij}+2h_{ij}({\bf x},\tau)\bigg\}dx^{i}dx^{j}\Bigg].
   \end{eqnarray}
   %To implement the imprints of the perturbation in a correct sense we choose the following gauge condition,    $\delta\phi({\bf x},\tau)=0$,
     % which is commonly known as the {\it unitary comoving gauge}.  
      In the above mentioned linearised version of the perturbed metric, two significant components appear,  which are scalar comoving curvature perturbation $\zeta({\bf x},\tau)$ and the transverse-traceless tensor perturbation $h_{ij}(({\bf x},\tau)$.  However, in this work, we are only restricted to the scalar perturbation, and for this reason, in the rest of the paper, we will not further refer to the tensor perturbation.
   
   %One can further write down the linearised approximated version of the scalar comoving curvature perturbation $\zeta({\bf x},\tau)$ in terms of the inflaton field perturbation by the following expression:
  % \begin{eqnarray}
  % \zeta({\bf x},\tau)=-\bigg(\frac{{\cal H}}{\overline{\phi}^{\,'}}\bigg)\delta\phi({\bf x},\tau).
  % \end{eqnarray}
  % This information is going to be extremely useful for the rest of our computation because instead of computing the correlation function for the scalar perturbations in terms of a gauge dependent object $\delta\phi({\bf x},\tau)$ we can compute all of these expressions in terms of the gauge fixed quantity scalar comoving curvature perturbation $\zeta({\bf x},\tau)$.
   
   Expanding the representative action for the scalar field as stated in equation(\ref{model}) up to the second order in the scalar co-moving curvature perturbation $\zeta({\bf x},\tau)$, gives rise to the following simplified action,
   \begin{eqnarray}
   S_{(2)}=M^2_{\rm pl}\int d\tau\;  d^3x\;  a^2\; \epsilon\; \bigg(\zeta^{'2}-\left(\partial_i\zeta\right)^2\bigg).
   \end{eqnarray}
   Next,  we introduce a new variable,   $v=zM_{\rm pl}\zeta$,  which is commonly known as {\it Mukhanov Sasaki } (MS) variable. In terms of the MS variable, the above mentioned second order perturbed action can be translated into the following canonically normalized form:
   \begin{eqnarray}
  \label{ms1} S_{(2)}=\frac{1}{2}\int d\tau\;  d^3x\;  \bigg(v^{'2}-\left(\partial_iv\right)^2+\frac{z^{''}}{z}v^{2}\bigg),
   \end{eqnarray}
   where, $z=a\sqrt{2\epsilon}.$  We then write the above action in the Fourier space by using the following {\it anstaz} for the Fourier transformation:
   \begin{eqnarray}
   v({\bf x},\tau):=\int\frac{d^3{\bf k}}{(2\pi)^3}\; e^{i{\bf k}.{\bf x}}\;v_{\bf k}(\tau).
   \end{eqnarray}
   In terms of the above mentioned Fourier transformed scalar modes, one can recast the action stated in equation(\ref{ms1}) in the following form,
     \begin{eqnarray}
  \label{ms2} S_{(2)}=\frac{1}{2}\int \frac{d^3{\bf k}}{(2\pi)^3}\;d\tau\; e^{i{\bf k}.{\bf x}}\; \bigg(|v^{'}_{\bf k}|^{2}-\left(k^2-\frac{z^{''}}{z}\right)|v_{\bf k}|^{2}\bigg).
   \end{eqnarray}
   Then the {\it Mukhanov Sasaki equation} for the scalar modes can be expressed as,
   \begin{eqnarray}
  \label{ms3} v^{''}_{\bf k}+\left(k^2-\frac{z^{''}}{z}\right)v_{\bf k}=0\,,
   \end{eqnarray}
   where the conformal time dependent part of the effective frequency can be expressed as,$\frac{z^{''}}{z}\approx\frac{2}{\tau^2}$.
  % Here it is important to note that equation(\ref{ms3}) is a second order differential equation of the scalar modes with respect to the conformal time coordinate.  
  We then use the following normalization condition in terms of the Klein Gordon product for the scalar modes to fix the mathematical structure of the general solution, $v^{'*}_{\bf k}v_{\bf k}-v^{'}_{\bf k}v^{*}_{\bf k}=i$.
The corresponding general solution of the {\it Mukhanov Sasaki equation} for the scalar mode during SR period is given by the following expression:
   \begin{eqnarray}
   v_{\bf k}(\tau)=\frac{\alpha^{\rm SR}_{\bf k}}{\sqrt{2k}}\left(1-\frac{i}{k\tau}\right)\; e^{-ik\tau}+\frac{\beta^{\rm SR}_{\bf k}}{\sqrt{2k}}\left(1+\frac{i}{k\tau}\right)\; e^{ik\tau},
   \end{eqnarray}
 where the coefficients $\alpha^{\rm SR}_{\bf k}$ and $\beta^{\rm SR}_{\bf k}$ are fixed by the appropriate choice of the quantum initial condition.   In the SR period, we fix initial condition choosing the Bunch Davies quantum vacuum state which gives the following constraint:
 \begin{eqnarray}\label {SRBogo}
 \alpha^{\rm SR}_{\bf k}=1\quad\quad\quad\quad {\rm and}\quad\quad\quad\quad\beta^{\rm SR}_{\bf k}=0.
 \end{eqnarray}
 Consequently,   the scalar mode is finally expressed in the following simplified form:
 \begin{eqnarray}
 v_{\bf k}(\tau)=\frac{1}{\sqrt{2k}}\left(1-\frac{i}{k\tau}\right)\; e^{-ik\tau}.
 \end{eqnarray}
   Using the above expression for the scalar mode in the SR regime at  early times, $\tau\leq \tau_s$,  one can write down the following expression for the gauge invariant co-moving curvature perturbation,
   \begin{eqnarray}
 \zeta_{\bf k}(\tau)&=&\frac{v_{\bf k}(\tau)}{zM_{\rm pl}}=\left(\frac{iH}{2M_{\rm pl}\sqrt{\epsilon}}\right)\frac{1}{k^{3/2}}\left(1+ik\tau\right)\; e^{-ik\tau}.
 \end{eqnarray}
 The above mentioned expression is the general solution in the SR regime and separately valid in the following regions\cite{Mukhanov:1990me,Polarski:1995jg,Kiefer:1998qe,Lesgourgues:1996jc}, Super-horizon region ($k\ll aH\rightarrow-k\tau\ll 1\rightarrow-k\tau\rightarrow 0\longrightarrow {\rm Classical \; regime}$), Sub-horizon region ($k\gg aH\rightarrow-k\tau\gg 1\rightarrow-k\tau\rightarrow \infty\longrightarrow {\rm Quantum \; regime}$), At the horizon crossing point ($k=aH\rightarrow-k\tau= 1\rightarrow {\rm Semi-classical \; regime}$).
% In the aforesaid regions, the expression for the gauge invariant co-moving curvature perturbation during SR period can be explicitly written as,
 %\be  \zeta_{\bf k}(\tau)\approx
%\left\{
%	\begin{array}{ll}
	%	\displaystyle \left(\frac{iH}{2M_{\rm pl}\sqrt{\epsilon}}\right)\frac{1}{k^{3/2}}\left(ik\tau\right)\; e^{-ik\tau}\; \quad\quad\quad\quad\quad\quad & \mbox{when sub-horizon}  \; \;(-k\tau\gg 1)  \\
%		\displaystyle \left(\frac{iH}{2M_{\rm pl}\sqrt{\epsilon}}\right)\frac{1}{k^{3/2}}\left(-i\right)\; e^{i} & \mbox{at horizon-crossing} \; \; (-k\tau= 1)  \\
%			\displaystyle  \left(\frac{iH}{2M_{\rm pl}\sqrt{\epsilon}}\right)\frac{1}{k^{3/2}}\; e^{-ik\tau}& \mbox{when super-horizon}  \; \;(-k\tau\ll 1)
%	\end{array}
%\right. \ee
Here we need to note down the following two important points,  which will be the necessary input for the further computation of this paper:
\begin{enumerate}
\item As we are interested in incorporating the quantum effects coming from the one loop correction at the level of the cosmological two-point correlation function and on its amplitude, the first two results appearing in the sub-horizon regime and at the horizon crossing point are physically relevant.  

\item As the scalar mode exits the cosmological horizon and becomes classical,  quantum effects become unimportant at super-horizon  scales. 

\end{enumerate}

In what follows, we shall extend our discussion to the USR region,\cite{Hirano:2016gmv,Dimopoulos:2017ged,Germani:2017bcs,Pattison:2018bct,Cheng:2018qof,Firouzjahi:2018vet,Cruces:2018cvq,Pattison:2019hef,Syu:2019uwx,Bhaumik:2019tvl,Ragavendra:2020sop,Ballesteros:2020sre,Pattison:2021oen,Cheng:2021lif},  which is appearing in the conformal time window $\tau_s\leq \tau\leq \tau_e$.  Here $\tau_s$ and $\tau_e$  correspond to the conformal time scale at the transition point from SR to USR and end of inflation respectively. In the USR regime, the conformal time dependence of the first slow-roll parameter can be written as, $\epsilon(\tau) \propto a^{-6}(\tau)$ which implies that, $\epsilon(\tau)=\epsilon  \;\left(\frac{\tau}{\tau_s}\right)^{6}$,
where $\epsilon$ on the right hand side corresponds to the value of the first slow-roll parameter in SR regime.  The above mathematical form suggests that at the transition point from SR to USR,  this parameter is approximately a constant quantity which will be important in the discussion to follow.

In the USR region,  the solution for the gauge invariant comoving curvature perturbation from the solution of the MS equation can be written as:
 \begin{eqnarray}
 \zeta_{\bf k}(\tau)=\left(\frac{iH}{2M_{\rm pl}\sqrt{\epsilon}}\right)\left(\frac{\tau_s}{\tau}\right)^{3}\frac{1}{k^{3/2}}\bigg[\alpha^{\rm USR}_{\bf k}\left(1+ik\tau\right)\; e^{-ik\tau}-\beta^{\rm USR}_{\bf k}\left(1-ik\tau\right)\; e^{ik\tau}\bigg],
 \end{eqnarray}
where $\alpha^{\rm USR}_{\bf k}$ and $\beta^{\rm USR}_{\bf k}$ are two Bogoliubov coefficients which actually connect the solution for the USR region to the SR region.  We have already mentioned earlier that in the SR region, these coefficients are fixed by the initial choice of the quantum vacuum to be Bunch Davies state. For the USR region, these coefficients can be fixed by the following two conditions: 
\begin{enumerate}
\item Solution for the gauge invariant comoving curvature perturbation for SR and USR region becomes continuous at the transition point, $\tau=\tau_s$,  i.e.$\left[\zeta_{\bf k}(\tau)\right]_{\rm SR, \tau=\tau_s}= \left[\zeta_{\bf k}(\tau)\right]_{\rm USR, \tau=\tau_s}$.
For the present computational purpose we have chosen instantaneous transition from SR to USR region,  as it confronts well with the numerical solutions. 

\item First derivative of the comoving curvature perturbation with respect to the conformal time scale,   which is commonly known as the canonically conjugate momenta of the scalar curvature perturbation field variable becomes continuous at the transition point $\tau=\tau_s$,  i.e.$\left[\zeta^{'}_{\bf k}(\tau)\right]_{\rm SR, \tau=\tau_s}= \left[\zeta^{'}_{\bf k}(\tau)\right]_{\rm USR, \tau=\tau_s}$.
\end{enumerate}
Applying the above mentioned two conditions at the transition point $\tau=\tau_s$, we obtain the following expressions for the two Bogoliubov coefficients:
\bea \label{bo1}\alpha^{\rm USR}_{\bf k}&=&1-\frac{3}{2ik^{3}\tau^{3}_s}\left(1+k^{2}\tau^{2}_s\right),\\
\label{bo2}\beta^{\rm USR}_{\bf k}&=&-\frac{3}{2ik^{3}\tau^{3}_s}\left(1+ik\tau_s\right)^{2}\; e^{-2ik\tau_s}.\eea
%Then the expression for the gauge invariant comoving curvature perturbation during USR period can be explicitly written as:
% \be  \zeta_{\bf k}(\tau)\approx
%\left\{
%	\begin{array}{ll}
	%	\displaystyle  \left(\frac{iH}{2M_{\rm pl}\sqrt{\epsilon}}\right)\left(\frac{\tau_s}{\tau}\right)^{3}\frac{1}{k^{3/2}}\left(ik\tau\right)\bigg[\alpha_{\bf k}\; e^{-ik\tau}+\beta_{\bf k}\; e^{ik\tau}\bigg]\; \quad\quad\quad& \mbox{when sub-horizon}  \; \;(-k\tau\gg 1)  \\
	%	\displaystyle \left(\frac{iH}{2M_{\rm pl}\sqrt{\epsilon}}\right)\left(\frac{\tau_s}{\tau}\right)^{3}\frac{1}{k^{3/2}}(-i)\bigg[\alpha_{\bf k}\; e^{i}+\beta_{\bf k}\; e^{-i}\bigg]& \mbox{at horizon-crossing} \; \; (-k\tau= 1)  \\
	%		\displaystyle \left(\frac{iH}{2M_{\rm pl}\sqrt{\epsilon}}\right)\left(\frac{\tau_s}{\tau}\right)^{3}\frac{1}{k^{3/2}}\bigg[\alpha_{\bf k}\; e^{-ik\tau}-\beta_{\bf k}\; e^{ik\tau}\bigg]& \mbox{when super-horizon}  \; \;(-k\tau\ll 1)
%	\end{array}
%\right. \ee
In case of USR regime similar to the SR, the first two solutions will be used to implement the one loop quantum correction as the last one is treated to be classical solution where no such effect appears.

To compute the expression for the two-point correlation function and the associated power spectrum in the Fourier space, we need to explicitly quantize the corresponding scalar modes, which is necessary to calculate the cosmological correlation functions. For this purpose we need to first of all define the creation $\hat{a}^{\dagger}_{\bf k}$ and annihilation $\hat{a}_{\bf k}$ operators, which will create an excited state or destroy it respectively out of the Bunch Davies quantum vacuum state.  Now we identify $|0\rangle$ as our Bunch Davies state which has to satisfy the constraint,
$\hat{a}_{\bf k}|0\rangle=0\quad\forall {\bf k}$.
The canonical quantization between the scalar mode and its associated conjugate momenta has to satisfy the following equal time commutation relation (ETCR):
\bea \left[\hat{\zeta}_{\bf k}(\tau),\hat{\zeta}^{'}_{{\bf k}^{'}}(\tau)\right]=i\;\delta^{3}\left({\bf k}+{\bf k}^{'}\right),\eea
where $\hat{\zeta}_{\bf k}(\tau)$ is the corresponding quantum operator for the scalar mode, and is given by the following expression:
\bea \hat{\zeta}_{\bf k}(\tau)&=&\bigg[{\zeta}_{\bf k}(\tau)\hat{a}_{\bf k}+{\zeta}^{*}_{\bf k}(\tau)\hat{a}^{\dagger}_{-{\bf k}}\bigg].\eea
This can further be translated into the language of the all possible commutation relations between the above mentioned creation and annihilation operators,  which are given by the following expressions:
\bea \left[\hat{a}_{\bf k},\hat{a}^{\dagger}_{{\bf k}^{'}}\right]&=&(2\pi)^{3}\;\delta^{3}\left({\bf k}+{\bf k}^{'}\right),\quad
 \left[\hat{a}_{\bf k},\hat{a}_{{\bf k}^{'}}\right]=0,\quad
  \left[\hat{a}^{\dagger}_{\bf k},\hat{a}^{\dagger}_{{\bf k}^{'}}\right]=0.\eea
  
  Then the corresponding tree level contribution to the two-point correlation function for the co-moving curvature perturbation at the late times $(\tau\rightarrow 0)$, can be expressed as follows,
\bea \langle \hat{\zeta}_{\bf k}\hat{\zeta}_{{\bf k}^{'}}\rangle_{{\bf Tree}} &=&(2\pi)^{3}\;\delta^{3}\left({\bf k}+{\bf k}^{'}\right)P^{\bf Tree}_{\zeta}(k),\eea
and the associated dimensionless power spectrum can be cast as:
\bea \Delta^{2}_{\zeta,{\bf Tree}}(k)=\frac{k^{3}}{2\pi^{2}}P^{\bf Tree}_{\zeta}(k) \quad\quad{\rm where}\quad\quad P^{\bf Tree}_{\zeta}(k)=\langle\langle \zeta_{\bf k}\zeta_{-{\bf k}}\rangle\rangle_{(0,0)}=|\zeta_{\bf k}(\tau\rightarrow 0)|^{2}_{{\bf Tree}}.\quad\quad\eea
Let us note that while computing the correlation, Vacuum Expectation Value (VEV) is taken with respect to the Bunch Davies quantum vacuum state. Now we will explicitly write down expression for the tree level contribution of the dimensionless power spectrum computed from the scalar mode:  
\be  \Delta^{2}_{\zeta,{\bf Tree}}(k)=\left(\frac{H^{2}}{8\pi^{2}M^{2}_{\rm pl}\epsilon}\right)\times
\left\{
	\begin{array}{ll}
		\displaystyle \left(1+k^{2}\tau^{2}\right)& \mbox{when} \; \; k\ll k_s\; \;(\rm SR)  \\ 
			\displaystyle \left(\frac{k_e}{k_s}\right)^{6}\left|\alpha^{\rm USR}_{\bf k}\left(1+ik\tau\right)\; e^{-ik\tau}-\beta^{\rm USR}_{\bf k}\left(1-ik\tau\right)\; e^{ik\tau}\right|^{2} & \mbox{when }  \; \;k_s\leq k\leq k_e\; \;(\rm USR)
	\end{array}
\right. \ee
where in the expressions appearing for $k_s\leq k\leq k_e$, we have used the fact that $-k_e\tau_e=1$ and $-k_s\tau_s=1$ specific to horizon crossing. 
%which is appearing actually at the horizon crossing point.  
It helps us to convert the factor $(\tau_s/\tau_e)^{6}$ to $(k_e/k_s)^{6}$. Additionally,   it is important to note that $k_e$ and $k_s$ represent the wave numbers associated with the conformal time scale $\tau_e$ and $\tau_s$. 

\section{Computation of one-loop corrected scalar power spectrum}
\label{oneloop}

Let us now extend our analysis to  explicit computation of  one-loop correction to power spectrum from the scalar modes of the perturbation.  To perform this computation let us consider the representative action as stated in equation(\ref{model}), expanded to the third order in the scalar comoving curvature perturbation:
\bea S_{(3)}&=&\int d\tau\;  d^3x\;  {\cal L}_{\rm int}(\tau) .\eea
where the interaction Lagrangian density for the third order perturbation can be expressed as \cite{Maldacena:2002vr}:
\bea \label{third}  &&{\cal L}_{\rm int}(\tau)=M^2_{\rm pl}a^2\; \bigg(\left(\epsilon^2-\frac{1}{2}\epsilon^3\right)\zeta^{'2}\zeta+\epsilon^2\left(\partial_i\zeta\right)^2\zeta-2\epsilon\zeta^{'}\left(\partial_i\zeta\right)\left(\partial_i\partial^{-2}\left(\epsilon\zeta^{'}\right)\right)+\frac{1}{2}\epsilon\zeta\left(\partial_i\partial_j\partial^{-2}\left(\epsilon\zeta^{'}\right)\right)^2+\frac{1}{2}\epsilon\eta^{'}\zeta^{'}\zeta^{2}\bigg).\quad\quad\quad\eea
This action is commonly used to compute three point function and to study the corresponding primordial non-Gaussian effects. See refs.\cite{Creminelli:2005hu,Creminelli:2006gc,Cheung:2007st,Cheung:2007sv,Baumann:2009ds,Baumann:2011su,Senatore:2013roa,Choudhury:2012whm,Lee:2016vti,Senatore:2016aui,Flauger:2016idt,Bravo:2017wyw,Choudhury:2017glj,Cai:2018dkf,Baumann:2018muz,Meerburg:2019qqi,Kitajima:2021fpq,Meng:2022ixx,Matsubara:2022nbr,Gow:2022jfb,Baumann:2022mni} for more details. To study the one-loop quantum effects on the two-point function and the associated power spectrum, the same third order action is used.
In the standard SR inflation and during PBH formation, first three and last two terms have the contributions ${\cal O}(\epsilon^{2})$ and ${\cal O}(\epsilon^{3})$ respectively.   The contribution from the last term in both of these cases are significantly different.  In this context,  during the standard SR inflation and during PBH formation, last term contributed as ${\cal O}(\epsilon^{3})$ and ${\cal O}(\epsilon)$ respectively.  The prime reason of this fact is that the second slow roll parameter changes from $\eta\sim 0$ to $\eta\sim -6$. 

We now explicitly compute the contribution from the quantum one-loop correction to the power spectrum of the scalar mode during PBH formation from the last term of the third order expanded action as stated in equation(\ref{third}).  For this purpose, we use the well known {\it in-in formalism},  which is actually motivated by the {\it Schwinger-Keldysh path integral formalism}.  Within the framework of {\it in-in formalism},  the two-point correlation function for comoving curvature perturbations at the fixed conformal time scale $\tau$ can be expressed as:
\bea\label{op} &&\textcolor{black}{\langle\hat{\zeta}_{\bf p}\hat{\zeta}_{-{\bf p}}\rangle:=\left\langle\bigg[\overline{T}\exp\bigg(i\int^{\tau}_{-\infty}d\tau^{'}\;H_{\rm int}(\tau^{'})\bigg)\bigg]\;\;\hat{\zeta}_{\bf p}(\tau)\hat{\zeta}_{-{\bf p}}(\tau)\;\;\bigg[{T}\exp\bigg(-i\int^{\tau}_{-\infty}d\tau^{'}\;H_{\rm int}(\tau^{'})\bigg)\bigg]\right\rangle_{\tau\rightarrow 0}}\quad\quad\nonumber\\
&&\quad\quad\quad\quad\textcolor{black}{=\underbrace{\langle\hat{\zeta}_{\bf p}\hat{\zeta}_{-{\bf p}}\rangle_{(0,0)}}_{\bf Tree\;level}+\underbrace{\langle\hat{\zeta}_{\bf p}\hat{\zeta}_{-{\bf p}}\rangle_{(0,2)}+\langle\hat{\zeta}_{\bf p}\hat{\zeta}_{-{\bf p}}\rangle^{\dagger}_{(0,2)}+\langle\hat{\zeta}_{\bf p}\hat{\zeta}_{-{\bf p}}\rangle_{(1,1)}}_{\bf One-loop\;level}}.\eea
where $\overline{T}$ and $T$ represent the anti-time and time ordering operation in the present context respectively.  \textcolor{black}{In this expression, for a given mode, $\tau\rightarrow 0$ represents supper-Hubble radius limit at the end of inflation where the quantum correlation is meaningfully computed in the {\it in-in formalism}.} The interaction Hamiltonian appearing in the above expression can be computed by the following {\it Legendre transformed} expression, where we are interested in only on the contribution from last term in the equation(\ref{third}), which physically represents the leading cubic self-interaction:
\bea H_{\rm int}(\tau)=-\frac{M^2_{\rm pl}}{2}\int d^3x\;  a^2\epsilon\eta^{'}\zeta^{'}\zeta^{2}.\eea
The tree-level contribution to the two-point correlation function and the corresponding power spectrum have already been computed in this paper. Our job is to fix the non-vanishing contributions from the last three terms in the above-given \textcolor{black}{expression at the one-loop level as appearing in equation (\ref{op}). The meaning of each of the terms as appealing in equation (\ref{op}) is further discussed in detail in the Appendix \ref{App:1}.}  
 
 By following the {\it in-in formalism} the one-loop contribution to the power spectrum of the scalar perturbation can be computed as:
 \bea \bigg[\Delta^{2}_{\zeta, {\bf One-loop}}(p)\bigg]_{\bf USR\;on\;SR}=\frac{1}{4}\bigg[\Delta^{2}_{\zeta,{\bf Tree}}(p)\bigg]^2_{\bf SR}\bigg(\left(\Delta\eta(\tau_e)\right)^2\int^{k_e}_{k_s}\frac{dk}{k}\;\left|{\cal V}_{\bf k}(\tau_e)\right|^{2}-\left(\Delta\eta(\tau_s)\right)^2\int^{k_e}_{k_s}\frac{dk}{k}\;\left|{\cal V}_{\bf k}(\tau_s)\right|^{2}\bigg),\quad\quad\eea
 where we have introduced a momentum and conformal time dependent function which captures the contribution from the USR period and defined by the following expression:
 \bea {\cal V}_{\bf k}(\tau)&=&\left(\frac{\tau_s}{\tau}\right)^{3}\bigg[\alpha^{\rm USR}_{\bf k}\left(1+ik\tau\right)\; e^{-ik\tau}-\beta^{\rm USR}_{\bf k}\left(1-ik\tau\right)\; e^{ik\tau}\bigg].%\nonumber\\
% &=&
%\left\{
%	\begin{array}{ll}
%		\displaystyle\left(\frac{\tau_s}{\tau}\right)^{3}(ik\tau)\bigg[\alpha^{\rm USR}_{\bf k}\; e^{-ik\tau}+\beta^{\rm USR}_{\bf k}\; e^{ik\tau}\bigg]\; \quad\quad\quad\quad\quad\quad & \mbox{when sub-horizon}  \; \;(-k\tau\gg 1)  \\
%		\displaystyle \left(\frac{\tau_s}{\tau}\right)^{3}(-i)\bigg[\alpha^{\rm USR}_{\bf k}\; e^{i}+\beta^{\rm USR}_{\bf k}\; e^{-i}\bigg]& \mbox{at horizon-crossing} \; \; (-k\tau= 1)  \\
%	\end{array}
%\right. 
\eea
\textcolor{black}{Here $\bigg[\Delta^{2}_{\zeta,{\bf Tree}}(p)\bigg]_{\bf SR}$ represents the tree-level contribution to the power spectrum as appearing in the SR period. Further, it is important to note that, the symbols $\Delta\eta(\tau_s)$ and $\Delta\eta(\tau_e)$ characterizing the values of the second slow-roll parameters at the SR to USR transition time scale $\tau=\tau_s$ and at the end of USR as well as the end of inflation scale $\tau=\tau_e$ respectively. Here $\tau=\tau_s$ is the most important scale where sharp transition is implemented to describe SR to USR phase change. We refer the reader to Appendix \ref{App:1}}  for details.
Here the expressions for the Bogoliubov coefficients $\alpha^{\rm USR}_{\bf k}$ and $\beta^{\rm USR}_{\bf k}$ are explicitly written in equation(\ref{bo1}) and equation(\ref{bo2}).  Additionally it is important to note that,  we have restricted the momentum integration within a window,  $k_s<k<k_e$ by introducing two cut-offs,  \textcolor{black}{ the Infra-Red (IR) cut-off $k_{\rm IR}=k_s$ and the Ultra-Violate (UV) } cut-off $k_{\rm UV}=k_e$,  to extract the finite contributions from each of the above mentioned integrals.  Our further job is to evaluate both of the integrals within the mentioned momentum window. Again it is important to note that, the integral in the sub horizon region and at the horizon crossing scale has the quantum effects and consequently the one-loop contributions become physically meaningful in these above mentioned regions. It is expected that whatever result we derive at the horizon exit point will further propagate to the super horizon region. Now instead of computing two independent integrals we will compute a single momentum integral at the arbitrary conformal time scale $\tau$ and in the final result, we will substitute the values of the conformal time scale $\tau=\tau_s$ and $\tau=\tau_e$ as required by the above mentioned structure of the momentum integrals.  To serve the purpose let us compute the following integrals:
\bea 
\underline{\mbox{Sub-horizon}(-k\tau\gg 1)}: {\cal I}(\tau):&=&\int^{k_e}_{k_s}\frac{dk}{k}\;\left|{\cal V}_{\bf k}(\tau)\right|^{2}\nonumber\\
&\approx & \left(\frac{\tau_s}{\tau}\right)^{6}\tau^2\int^{k_e}_{k_s}dk\;k\;\bigg(1+\frac{9}{2}\frac{(1+k^2\tau^2_s)^2}{k^6\tau^6_s}\bigg)\nonumber\\
		&\approx& \left(\frac{\tau_s}{\tau}\right)^{6}\bigg[\frac{(k^2_e-k^2_s)\tau^2}{2}+\frac{9}{2}\left(\frac{\tau}{\tau_s}\right)^2\ln \left(\frac{k_e}{k_s}\right)\bigg].\;\quad\quad\\
		\nonumber\\ 		
\underline{\mbox{Horizon-crossing}(-k\tau= 1)}: {\cal I}(\tau):&=&\int^{k_e}_{k_s}\frac{dk}{k}\;\left|{\cal V}_{\bf k}(\tau)\right|^{2}\nonumber\\
&\approx & \left(\frac{\tau_s}{\tau}\right)^{6}\int^{k_e}_{k_s}\frac{dk}{k}\;\bigg(1+\frac{9}{2}\frac{(1+k^2\tau^2_s)^2}{k^6\tau^6_s}\bigg)\nonumber\\
		&\approx& \left(\frac{\tau_s}{\tau}\right)^{6}\ln \left(\frac{k_e}{k_s}\right),\;\eea
where we have considered the contributions which will only give rise to significant cut off dependent divergent contributions in the loop integral.  Additionally,  two more terms appear which captures the interference between the Bogoliubov coefficients,  giving rise to oscillating contribution and no significant divergences.  For this reason we have neglected these two contributions. \textcolor{black}{However, to maintain the overall clarity in the computation the appearance of all such contributions in the momentum dependent one-loop contribution in the USR phase are explicitly computed in Appendix \ref{App:B} from which one can clearly observe the corresponding suppression.}
 
Finally, the one-loop contribution to the power spectrum of the scalar perturbation due to the USR period on SR contribution can be expressed in the following simplified form in the sub horizon region considering the UV as well IR mode contributions:
%\bea \bigg[\Delta^{2}_{\zeta, {\bf One-loop}}(p)\bigg]_{\bf USR\;on\;SR}&=&\frac{1}{8}\bigg[\Delta^{2}_{\zeta,{\bf Tree}}(p)\bigg]^2_{\bf SR}\nonumber\\
%&& \quad\quad\times\Bigg(\left(\Delta\eta(\tau_e)\right)^2\left(\frac{k_e}{k_s}\right)^{6}\Bigg[\left(1-\left(\frac{k_s}{k_e}\right)^2\right)+ 9\left(\frac{k_s}{k_e}\right)^2\ln \left(\frac{k_e}{k_s}\right)\Bigg]\nonumber\\
% &&\quad\quad\quad\quad\quad+\left(\Delta\eta(\tau_s)\right)^2\Bigg[\left(1-\left(\frac{k_e}{k_s}\right)^2\right)- 9\ln \left(\frac{k_e}{k_s}\right)\Bigg]\Bigg),\eea 
% and at the  horizon crossing point as well as in the super horizon scale we have following expression:
 \bea \bigg[\Delta^{2}_{\zeta, {\bf One-loop}}(p)\bigg]_{\bf USR\;on\;SR}&=&\frac{1}{4}\bigg[\Delta^{2}_{\zeta,{\bf Tree}}(p)\bigg]^2_{\bf SR}\nonumber\\
&&\quad\quad\quad\times\Bigg(\Bigg(\left(\Delta\eta(\tau_e)\right)^2\left(\frac{k_e}{k_s}\right)^{6}-\left(\Delta\eta(\tau_s)\right)^2\Bigg)\Bigg(\ln \left(\frac{k_e}{k_s}\right)\textcolor{black}{+\frac{1}{2}\left(\frac{k_e}{k_s}\right)^2-\frac{1}{2}}\Bigg)-\textcolor{black}{c_{\bf USR}}\Bigg).\quad\quad\quad\eea
 Here we consider the wave number is lying within the window $k_s\leq p\leq k_e$.  \textcolor{black}{Also it is important to note that here $c_{\bf USR}$ is an arbitrary renormalization scheme dependent parameter which is appearing after the cancellation of the contributions from the UV divergences in the underlying theory in USR period.   In the next section once we fix the scheme of renormalization one can able to explicitly compute the expression for the expression for the parameter $c_{\bf USR}$ in the USR period.  Most importantly,  it is important to mention that this parameter $c_{\bf USR}$ will going to mimic the role of counter term during performing the renormalization for a specified given scheme.  In the next section we will evaluate this counter term to remove the contribution of the UV divergence from the short range mode in the one-loop contribution to the primordial power spectrum for scalar modes in the USR period. }
 
 Further considering the rest of the contributions in the interaction Hamiltonian and computed mode function for the scalar modes in the SR period, one can consider the following momentum integral,  in which after substituting the appropriate IR cut-off scale $k_{\rm IR}=p_{*}$ and UV cut-off scale $k_{\rm UV}=k_e$ we have the following regulated closed expression:
 \bea \int^{k_e}_{p_*}\frac{d^3k}{(2\pi)^3}\frac{1}{k^3}\left(1+k^2\tau^2\right)&=&\frac{1}{2\pi^2}\Bigg[\int^{k_e}_{p_*}\frac{dk}{k}+\tau^2\int^{k_e}_{p_*}dk \; k\Bigg]\nonumber\\
 &=&\frac{1}{2\pi^2}\Bigg[\ln\left(\frac{k_e}{p_*}\right)+\frac{\tau^2}{2}\left(k^2_e-p^2_*\right)\Bigg]\nonumber\\
&=&\frac{1}{2\pi^2}\Bigg[\ln\left(\frac{k_e}{p_*}\right)\textcolor{black}{+\frac{1}{2}\left(\frac{k_e}{p_*}\right)^2-\frac{1}{2}}\Bigg].\eea 
 Here $p_*$ represents the pivot scale which is expected to be $p_*\ll k_s$.  The last term is the outcome of the sub horizon scale computation.  \textcolor{black}{Now in the  limit,  $\tau\rightarrow 0$ which is actually representing the super-horizon scale $k\tau\ll -1$ (which corresponds to the small momentum modes or commonly identified as the IR modes),} the last term vanishes for which we get only logarithmically divergent contribution which will going to survive from this computation at the end and contribute to the one-loop correction to the scalar power spectrum in the SR period.  \textcolor{black}{However,  since inflation ends at the finite conformal time scale value it is expected to have some small contribution from the UV modes which basically represents the sub horizon fluctuations in the present context. For this reason such UV divergent contributions we have kept intact and in the later part of the paper by doing technical computation we will explicitly show that one can easily remove such contributions from the one-loop corrected power spectrum both from the SR and USR periods of inflation separately by implementing a special renormalization  scheme,  commonly known as the adiabatic or wave function renormalization technique,  which has the inherent power to remove the UV divergences from the theory.  However,  the IR logarthmic divergences cannot be completely removed after renormalization at the one-loop level.  We will further show by implementing the power spectrum renormalization method that in the present context the IR divergent contribution will be shifted to the next-to-next leading order,  which is exactly equivalent of performing the computation at the two-loop level.  In support of the previously mentioned statement regarding the softening of the IR divergent contribution from the one-loop result is quite understandable from the perspective of the Quantum Field Theory of de Sitter space. It is well known and established fact that de Sitter or its cosmologically relevant version (quasi de Sitter) is IR divergent always and it is almost impossible to remove such divergences from the underlying Quantum Field Theory with the help of presently available tools and techniques.  }The one-loop correction in the SR phase can be exprerssed by the following expression:
  \bea \bigg[\Delta^{2}_{\zeta, {\bf One-loop}}(p)\bigg]_{\bf SR}&=&\bigg[\Delta^{2}_{\zeta,{\bf Tree}}(p)\bigg]^2_{\bf SR}\Bigg(c_{\bf SR}-\frac{4}{3}\bigg(\ln \left(\frac{k_e}{p_*}\right)\textcolor{black}{+\frac{1}{2}\left(\frac{k_e}{p_*}\right)^2-\frac{1}{2}}\bigg)\Bigg),\eea
  \textcolor{black}{where $c_{\bf SR}$ is an arbitrary renormalization scheme dependent parameter which is appearing after the cancellation of the contributions from the UV divergences in the underlying theory in SR period.  Once we fix the scheme of renormalization one can able to explicitly compute the expression for the expression for the parameter $c_{\bf SR}$ in the SR period.  Most importantly,  it is important to mention that this parameter $c_{\bf SR}$ will going to mimic the role of counter term during performing the renormalization for a specified given scheme.  In the next section we will evaluate this counter term to remove the contribution of the UV divergence from the short range mode in the one-loop contribution to the primordial power spectrum for scalar modes in the SR period.  In order for the renormalization counter terms to effectively cancel the UV divergences that are encountered when calculating quantum corrections, the effective field theory must be defined to include renormalization counter terms in the bare effective Lagrangian.} Here in the SR region we consider the wave number $p\leq k_s$, which will be useful for rest of the analysis.  For more details see refs. \cite{Sloth:2006az,Seery:2007we,Seery:2007wf,Bartolo:2007ti,Senatore:2009cf,Seery:2010kh,Bartolo:2010bu,Senatore:2012ya,Senatore:2012nq,Pimentel:2012tw,Chen:2016nrs,Markkanen:2017rvi,Higuchi:2017sgj,Syu:2019uwx,Rendell:2019jnn,Cohen:2020php,Green:2022ovz,Premkumar:2022bkm}.
  
\textcolor{black}{Here it is important to note that the impacts of short range modes are contained in the UV component, which persists through the brief USR phase of inflation in a thermal vacuum inside a sub horizon domain. The related UV divergences are anticipated to be absorbed into relevant, physically observable amounts, and these modes should not contribute to the dynamics of the USR period during inflation.  }  
  
From the above calculations, one can observe the following points:
\begin{itemize}
\item In the sub horizon scale we have three possible contributions from the pure quantum effects,  scale independent terms, dimensionless quadratic divergent terms and the logarithmic divergent contributions,  which are the outcome of the one loop effects in the scalar power spectrum in the USR period.

\item \textcolor{black}{The UV divergences appearing in the one-loop contributions appearing in both SR and USR period due to having sub horizon quantum fluctuations of the short range modes.  Such contributions will be removed fully from the underlying Quantum Field Theoretic setup by implementing the adiabatic or wave function renormalization technique which we will explicitly demonstrate in the upcoming section of this paper.}
%At the horizon crossing point and in the super horizon region,  we have found that only one logarithmic divergent contribution is appearing in the USR period.

%\item The absence of the quadratically divergent UV contribution at the horizon crossing point in the USR period is the direct outcome of matching the spectrum obtained from sub horizon and super horizon region at that point.  \textcolor{red}{We will going to clarify this removal UV divergences in detail in the next section.}

\item The logarithmically divergent IR contribution appearing in the USR period is also appearing in the super horizon scale because the result obtained at the horizon crossing point is propagating outside the horizon and the corresponding modes become frozen and thus classical.  

\item From the observational perspective, the result obtained from the horizon crossing point is most significant as all the signatures from the quantum loop corrections appearing in the sub horizon scale can be directly tested.  Since, we have found that due to the matching condition at the horizon crossing, only the logarithmically divergent IR contributions survive. We will use this result to compute the other relevant quantities from the one loop corrected scalar power spectrum including the correction from the USR period \textcolor{black}{after performing adiabatic renormalization using which we completely remove UV divergences and power spectrum renormalization which smoothing out IR contributions by shifting it at the higher order in the perturbative expansion.}

\item %Matching the boundary condition at the horizon crossing scale also demands that the one loop correction to the scalar power spectrum during SR period gets sole contribution in the form of logarithmic divergence at the IR end. 
\textcolor{black}{The overall coefficient in front of this contribution and one additional additive parameter is fixed by the proper choice of the renormalization scheme,  which allows us to remove such UV divergent contribution by proper choice of the counter terms. This scheme is applicable to the one-loop contribution coming from the SR and USR period and such UV divergences can be removed by following the mentioned procedure.}

\item In the SR region the tree level and the one-loop contributions are computed for the wave number $p\leq k_s$.  On the other hand, in the USR region the tree level and one-loop contribution is computed within the window $k_s\leq p\leq k_e$. This is an extremely useful information to write down the expression for the total contribution of the one-loop corrected power spectrum for the scalar modes considering both the effects from SR and USR region respectively. The {\it raison d'etre} of this paper is: {\it  except logarithmically divergent contributions, no other contribution survive in the one loop contribution of the scalar power spectrum at the super horizon scale}.  This contribution is generated due to the quantum loop effects as appearing in the sub horizon scale and is going to survive at the horizon crossing point as well as in the super horizon scale. As an immediate consequence,  such contribution will appear in the computation of scalar spectral tilt,  which we are going to explicitly show in the section to follow. 

\end{itemize}

Before going to discuss any further issues, let us now write down the total expression for the one loop corrected scalar power spectrum, which will be helpful for the further discussions. The corresponding expression for the scalar power spectrum in the USR period when the PBH formation occurs can be written as:
\bea \Delta^{2}_{\zeta, {\bf Total}}(p)&=&  \Delta^{2}_{\zeta, {\bf Tree}}(p)+\Delta^{2}_{\zeta, {\bf One-loop}}(p)\nonumber\\
&=& \Bigg\{ \underbrace{\bigg[\Delta^{2}_{\zeta,{\bf Tree}}(p)\bigg]_{\bf SR}}_{\bf SR\;contribution\;for\;inflation}\nonumber\\
&&+ \underbrace{\bigg[\Delta^{2}_{\zeta, {\bf One-loop}}(p)\bigg]_{\bf SR}}_{\bf Sub-leading\;one-loop\;correction\;due\;to\;SR}\nonumber\\
&&+\underbrace{\bigg[\Delta^{2}_{\zeta,{\bf One-loop}}(p)\bigg]^{2}_{\bf USR\;on\;SR}}_{\bf Sub-leading\;one-loop\;correction\;due\;to\;USR\;on\;SR}\Theta(p-k_s)\Bigg\}.\nonumber\\
&=&\bigg[\Delta^{2}_{\zeta,{\bf Tree}}(p)\bigg]_{\bf SR}\Bigg\{1+\bigg[\Delta^{2}_{\zeta,{\bf Tree}}(p)\bigg]_{\bf SR}\Bigg(c_{\bf SR}-\frac{4}{3}\bigg(\ln \left(\frac{k_e}{p_*}\right)\textcolor{black}{+\frac{1}{2}\left(\frac{k_e}{p_*}\right)^2-\frac{1}{2}}\bigg)\bigg)\nonumber\\
&&+\frac{1}{4}\bigg[\Delta^{2}_{\zeta,{\bf Tree}}(p)\bigg]_{\bf SR}\Bigg(\Bigg(\left(\Delta\eta(\tau_e)\right)^2\left(\frac{k_e}{k_s}\right)^{6}-\left(\Delta\eta(\tau_s)\right)^2\Bigg)\Bigg(\ln \left(\frac{k_e}{k_s}\right)\textcolor{black}{+\frac{1}{2}\left(\frac{k_e}{k_s}\right)^2-\frac{1}{2}}\Bigg)-\textcolor{black}{c_{\bf USR}}\Bigg)\textcolor{black}{\Theta(p-k_s)}\Bigg\},\quad\quad\quad\eea
where the slow-roll (SR) contribution to the scalar power spectrum can be expressed as:
\bea \bigg[\Delta^{2}_{\zeta,{\bf Tree}}(p)\bigg]_{\bf SR}&=&\left(\frac{H^{2}}{8\pi^{2}M^{2}_{\rm pl}\epsilon}\right)\left(1+(p/k_s)^2\right)\quad\quad {\rm where}\quad p\leq k_s.\eea
\textcolor{black}{Here we have introduced a Heaviside Theta function at the transition point $\tau=\tau_s$ (SR to USR) \textcolor{black}{to describe the sharp transition}:
\bea  \Theta(p-k_s)
&=& \left\{
	\begin{array}{ll}
		0\quad\quad\quad\quad\quad\quad\quad\quad & \mbox{when }  p<k_s  \;(\rm SR)\\ \\
   \displaystyle 
			\displaystyle 1 & \mbox{when }  k_s\leq p< k_e  \;(\rm USR) 
	\end{array}
\right. \eea}

\section{Renormalization of one-loop corrected scalar power spectrum} 
\label{ren}

\subsection{Wave function/ Adiabatic renormalization method: Complete removal of quadratic UV divergence}

\textcolor{black}{In this subsection our prime objective is to remove the contributions from quadratic UV divergences as appearing from the one-loop computation of the primordial power spectrum for the comoving curvature perturbation both for the SR and USR periods.  In a more technical sense such UV divergences are originated from the sub horizon region ($-k\tau\gg 1$) where the effects of quantum mechanical fluctuations are dominant.  It is quite obvious that in the quasi de Sitter background only logarithmic divergent contributions survive at the late time limit where the scalar modes exited the horizon and goes to the super horizon region.  For this reason one can drop the contributions of the quadratic UV divergences at late times.  However,  it is important to systematically provide a technical scheme which allows us to do so in the present context.  In this article,  we have not dropped the contribution of quadratic UV divergence at the late time limit.  This section is devoted to provide the technical computation of a renormalization scheme which allows us automatically cancel the contribution from the quadratic divergence by adding an appropriate counter term.  Moreover,  since we have quantized the comoving curvature perturbation and treating it at the same footage of the scalar quantum field,  removal of the UV divergences of the underlying cosmological perturbation theory has to be done by implementing a smoothing scheme and adiabatic renormalization exactly serve the purpose of smoothing the effect of quantum fluctuations at the sub-horizon region where the short range UV modes are dominant.}

\textcolor{black}{In the present context we are dealing with the Quantum Field Theory of quasi de Sitter space where to remove the contributions from the UV divergences appearing at different powers the well known adiabatic regularization and the corresponding renormalization scheme play significant role.  For more details see the refs. \cite{Durrer:2009ii,Wang:2015zfa,Boyanovsky:2005sh,Marozzi:2011da,Finelli:2007fr,Ford:1977in,Parker:1974qw,Fulling:1974pu,Parker:2012at,Ford:1986sy} where the physical implications and various applications was explicitly discussed in detail.  Applying the minimal subtraction rule on the corresponding UV modes captures the sub horizon quantum fluctuations one can easily remove the contributions of the UV divergences from the the underlying Quantum Field Theoretic set up.  Here the minimal subtraction technically corresponds to the adding a counter term in the underlying theory to remove the UV divergent contribution from the short range UV modes dominant in the sub horizon region.   In this particular computation the second order subtraction is sufficient enough to remove the effects from quadratic UV divergences as appearing from both the SR and USR periods.  Going beyond the second order minimal subtraction of the counter terms are important only when we are dealing with the contributions of the UV divergent terms appearing with higher powers.  Fortunately in the present computation we don't need to think about beyond second order counter terms as it has the sufficient power to remove the contribution of the quadratic divergences by applying the adiabatic renormalization scheme.  Within the framework of Quantum Field Theory curved space-time,  particularly in quasi de Sitter background the adiabatic renormalization technique has been designed aiming to remove the UV divergences only,  and using such scheme IR divergences cannot be treated at all.  See more refs. \cite{Durrer:2009ii,Wang:2015zfa,Boyanovsky:2005sh,Marozzi:2011da,Finelli:2007fr,Ford:1977in,Parker:1974qw,Fulling:1974pu,Parker:2012at,Ford:1986sy} on this issue where such possibilities have been discussed in detail.  Here it is important to note that,  the  adiabatic renormalization scheme actually renormalizes the comoving curvature perturbation modes directly in the adiabatic limit of the cosmological perturbation.  This is exactly similar like wave function renormalization which we usually perform within the framework of Quantum Field Theory.  Such renormalised modes or the wave function further renormalizes the one-loop contribution to the cosmological power spectrum in the present context.  Another important thing we need to point in the present context that,  WKB approximation method play significant role which helps to construct a general form of the mode function or more precisely the regularized wave function in the adiabatic limit.  Consequently,  the regularized version of the mode function can able to completely remove the contributions of UV divergences from the short range modes.  Such a structure of the regularized modes are general and designed in such a fashion that it can able to perfectly take care of UV divergences at any arbitrary order without loosing any generality of the underlying Quantum Field Theoretical set up.}

\textcolor{black}{Let us first consider that in the adiabatic limiting situation the UV divergences are appearing at the $n$-th power which basically represents the $n$-th order UV divergences appearing from short range scalar modes.  With the help of the previously mentioned well known WKB approximation method the renormalized mode function for the comoving scalar curvature perturbation for the SR period can be expressed by the following expressions:}
\bea &&\textcolor{black}{\zeta^{(n),{\rm SR}}_{k}(\tau)=-\frac{H\tau}{M_{\rm pl}\sqrt{2\epsilon}}\frac{1}{\sqrt{2{\cal W}^{(n)}_{k}(\tau)}}\exp\bigg(-i\int^{\tau}_{\tau_0}d\tau_1\;{\cal W}^{(n)}_{k}(\tau_1)\bigg)}\nonumber\\
&&\quad\quad\quad\quad\; \textcolor{black}{=-\frac{H\tau}{2M_{\rm pl}\sqrt{\epsilon}\sqrt{{\cal W}^{(n)}_{k}(\tau)}}\exp\bigg(-i\int^{\tau}_{\tau_0}d\tau_1\;{\cal W}^{(n)}_{k}(\tau_1)\bigg)},\eea
\textcolor{black}{where ${\cal W}^{(n)}_{k}(\tau)$ is the characteristic function which is defined for the adiabatic $n$-th order by the following expression \cite{Wang:2015zfa}:}
\bea \textcolor{black}{{\cal W}^{(n)}_{k}(\tau)=\sqrt{\bigg(k^2-\frac{z^{''}}{z}\bigg)-\frac{1}{2}\Bigg[\frac{{\cal W}^{''(n-2)}_{k}(\tau)}{{\cal W}^{(n-2)}_{k}(\tau)}-\frac{3}{2}\left(\frac{{\cal W}^{'(n-2)}_{k}(\tau)}{{\cal W}^{(n-2)}_{k}(\tau)}\right)^2\Bigg]}\quad\quad{\rm where}\quad\quad\frac{z^{''}}{z}\approx \frac{2}{\tau^2}}.\eea
\textcolor{black}{Here $'$ denotes the conformal time derive.} \textcolor{black}{The above mentioned characteristic function physically represents the adibaticallay regularized conformal time dependent frequency factor appearing at the $n$-th order.  Additionally,  it is important to mention that to define the adiabatic mode at the $n$-th order in the SR period of inflation for comoving curvature perturbation we have used the Bunch Davies quantum initial condition which fixes the Bogoliubov coefficients at the values as stated in equation (\ref{SRBogo}).  }

\textcolor{black}{By following the same logical argument one can further write down the expression for the adiabatic mode at the $n$-th order in the USR period of inflation for comoving curvature perturbation,  which is given by:}
\bea &&\textcolor{black}{\zeta^{(n),{\rm USR}}_{k}(\tau)=-\frac{H\tau}{M_{\rm pl}\sqrt{2\epsilon}}\left(\frac{\tau_0}{\tau}\right)^3\frac{1}{\sqrt{2{\cal W}^{(n)}_{k}(\tau)}}\Bigg[\alpha^{\rm USR}_{\bf k}\exp\bigg(-i\int^{\tau}_{\tau_0}d\tau_1\;{\cal W}^{(n)}_{k}(\tau_1)\bigg)+\beta^{\rm USR}_{\bf k}\exp\bigg(i\int^{\tau}_{\tau_0}d\tau_1\;{\cal W}^{(n)}_{k}(\tau_1)\bigg)\Bigg]}\nonumber\\
&&\quad\quad\quad\quad\;\;\; \textcolor{black}{=-\frac{H\tau}{2M_{\rm pl}\sqrt{\epsilon}\sqrt{{\cal W}^{(n)}_{k}(\tau)}}\left(\frac{\tau_0}{\tau}\right)^3\Bigg[\alpha^{\rm USR}_{\bf k}\exp\bigg(-i\int^{\tau}_{\tau_0}d\tau_1\;{\cal W}^{(n)}_{k}(\tau_1)\bigg)+\beta^{\rm USR}_{\bf k}\exp\bigg(i\int^{\tau}_{\tau_0}d\tau_1\;{\cal W}^{(n)}_{k}(\tau_1)\bigg)\Bigg]},\quad\quad\eea
\textcolor{black}{In the above mentioned expression, the adiabatically regularized modes at the $n$-th order Bogoliubov coefficients in the USR period are characterized by $\alpha^{\rm USR}_{\bf k}$ and $\beta^{\rm USR}_{\bf k}$.  One may think that due to the adiabaticity in the scalar modes, the expressions for the Bogolibov coefficients may differ from the results explicitly computed in equation (\ref{bo1}) and equation (\ref{bo2}) respectively in the USR period.  However,  it is also expected that due to the very small change in the adiabatic limit these Bogoliubov coefficients in the USR period will not significantly differ from the results that we have computed in equations (\ref{bo1}) and equation (\ref{bo2}).  This is quite physically justifiable and fully acceptable as far as the validity of the adiabatic regularization in the corresponding mode at the $n$-th order is concerned. Though we are not going to use the explicit expression for the Bogoliubov coefficients in the USR period, before carrying forward the rest of the computation, it is better to clarify this point.  During the computation of the counter term as appearing in the renormalized version of the one-loop power spectrum in the USR period of inflation we will explicitly show that the final result will be insensitive to the explicit structure of the Bogoliubov coefficients $\alpha^{\rm USR}_{\bf k}$ and $\beta^{\rm USR}_{\bf k}$ as it will not affect the short range UV modes in the corresponding computation much.  We will clarify this point in the rest of the computation performed in this section for the USR period. For more physical justification, one can argue here from the point of view of the shifting of the quantum initial condition and the shifting of the corresponding quantum vacuum state in the USR period compared to the Bunch Davies initial state that we have in the he SR period of inflation.  If the shifted initial vacuum as appearing in the USR period is significantly different in the adiabatic limiting situation from that we obtained originally by exactly solving the {\it Mukhanov Sasaki equation}, then it will go against the basic understanding and the applicability of the adiabatic regularization scheme itself.  The adiabatic limiting approximation in therms of the WKB regularized modes are implemented by imposing the constraint that adiabaticity will not shift the structure of the underlying quantum vacuum state in the USR period of inflation.  Strong adiabaticity limit confirms the uniqueness of the quantum vacuum state and the underlying quantum initial condition in the USR period of inflation in the present computation.  For this reason, one can take the structure of the Bogoliubov coefficients $\alpha^{\rm USR}_{\bf k}$ and $\beta^{\rm USR}_{\bf k}$ as derived in equation (\ref{bo1}) and equation (\ref{bo2}) respectively, without having any further confusion from the underlying physical point of view.  This is the inherent power of the Quantum Field Theory of curved space-time,  which is written in the background of quasi de Sitter space, and in the present computation we are fully utilizing this fact.  Additionally, we need to point out that such a possibility does not appear in the context of the SR period of inflation as we have fixed the quantum initial condition to be Bunch Davies and due to this fixed choice, the structure of the Bogolibov coefficient is also fixed,  which will not change if we consider the adiabatically regularized mode or the exact mode that we have computed by solving the  {\it Mukhanov Sasaki } equation.}

\textcolor{black}{Before going into the further details of the technical computation, let us first mention some of the important facts that will be extremely useful for the rest of analysis performed in this section as well as in the rest of the paper.  In the previous section,  during the computation of on-loop effects in the primordial power spectrum for the comoving curvature perturbation, we introduced two renormalization scheme dependent parameters,  $c_{\bf SR}$ and $c_{\bf USR}$,  which we have not been able to fix as the renormalization scheme was not fixed before. It can be understood from the computed structure of the one-loop correction to the primordial power spectrum for scalar modes that it will play the role of counter terms, which helps us to nullify the contribution of the UV and IR divergences.  Once we fix the renormalization scheme to be adiabatic in the present context of discussion,  then automatically one can explicitly compute the expression for the counter terms $c_{\bf SR}$ and $c_{\bf USR}$ in the SR and USR periods of inflation. We will explicitly perform this computation in this section, using which we can show the complete removal of the contributions of the short range modes in the quadratic UV divergence. Using the present scheme of renormalization, IR logarithmic divergences cannot be removed from the underlying theory, and we will demonstrate this possibility in the computation performed in this section.} 

\textcolor{black}{Initially, we demonstrated the generalized structure of the scalar modes in the SR as well as in the USR periods of inflation by considering the fact that the adiabatic limit is applicable to the $n$-th order and that it can be used to remove the $n$-th power of UV divergences from the short range mode contributions.  But we already know from the computation performed in the previous section to demonstrate the total one-loop corrected primordial power spectrum  for scalar modes that UV divergences appear in the SR and USR periods of inflation with quadratic power. For this reason, utilising the technical structure of adiabatic regularization we restrict our rest of the computation to the second order,  i.e.  we will fix $n=2$ which will be helpful to remove the contribution of the quadratic divergences from the one-loop correction terms fully both from the SR and USR periods of inflation.  With the above mentioned restrictions, the adiabatically regularized WKB approximated modes for comoving scalar curvature perturbation can be further simplified in the SR and USR periods of inflation by the following expressions:}
\bea &&\textcolor{black}{\zeta^{(2),{\rm SR}}_{k}(\tau)=-\frac{H\tau}{2M_{\rm pl}\sqrt{\epsilon}\sqrt{{\cal W}^{(2)}_{k}(\tau)}}\exp\bigg(-i\int^{\tau}_{\tau_0}d\tau_1\;{\cal W}^{(2)}_{k}(\tau_1)\bigg)},\\
&&\textcolor{black}{\zeta^{(2),{\rm USR}}_{k}(\tau)=-\frac{H\tau}{2M_{\rm pl}\sqrt{\epsilon}\sqrt{{\cal W}^{(2)}_{k}(\tau)}}\left(\frac{\tau_0}{\tau}\right)^3\Bigg[\alpha^{\rm USR}_{\bf k}\exp\bigg(-i\int^{\tau}_{\tau_0}d\tau_1\;{\cal W}^{(2)}_{k}(\tau_1)\bigg)+\beta^{\rm USR}_{\bf k}\exp\bigg(i\int^{\tau}_{\tau_0}d\tau_1\;{\cal W}^{(2)}_{k}(\tau_1)\bigg)\Bigg]},\quad\quad\eea
\textcolor{black}{where ${\cal W}^{(2)}_{k}(\tau)$ is the characteristic conformal time dependent frequency function which is defined for the adiabatic $2$-nd order by the following expression:}
\bea \textcolor{black}{{\cal W}^{(2)}_{k}(\tau)=\sqrt{k^2-\frac{z^{''}}{z}}\approx\sqrt{k^2-\frac{2}{\tau^2}}\approx k}.\eea  
\textcolor{black}{In the above mentioned expression,  in the last step, we have considered the contribution from short range UV mode frequency by taking the limit $-k\tau\rightarrow \infty$.  This approximation is quite justifiable in the limit where adiabatic regularization holds perfectly. In the presence of the above mentioned adiabatically regularized scalar mode function for the comoving curvature perturbation, one can further compute the expression for the counter terms for the SR and USR periods of inflation explicitly.  Before going to discuss the technical computation,  let us finally write down the expression for the scalar modes for comoving curvature perturbation in the SR and USR periods of inflation, which we will use to compute the counter terms:}    
\bea &&\textcolor{black}{\zeta^{(2),{\rm SR},{\bf UV}}_{k}(\tau)\approx-\frac{H\tau}{2M_{\rm pl}\sqrt{\epsilon}\sqrt{k}}\exp\bigg(-ik\left(\tau-\tau_0\right)\bigg)},\\
&&\textcolor{black}{\zeta^{(2),{\rm USR},{\bf UV}}_{k}(\tau)\approx-\frac{H\tau}{2M_{\rm pl}\sqrt{\epsilon}\sqrt{k}}\left(\frac{\tau_0}{\tau}\right)^3\Bigg[\alpha^{\rm USR}_{\bf k}\exp\bigg(-ik\left(\tau-\tau_0\right)\bigg)+\beta^{\rm USR}_{\bf k}\exp\bigg(ik\left(\tau-\tau_0\right)\bigg)\Bigg]}.\quad\quad\eea
\textcolor{black}{Now in the SR and USR periods of inflation we introduce two counter terms $Z^{\bf SR,  UV}_{\zeta}(\mu,\mu_0)$ and $Z^{\bf USR, UV}_{\zeta}(\mu,\mu_0)$  related to the adiabatic renormalization scheme dependent parameters $c_{\bf SR}(\mu,\mu_0)$ and $c_{\bf USR}(\mu,\mu_0)$ by the following expressions\footnote{\textcolor{black}{ Here $\mu~\&~\mu_0$ refer to the points of subtraction.}}:}
\bea &&\label{co1}\textcolor{black}{Z^{\bf SR,  UV}_{\zeta}(\mu,\mu_0):=\bigg[\Delta^{2}_{\zeta,{\bf Tree}}(p)\bigg]^2_{\bf SR}\times c_{\bf SR}(\mu,\mu_0)},\\
&&\label{co2}\textcolor{black}{Z^{\bf USR,  UV}_{\zeta}(\mu,\mu_0):=\frac{1}{4}\bigg[\Delta^{2}_{\zeta,{\bf Tree}}(p)\bigg]^2_{\bf SR}\times c_{\bf USR}(\mu,\mu_0)} ,\eea
\textcolor{black}{where the prefactors are computed by collecting the terms in a controlled fashion and by performing all possible wick contractions as appearing in the one-loop correction of the primordial power spectrum. Our job is to further determine the expressions for $c_{\bf SR}(\mu,\mu_0)$ and $c_{\bf USR}(\mu,\mu_0)$,  which can be fixed by computing the following expressions in the SR and USR periods of inflation. Here it is additionally important to note that the counter terms as well as the adiabatic renormalization scheme dependent parameters are all functions of a new mass scale $\mu$,  which is in principle is interpreted as the renormalization scale of the underlying quantum field theory on the backdrop of which we are performing all the computation.  Additionally, it is important to note that the other mass scale $\mu_0$ appearing at the conformal time scale $\tau_0$ is interpreted at the reference level with respect which we implement the adiabatic regularization at the second order.  Next, we will explicitly compute the dependence on the renormalization scale $\mu$ by evaluating the following useful expressions at the SR and USR periods, respectively:} 
\bea && \textcolor{black}{ c_{\bf SR}(\mu,\mu_0):=\int^{\mu}_{\mu_0}dk\; k^2\; \frac{\tau^2}{k}=\frac{1}{2}\left(\mu^2-\mu^2_0\right)\tau^2=\frac{1}{2}\bigg[\left(\frac{\mu}{\mu_0}\right)^2-1\bigg]},\\
 && \textcolor{black}{ c_{\bf USR}(\mu,\mu_0):=\Bigg(\left(\Delta\eta(\tau)\right)^2\left(\frac{\mu}{\mu_0}\right)^{6}-\left(\Delta\eta(\tau_0)\right)^2\Bigg)\int^{\mu}_{\mu_0}dk\; k^2\; \frac{\tau^2}{k}=\frac{1}{2}\Bigg(\left(\Delta\eta(\tau)\right)^2\left(\frac{\mu}{\mu_0}\right)^{6}-\left(\Delta\eta(\tau_0)\right)^2\Bigg)\bigg[\left(\frac{\mu}{\mu_0}\right)^2-1\bigg]}.\quad\quad\quad
\eea  
\textcolor{black}{After substituting the above mentioned computed expressions for $c_{\bf SR}(\mu,\mu_0)$ and $c_{\bf USR}(\mu,\mu_0)$ at arbitrary renormalization scale $\mu$ in equation (\ref{co1}) and equation (\ref{co2}),  the corresponding counter terms of the one-loop primordial power spectrum for the scalar modes can be further expressed as:}
\bea &&\label{co11}\textcolor{black}{Z^{\bf SR,  UV}_{\zeta}(\mu,\mu_0):=\frac{1}{2}\times\bigg[\Delta^{2}_{\zeta,{\bf Tree}}(p)\bigg]^2_{\bf SR}\times \bigg[\left(\frac{\mu}{\mu_0}\right)^2-1\bigg]},\\
&&\label{co22}\textcolor{black}{Z^{\bf USR,  UV}_{\zeta}(\mu,\mu_0):=\frac{1}{8}\times\bigg[\Delta^{2}_{\zeta,{\bf Tree}}(p)\bigg]^2_{\bf SR}\times \Bigg(\left(\Delta\eta(\tau)\right)^2\left(\frac{\mu}{\mu_0}\right)^{6}-\left(\Delta\eta(\tau_0)\right)^2\Bigg)\bigg[\left(\frac{\mu}{\mu_0}\right)^2-1\bigg]}.\eea
\textcolor{black}{HHere it is important to note that the renormalization scale $\mu$ and the reference level $\mu_0$ can be anything that is physically viable and properly justifiable in the present scenario,  particularly within the framework of Quantum Field Theory curved space-time written in the backdrop of de Sitter space.  In the presence of the above mentioned arbitrary renormalization scale $\mu$ and the corresponding reference scale $\mu_0$ the adiabatically renormalized one-loop amplitude of the primordial power spectrum of the scalar modes corresponding to the comoving scalar curvature perturbation after adding the counter terms can be expressed in the SR and USR periods of inflation as:}
 \bea &&\textcolor{black}{\bigg[\Delta^{2}_{\zeta, {\bf One-loop}}(p,\mu,\mu_0)\bigg]_{\bf SR}=\bigg[\Delta^{2}_{\zeta,{\bf Tree}}(p)\bigg]^2_{\bf SR}\Bigg(\frac{1}{2}\bigg[\left(\frac{\mu}{\mu_0}\right)^2-\left(\frac{k_e}{p_*}\right)^2\bigg]-\frac{4}{3}\ln \left(\frac{k_e}{p_*}\right)\Bigg)},\\
 &&\textcolor{black}{\bigg[\Delta^{2}_{\zeta, {\bf One-loop}}(p,\mu,\mu_0)\bigg]_{\bf USR\;on\;SR}=\frac{1}{4}\bigg[\Delta^{2}_{\zeta,{\bf Tree}}(p)\bigg]^2_{\bf SR}}\nonumber\eea\bea
 &&\quad\quad\quad\quad\quad\quad\quad\quad\quad\quad\quad\quad\quad
 \quad\quad\textcolor{black}{\times\Bigg\{\Bigg(\left(\Delta\eta(\tau_e)\right)^2\left(\frac{k_e}{k_s}\right)^{6}-\left(\Delta\eta(\tau_s)\right)^2\Bigg)\Bigg(\ln \left(\frac{k_e}{k_s}\right)+\frac{1}{2}\bigg[\left(\frac{k_e}{k_s}\right)^2-1\bigg]\Bigg)}\nonumber\\
&&\quad\quad\quad\quad\quad\quad\quad\quad\quad\quad\quad\quad\quad\quad\quad\quad\quad\quad\quad\quad\quad\quad\quad\quad
 \quad\quad\textcolor{black}{-\frac{1}{2}\Bigg(\left(\Delta\eta(\tau)\right)^2\left(\frac{\mu}{\mu_0}\right)^{6}-\left(\Delta\eta(\tau_0)\right)^2\Bigg)\bigg[\left(\frac{\mu}{\mu_0}\right)^2-1\bigg]\Bigg\}}.\quad\quad\quad\eea
 \textcolor{black}{which will further give rise to the following expression for the total one-loop corrected primordial power spectrum for the scalar modes:}
 \bea &&\textcolor{black}{\Delta^{2}_{\zeta, {\bf Total}}(p,\mu,\mu_0)
=\bigg[\Delta^{2}_{\zeta,{\bf Tree}}(p)\bigg]_{\bf SR}\Bigg\{1+\bigg[\Delta^{2}_{\zeta,{\bf Tree}}(p)\bigg]_{\bf SR}\Bigg(\frac{1}{2}\bigg[\left(\frac{\mu}{\mu_0}\right)^2-\left(\frac{k_e}{p_*}\right)^2\bigg]-\frac{4}{3}\ln \left(\frac{k_e}{p_*}\right)\Bigg)}\nonumber\\
&&\quad\quad\quad\quad\quad\quad\quad\quad\quad\quad\quad\quad\textcolor{black}{+\frac{1}{4}\bigg[\Delta^{2}_{\zeta,{\bf Tree}}(p)\bigg]_{\bf SR}\Bigg(\Bigg(\left(\Delta\eta(\tau_e)\right)^2\left(\frac{k_e}{k_s}\right)^{6}-\left(\Delta\eta(\tau_s)\right)^2\Bigg)\Bigg(\ln \left(\frac{k_e}{k_s}\right)+\frac{1}{2}\bigg[\left(\frac{k_e}{k_s}\right)^2-1\bigg]\Bigg)}\nonumber\\
&&\quad\quad\quad\quad\quad\quad\quad\quad\quad\quad\quad
 \quad\quad\textcolor{black}{-\frac{1}{2}\Bigg(\left(\Delta\eta(\tau)\right)^2\left(\frac{\mu}{\mu_0}\right)^{6}-\left(\Delta\eta(\tau_0)\right)^2\Bigg)\bigg[\left(\frac{\mu}{\mu_0}\right)^2-1\bigg]\Bigg)\Theta(p-k_s)\Bigg\}},\quad\quad\quad\eea

\textcolor{black}{Here it is important to note that the mentioned adiabatically renormalized results obtained for one-loop correction of the primordial power spectrum of the scalar modes clearly suggest that it basically smooths the behaviour of quadratic UV divergence originated from short range modes, and IR logarithmic divergences, which capture the long range behaviour of scalar modes, are completely unaffected due to the inclusion of such techniques in both SR and USR periods of inflation.  As a result, we have obtained a UV protected coarse-grained version of quantum field theory that is IR sensitive.  In the present computation, such IR behaviour of the underlying theory is strongly associated with the validity of the perturbativity, which we need to ensure always holds perfectly to perform the rest of the computation performed in this paper and to extract the physically relevant information from the present analysis.  Though some arbitrariness is inherent in the choice of the renormalization scale $\mu$ and associated reference scale $\mu_0$,  but for the practical purposes to validate the applicability of course graining and perturbativity throughout the computation performed in this paper it is always preferred to fix such scales in the vicinity of UV cut off scale $\Lambda_{\rm UV}=k_e$ for $\mu$ (both in the phases SR and USR) and IR cut-off scale $\Lambda_{\rm IR}=p_*$ (in SR) and $\Lambda_{\rm IR}=k_s$ (in USR) for $\mu_0$.  By following this requirement we consider a specific situation where we fix the UV cut-off scale at $\Lambda_{\rm UV}=k_e=\mu$ (for SR and USR) and IR cut-off scale at $\Lambda_{\rm IR}=p_*=\mu_0$ (for SR) and $\Lambda_{\rm IR}=k_s=\mu_0$ (for USR). Consequently, the adiabatically renormalized one-loop correction to the power spectrum in the SR and USR periods of inflation can be further simplified as:}
 \bea &&\textcolor{black}{\bigg[\Delta^{2}_{\zeta, {\bf One-loop}}(p,\mu=\Lambda_{\rm UV}=k_e,\mu_0=\Lambda_{\rm IR}=p_*)\bigg]_{\bf SR}=-\frac{4}{3}\bigg[\Delta^{2}_{\zeta,{\bf Tree}}(p)\bigg]^2_{\bf SR}\ln \left(\frac{k_e}{p_*}\right)},\\
 &&\textcolor{black}{\bigg[\Delta^{2}_{\zeta, {\bf One-loop}}(p,\mu=\Lambda_{\rm UV}=k_e,\mu_0=\Lambda_{\rm IR}=k_s)\bigg]_{\bf USR\;on\;SR}=\frac{1}{4}\bigg[\Delta^{2}_{\zeta,{\bf Tree}}(p)\bigg]^2_{\bf SR}}\nonumber\\
 &&\quad\quad\quad\quad\quad\quad\quad\quad\quad\quad\quad
 \quad\quad\quad\quad\quad\quad\quad\quad\quad
 \quad\quad\quad\quad\quad\quad\quad\quad\quad\quad\textcolor{black}{\times\Bigg(\left(\Delta\eta(\tau_e)\right)^2\left(\frac{k_e}{k_s}\right)^{6}-\left(\Delta\eta(\tau_s)\right)^2\Bigg)\ln \left(\frac{k_e}{k_s}\right)}.\quad\quad\quad\eea
 \textcolor{black}{Hence the final form of the total one-loop corrected primordial power spectrum for the scalar modes can be further simplified as:}
 \bea &&\textcolor{black}{\Delta^{2}_{\zeta, {\bf Total}}(p)
=\bigg[\Delta^{2}_{\zeta,{\bf Tree}}(p)\bigg]_{\bf SR}\Bigg\{1-\frac{4}{3}\bigg[\Delta^{2}_{\zeta,{\bf Tree}}(p)\bigg]_{\bf SR}\ln \left(\frac{k_e}{p_*}\right)}\nonumber\\
&&\quad\quad\quad\quad\quad\quad\quad\quad\quad\quad\quad\quad\textcolor{black}{+\frac{1}{4}\bigg[\Delta^{2}_{\zeta,{\bf Tree}}(p)\bigg]_{\bf SR}\Bigg(\left(\Delta\eta(\tau_e)\right)^2\left(\frac{k_e}{k_s}\right)^{6}-\left(\Delta\eta(\tau_s)\right)^2\Bigg)\ln \left(\frac{k_e}{k_s}\right)\Theta(p-k_s)\Bigg\}},\quad\quad\quad\eea
 
 \textcolor{black}{The above quoted results obtained in both the SR and USR periods of inflation suggest that in the final expressions for the one-loop correction, the quadratic UV divergence is completely removed due to having maximum course graining on the renormalization scales $\mu$ and $\mu_0$,  which further helps us to maintain the maximum amount of perturbativity in our computation. However,  such course graining,  which is the natural biproduct of the adiabatic renormalization scheme, could not soften or remove the sensitivity of the logarithmic IR divergence. In the next section, we will discuss all the technical details of the procedure for softening logarithmic IR divergence both in the context of Sr and USR periods of inflation. }

\subsection{Power spectrum renormalization method: Softening of logarithmic IR divergence}

\textcolor{black}{Now to softening all the effects coming from one-loop logarithmic IR divergences from SR as well as the USR effects, on SR period, we further define the following renormalized power spectrum for the scalar perturbation} \footnote{\textcolor{black}{See the appendix \ref{App:R} for the more details on the justification of using two renormalization schemes under a single theoretical framework.}}:
\bea  \overline{\Delta^{2}_{\zeta,{\bf Total}}(p)}&=&{\bf Z}^{\rm IR}\Delta^{2}_{\zeta, {\bf Total}}(p),\eea
where ${\bf Z}^{\bf IR}$ is the renormalization factor,  commonly known as the {\it counter-term} and is determined by the explicit renormalization condition.  \textcolor{black}{This another multiplicative counter term is introduced to softening the contribution of logarithmic IR divergence in the present computation.  Here one need to compute the expression of the {\it counter term} from the underlying theoretical set up.  In the present framework the corresponding renormalization condition is fixed at the pivot scale $p_*$,  which is given by the following expression which helps us to soften the sensitivity of IR divergence:}
\bea \overline{\Delta^{2}_{\zeta,{\bf Total}}(p_*)}&=& \bigg[\Delta^{2}_{\zeta,{\bf Tree}}(p_*)\bigg]_{\bf SR},\eea
using which the {\it counter term},  ${\bf Z}^{\bf IR}$ can be computed as:
\bea  &&\textcolor{black}{{\bf Z}^{\bf IR}=\Bigg\{1-\frac{4}{3}\bigg[\Delta^{2}_{\zeta,{\bf Tree}}(p_*)\bigg]_{\bf SR}\ln \left(\frac{k_e}{p_*}\right)}\nonumber\\
&&\quad\quad\quad\quad\quad\quad\quad\quad\textcolor{black}{+\frac{1}{4}\bigg[\Delta^{2}_{\zeta,{\bf Tree}}(p_*)\bigg]_{\bf SR}\Bigg(\left(\Delta\eta(\tau_e)\right)^2\left(\frac{k_e}{k_s}\right)^{6}-\left(\Delta\eta(\tau_s)\right)^2\Bigg)\ln \left(\frac{k_e}{k_s}\right)\Bigg\}^{-1}}.\quad\quad\quad\eea
Then the corresponding one-loop corrected \textcolor{black}{IR softened one-loop renormalized power spectrum} for the scalar modes can be expressed as:
\bea  &&\textcolor{black}{\overline{\Delta^{2}_{\zeta,{\bf Total}}(p)}=\bigg[\Delta^{2}_{\zeta,{\bf Tree}}(p)\bigg]_{\bf SR}\Bigg\{1-\frac{4}{3}\bigg[\Delta^{2}_{\zeta,{\bf Tree}}(p_*)\bigg]_{\bf SR}\ln \left(\frac{k_e}{p_*}\right)}\nonumber\\
&&\quad\quad\quad\quad\quad\quad\quad\quad\textcolor{black}{+\frac{1}{4}\bigg[\Delta^{2}_{\zeta,{\bf Tree}}(p_*)\bigg]_{\bf SR}\Bigg(\left(\Delta\eta(\tau_e)\right)^2\left(\frac{k_e}{k_s}\right)^{6}-\left(\Delta\eta(\tau_s)\right)^2\Bigg)\ln \left(\frac{k_e}{k_s}\right)\Bigg\}^{-1}}\nonumber\\
&&\quad\quad\quad\quad\quad\quad\quad\quad\textcolor{black}{\times\Bigg\{1-\frac{4}{3}\bigg[\Delta^{2}_{\zeta,{\bf Tree}}(p)\bigg]_{\bf SR}\ln \left(\frac{k_e}{p_*}\right)}\nonumber\\
&&\quad\quad\quad\quad\quad\quad\quad\quad\textcolor{black}{+\frac{1}{4}\bigg[\Delta^{2}_{\zeta,{\bf Tree}}(p)\bigg]_{\bf SR}\Bigg(\left(\Delta\eta(\tau_e)\right)^2\left(\frac{k_e}{k_s}\right)^{6}-\left(\Delta\eta(\tau_s)\right)^2\Bigg)\ln \left(\frac{k_e}{k_s}\right)\Bigg\}}.\eea
From the derived structure of the one-loop corrected renormalized power spectrum for the scalar modes one can immediate conclude that no information of the quantum loop correction will be propagated to the pivot scale $p_*$,  where CMB observation takes place.  As an immediate consequence in any of the flow of the power spectrum (in the language of quantum field theory,  one can say the Renormalization Group (RG) flow equations and the corresponding $\beta$- functions),  which are spectral tilt,  running and running of the running of the tilt,  are going to be completely independent of the quantum loop effects at the pivot scale $p_*$.  \textcolor{black}{This implies the constraints on the following cosmological $\beta$- functions:}
\bea &&\overline{\beta_1(p_*)}=\Bigg(\frac{d\ln\overline{\Delta^{2}_{\zeta,{\bf Total}}(p)}}{d\ln p}\Bigg)_{p=p_*}=\Bigg(\frac{d\ln{\bigg[\Delta^{2}_{\zeta,{\bf Tree}}(p)\bigg]_{\bf SR}}}{d\ln p}\Bigg)_{p=p_*}=\beta_1(p_*)\nonumber\\
\longrightarrow&&\overline{n_{\zeta,{\bf Total}}(p_*)}-1= n_{\zeta,{\bf SR}}(p_*)-1,\quad\quad\eea\bea
&&\overline{\beta_2(p_*)}=\Bigg(\frac{d^2\ln\overline{\Delta^{2}_{\zeta,{\bf Total}}(p)}}{d\ln p^2}\Bigg)_{p=p_*}=\Bigg(\frac{d^2\ln{\bigg[\Delta^{2}_{\zeta,{\bf Tree}}(p)\bigg]_{\bf SR}}}{d\ln p^2}\Bigg)_{p=p_*}=\beta_2(p_*)\nonumber\\
\longrightarrow&&\overline{\alpha_{\zeta,{\bf Total}}(p_*)}= \alpha_{\zeta,{\bf SR}}(p_*),\quad\quad\\
&&\overline{\beta_3(p_*)}=\Bigg(\frac{d^3\ln\overline{\Delta^{2}_{\zeta,{\bf Total}}(p)}}{d\ln p^3}\Bigg)_{p=p_*}=\Bigg(\frac{d^3\ln{\bigg[\Delta^{2}_{\zeta,{\bf Tree}}(p)\bigg]_{\bf SR}}}{d\ln p^3}\Bigg)_{p=p_*}=\beta_2(p_*)\nonumber\\
\longrightarrow&&\overline{\beta_{\zeta,{\bf Total}}(p_*)}= \beta_{\zeta,{\bf SR}}(p_*).\eea
These are the outcome of very clever yet extremely logical choice of the {\it counter-term} determining renormalization condition.  It becomes theoretically justifiable and logically consistent because of the fact that,  there should not be any effect appearing from the quantum loops due to having effects from both the SR and USR region, at the pivot scale $p_*$. Otherwise, the loop effects has to be seen and directly tested by the observational probes available till date.   

\textcolor{black}{Now we are going to explicitly check that after inclusion of the {\it counter term},  ${\bf Z}^{\bf IR}$ in the renormalized power spectrum of scalar perturbation whether the IR quantum loop effects are completely removed or softened as well as shifted to the next to leading order for the other wave numbers which is bigger than the pivot scale $p_*$}.  To understand this effect clearly, let us Taylor series expand the contribution obtained for the {\it counter term},  ${\bf Z}^{\bf IR}$ at the pivot scale $p_*$:
\bea  &&\textcolor{black}{{\bf Z}^{\bf IR}\approx\Bigg\{1+\frac{4}{3}\bigg[\Delta^{2}_{\zeta,{\bf Tree}}(p_*)\bigg]_{\bf SR}\ln \left(\frac{k_e}{p_*}\right)}\nonumber\\
&&\quad\quad\quad\quad\quad\quad\quad\quad\textcolor{black}{-\frac{1}{4}\bigg[\Delta^{2}_{\zeta,{\bf Tree}}(p_*)\bigg]_{\bf SR}\Bigg(\left(\Delta\eta(\tau_e)\right)^2\left(\frac{k_e}{k_s}\right)^{6}-\left(\Delta\eta(\tau_s)\right)^2\Bigg)\ln \left(\frac{k_e}{k_s}\right)\Bigg\}}.\quad\quad\quad\eea
Here we have truncated the the above mentioned expression during the expansion in the Taylor series by considering up to the contribution in first order due to fact that except the tree level contribution in the SR region all other contributions appearing in the above expression are extremely small and consequently negligible.  Now we plug it back the expression for the renormalized power spectrum for the scalar modes,  which gives the following outcome:
\bea  \overline{\Delta^{2}_{\zeta,{\bf Total}}(p)}&=&\bigg[\Delta^{2}_{\zeta,{\bf Tree}}(p)\bigg]_{\bf SR}\Bigg\{1+ \sum^3_{i=1}{\cal Q}_i(p,p_*,k_e,k_s)\Bigg\},\eea
where we introduce three momentum dependent functions ${\cal Q}_i(p,p_*,k_e,k_s)\forall i=1,2,3$, and are defined as:
\bea 
&&\textcolor{black}{{\cal Q}_1(p,p_*,k_e,k_s)=-\frac{4}{3}\Bigg(\bigg[\Delta^{2}_{\zeta,{\bf Tree}}(p)\bigg]_{\bf SR}-\bigg[\Delta^{2}_{\zeta,{\bf Tree}}(p_*)\bigg]_{\bf SR}\Bigg)\ln \left(\frac{k_e}{p_*}\right)},\\
&&\textcolor{black}{{\cal Q}_2(p,p_*,k_e,k_s)=\frac{1}{4}\Bigg(\bigg[\Delta^{2}_{\zeta,{\bf Tree}}(p)\bigg]_{\bf SR}-\bigg[\Delta^{2}_{\zeta,{\bf Tree}}(p_*)\bigg]_{\bf SR}\Bigg)}\nonumber\\
&&\quad\quad\quad\quad\quad\quad\quad\quad\quad\quad\quad\quad\textcolor{black}{\times\Bigg(\left(\Delta\eta(\tau_e)\right)^2\left(\frac{k_e}{k_s}\right)^{6}-\left(\Delta\eta(\tau_s)\right)^2\Bigg)\ln \left(\frac{k_e}{k_s}\right)},\\
&&\textcolor{black}{{\cal Q}_3(p,p_*,k_e,k_s)=-\bigg[\Delta^{2}_{\zeta,{\bf Tree}}(p)\bigg]_{\bf SR}\bigg[\Delta^{2}_{\zeta,{\bf Tree}}(p_*)\bigg]_{\bf SR}\times\Bigg\{\frac{16}{9}\ln^2 \left(\frac{k_e}{p_*}\right)}\nonumber\\
&&\quad\quad\quad\quad\quad\quad\quad\quad\quad\quad\quad\quad\textcolor{black}{+\frac{1}{16}\Bigg(\left(\Delta\eta(\tau_e)\right)^2\left(\frac{k_e}{k_s}\right)^{6}-\left(\Delta\eta(\tau_s)\right)^2\Bigg)^{2}\ln^{2} \left(\frac{k_e}{k_s}\right)}\nonumber\\
&&\quad\quad\quad\quad\quad\quad\quad\quad\quad\quad\quad\quad\textcolor{black}{+{\rm higher \; even \; order \; terms}\Bigg\}}.\eea
Now in the late time scale in the SR region where the cosmological observation takes place it is always expected that:
\bea \bigg[\Delta^{2}_{\zeta,{\bf Tree}}(p)\bigg]_{\bf SR}\approx \bigg[\Delta^{2}_{\zeta,{\bf Tree}}(p_*)\bigg]_{\bf SR},\eea
which immediate implies the following facts:

\begin{itemize}
\item One of the findings of our compution is:
\bea {\cal Q}_1(p,p_*,k_e,k_s)\approx0,   \quad{\rm and}\quad {\cal Q}_2(p,p_*,k_e,k_s)\approx 0.\eea

\item Contribution coming from ${\cal Q}_4(p,p_*,k_e,k_s)$ is non zero but small.  All the logarithmically divergent terms appearing in quadratic or more higher order. The first term in ${\cal Q}_4(p,p_*,k_e,k_s)$ is more dominant than the other higher order terms.

\item In the linear order all the logarithmically divergent contributions are completely removed after renormalization and the quadratic or more higher order contributions are appearing in the final expression.  Up to the linear sub leading order, the renormalized power spectrum for the scalar mode is completely free from logarithmically divergent terms.  Next to sub leading order or more higher order terms appearing from one loop contributions cannot be removed from the final result.

\item One-loop contribution to the SR region is completely removed from the linear order and shifted to the next order. On top of the tree level contribution this additional effect turns out to be negligibly small. Now for the USR contribution we have also found that the leading order effect is shifted to the next order for which the corresponding one-loop contribution turns out to have very small effect on top of the tree level SR contribution. 

\end{itemize}

Finally, the one-loop renormalized power spectrum for the scalar modes can be simplified as:
\bea  \boxed{\boxed{\overline{\Delta^{2}_{\zeta,{\bf Total}}(p)}=\bigg[\Delta^{2}_{\zeta,{\bf Tree}}(p)\bigg]_{\bf SR}\Bigg\{1+\underbrace{{\cal Q}_3(p,p_*,k_e,k_s)}_{\bf  Quadratic/higher \;log \;divergence\;}\Bigg\}}}.\eea
The final derived result for the renormalized power spectrum for the scalar modes implied the following immediate consequences:
\begin{enumerate}

    \item Only logarithmic correction appears at the one-loop level computation and no other divergences appearing at the horizon crossing and super horizon scale where the cosmological observation takes place.

    \item At the pivot scale, all one-loop effects completely disappear from the renormalized power spectrum. Consequently, all the derived quantities from the spectrum, such as, spectral tilt, running and running of the running of the tilt, are free from all one-loop effects at the pivot scale.

    \item By choosing the appropriate counter term it is possible to shift the first order one loop contribution to the next order in the renormalized power spectrum. As an immediate consequence, spectral tilt, running and running of the running of the tilt are free from all one-loop effects at other wave numbers away from the pivot scale up to the first order.
\end{enumerate}

\section{Dynamical Renormalization Group (DRG) resummed scalar power spectrum}
\label{DRG}
\textcolor{black}{Let's make it very clear before we go into the technical aspects of the calculation in this part that the current computation is not dependent on the explicit structure of the prime component of the one-loop correction. The overall size of this component must be kept inside the perturbative limit in order to execute the resummation and ultimately provide a finite outcome. We cite the result in a way that shows this method performs effectively for the aforementioned framework.}

Further, we briefly discuss the  Dynamical Renormalization Group
 (DRG) method \cite{Boyanovsky:1998aa,Boyanovsky:2001ty,Boyanovsky:2003ui,Burgess:2009bs,Dias:2012qy,Chen:2016nrs,Baumann:2019ghk}, which allows us to resum over all the logarithmically divergent contributions in all the loop orders. This is technically possible, provided the corresponding resummation infinite series is strictly convergent at the late time scales. Each of the terms in this infinite series is the direct artefact of the perturbative expansion in all possible loop orders. In general, DRG is treated as the natural mechanism using which the validity of secular time-dependent, momentum-scale-dependent, and energy-scale-dependent contributions can be easily justified in perturbative expansion in the late time scale, which we are using within the framework of cosmology. This technique helps to extract the late time limiting behaviour instead of knowing the full behaviour from the perturbative expansion after performing the resummation. In more technical language, DRG is interpreted as the logarithmically divergent contributions of scattering amplitudes computed at a given renormalization scale to any arbitrary energy scale. The basic strategy for using this type of technique is to absorb the contributions from the energy into the expressions for the background scale dependent running couplings of the underlying theory, which is commonly known as the {\it Renomrmalization Group} (RG) resummation technique. In the small coupling regime, one can use this result for a wide range of running energy scales. In the present cosmological framework, the running coupling can be easily understood in the language of three $\beta$ functions, which are directly related to spectral tilt running  and running of the running of the scalar modes. It is important to note that, DRG resummation is a much improved version of the well known RG resummation,, in which we are doing the same job at the late time scale of the smological evolution. It is a  conceptually very richer version, technically correct, and sometimes in literature it is commonly referred to as {\it resummation by exponentiation} at the late time scale. However, the diagrammatic realization of such a technique is yet to be understood clearly, though the outcome is extremely impressive as we have a convergent all loop order resummed finite result. \textcolor{black}{More specifically, this resummed finding is true for any perturbative computing loop order that can accurately capture quantum effects. This is only practical, though, if the resummed infinite series adheres to the tight convergence standards at the horizon crossing and super-horizon scales. These components in the convergent series are all by-products of the cosmological perturbation theory of scalar modes in all possible loop orders. DRG is typically viewed as the natural process that makes it simple to justify secular momentum-dependent contributions to the convergent infinite series at horizon crossing and super-horizon scales, which we are using within the context of  the primordial single field driven cosmological set-up. This approach enables accurate knowledge of the behaviour at the horizon crossing and super-horizon scales rather than knowing the entire behaviour from the series expansion term by term after the resummation procedure has been completed. The idea of the Renomrmalization Group (RG) resummation approach initially entered the scene, which entails momentum-dependent contributions to the equations for scale-dependent running couplings of the underlying theory in terms of $\beta$ functions. The improved version of the RG resummation method is the DRG resummation methodology. This finding is applicable to a specific, observationally feasible extended range of running momentum scales in the minuscule coupling domain where the perturbative approximation absolutely holds true within the context of the underlying single field driven set-up. 
}

The final form of the resummed dimensionless power spectrum for the scalar modes can be stated in the following way using the DRG approach \footnote{\textcolor{black}{Note: The uncontrollable increase of the secular terms generally becomes quite troublesome since it invalidates the perturbative computation at the late time scale, especially at the horizon crossing and super-horizon scales. This problem in primordial cosmology can be solved using the DRG approach.}
}:
\bea  \label{res1}\textcolor{black}{\boxed{\boxed{\overline{\overline{\Delta^{2}_{\zeta,{\bf Total}}(p)}}=\bigg[\Delta^{2}_{\zeta,{\bf Tree}}(p)\bigg]_{\bf SR}\exp\bigg({\cal Q}_3(p,p_*,k_e,k_s)\bigg)\times\bigg\{1+{\cal O}\bigg(\bigg[\Delta^{2}_{\zeta,{\bf Tree}}(p_*)\bigg]^2_{\bf SR}\bigg)\bigg\}}}},\eea
where we have:
\bea \exp\bigg({\cal Q}_3(p,p_*,k_e,k_s)\bigg):&=&1+\sum^{\infty}_{n=1}\frac{1}{n!}\bigg({\cal Q}_3(p,p_*,k_e,k_s)\bigg)^{n}\nonumber\\
&=&\underbrace{1}_{\bf Tree-level}+\underbrace{{\cal Q}_3(p,p_*,k_e,k_s)}_{\bf Two-loop}+\underbrace{\frac{1}{2!}\bigg({\cal Q}_3(p,p_*,k_e,k_s)\bigg)^{2}}_{\bf Four-loop}+\cdots.\quad\quad\quad\eea
\textcolor{black}{Thus, after combining the contributions from the secular terms, it really reflects behaviour on a wide scale. In this case, the DRG resummed version of the result is valid for all orders of the ${\cal Q}_3(p,p_*,k_e,k_s)$ function. It should be emphasised that the convergence condition in this case strictly requires$|{\cal Q}_3(p,p_*,k_e,k_s)|\ll 1$, which is fully met in the discussion's current setting. The most important result of the DRG resummed version of the one-loop corrected primordial power spectrum for the scalar modes is that it creates a controlled version of the dimensionless primordial power spectrum where the behaviour of the logarithmically divergent contribution is more softened than in the renormalized version of the one-loop power spectrum explicitly derived in the previous section. }

\textcolor{black}{The dominance of all chain diagrams over the other conceivable diagrams included in the calculation was not necessary for the application of the DRG resummation, but it will undoubtedly add to the leading contributions made by these logarithmically dependent components. The higher-order convergent components in the infinite series in the aforementioned result perfectly match the function of higher-loop contributions in the current context of the debate. It is truly amazing that, even without explicitly applying the higher-loop correction to the primordial power spectrum for the scalar modes, we can study the behaviour of each correction term in all-loop order. This allows us to examine the non-perturbative but convergent behaviour of the spectrum in this situation, where the sum of all-loop contributions is finite and expressed in terms of an exponential function.}

Additionally, it is important to note that, the all loop DRG resummed result can also be connected to the renormalized one-loop corrected total power spectrum by the following expression:
\bea \label{res2} \boxed{\boxed{\overline{\overline{\Delta^{2}_{\zeta,{\bf Total}}(p)}}=\bigg[\Delta^{2}_{\zeta,{\bf Tree}}(p)\bigg]_{\bf SR}\exp\Bigg(\frac{\overline{\Delta^{2}_{\zeta,{\bf Total}}(p)}}{\bigg[\Delta^{2}_{\zeta,{\bf Tree}}(p)\bigg]_{\bf SR}}-1\Bigg)}},\eea
where in the one-loop we have:
\bea \exp\Bigg(\frac{\overline{\Delta^{2}_{\zeta,{\bf Total}}(p)}}{\bigg[\Delta^{2}_{\zeta,{\bf Tree}}(p)\bigg]_{\bf SR}}-1\Bigg)&=&1+\sum^{\infty}_{n=1}\frac{1}{n!}\Bigg(\Bigg(\frac{\overline{\Delta^{2}_{\zeta,{\bf Total}}(p)}}{\bigg[\Delta^{2}_{\zeta,{\bf Tree}}(p)\bigg]_{\bf SR}}-1\Bigg)\Bigg)^{n}\nonumber\\
&=&\underbrace{1}_{\bf Tree-level}+\underbrace{\Bigg(\frac{\overline{\Delta^{2}_{\zeta,{\bf Total}}(p)}}{\bigg[\Delta^{2}_{\zeta,{\bf Tree}}(p)\bigg]_{\bf SR}}-1\Bigg)}_{\bf One-loop\;quadratic\;log \;term}+\underbrace{\frac{1}{2!}\Bigg(\Bigg(\frac{\overline{\Delta^{2}_{\zeta,{\bf Total}}(p)}}{\bigg[\Delta^{2}_{\zeta,{\bf Tree}}(p)\bigg]_{\bf SR}}-1\Bigg)\Bigg)^{2}}_{\bf One-loop\;quartic\;log \;term}+\cdots.\quad\quad\quad\nonumber\\
&=&\underbrace{1}_{\bf Tree-level}+\underbrace{{\cal Q}_3(p,p_*,k_e,k_s)}_{\bf One-loop\;quadratic\;log \;term}+\underbrace{\frac{1}{2!}\bigg({\cal Q}_3(p,p_*,k_e,k_s)\bigg)^{2}}_{\bf One-loop\;quartic\;log \;term}+\cdots.\quad\quad\quad\eea
Finally after comparing equation(\ref{res1}) and equation(\ref{res2}), it is obvious, the one-loop quadratic log term is equivalent to the two-loop contribution and one-loop quartic log term is equivalent to the four-loop contribution. This is extremely pivotal finding on which our claims hinge.

\textcolor{black}{Now, we use the fact that the contribution from the SR phase is very minor compared to the USR contribution since it contains a factor $(k_e/k_s)^6\ln(k_e/k_s)$ to comprehend and easily visualise the softening of the behaviour of the logarithmic divergence. Because of this, in the SR contribution, the factor ${\cal Q}_3$ can be temporarily disregarded. As a result, we may write the following approximation of the outcome:}

\textcolor{black}{Here we use the following fact, which is going to be extremely useful for simplification purposes in the perturbative regime of the computation:}
\bea \textcolor{black}{\left(\Delta\eta(\tau_e)\right)^2\left(\frac{k_e}{k_s}\right)^{6}\gg\left(\Delta\eta(\tau_s)\right)^2}.\eea
\textcolor{black}{Here we choose, $\Delta\eta(\tau_e)=1$, $k_e=10^{22}{\rm Mpc}^{-1}$ and $k_s=10^{21}{\rm Mpc}^{-1}$ which is going to be helpful for the further estimation purpose. Hence the factor ${\cal Q}_3($ can be further recast in the following simplified form:}
\bea \textcolor{black}{{\cal Q}_3\approx  -{\cal I}_*\ln^2\left(\frac{k_e}{k_s}\right)},\eea
\textcolor{black}{where the pre-factor ${\cal I}_*$ at the pivot scale is defined as:}
\bea \textcolor{black}{{\cal I}_*=\frac{1}{4}\bigg[\Delta^{2}_{\zeta,{\bf Tree}}(p_*)\bigg]_{\bf SR}\times \left(\Delta\eta(\tau_e)\right)^2 \left(\frac{k_e}{k_s}\right)^{6}\approx {\cal O}(10^{-2})\ll 1},\eea
\textcolor{black}{which further implies that:}
\bea \textcolor{black}{{\cal Q}_3\approx -0.053\ll 1}.\eea
\textcolor{black}{This means that the perturbativity argument holds perfectly in the present computation, and no such question appears regarding the justifiability and physical applicability of the DRG resummation technique as far as the present computation is concerned.}

\textcolor{black}{The IR softened and smoothed version of the DRG resummed dimensionless primordial power spectrum for the scalar modes can be further recast into the following simplified form:}
 %\begin{widetext}
\bea \textcolor{black}{\boxed{\boxed{\overline{\overline{\Delta^{2}_{\zeta,{\bf Total}}(p)}}
\approx\bigg[\Delta^{2}_{\zeta,{\bf Tree}}(p)\bigg]_{\bf SR}\left(\frac{k_e}{k_s}\right)^{-{\cal I}_*}\times\bigg\{1+{\cal O}\bigg(\bigg[\Delta^{2}_{\zeta,{\bf Tree}}(p_*)\bigg]^2_{\bf SR}\bigg)\bigg\}}}}.\eea
\textcolor{black}{The aforementioned result indicates unequivocally that following exponentiation and therefore its subsequent softening version, the IR logarithmic divergence is fully smoothed out after executing DRG resummation. The softening of the behaviour of the divergences after DRG resummation may not always be achieved if we have other types of variations of IR divergences besides logarithmic divergence, which is another crucial point to keep in mind. Fortunately, none of the other IR divergences included in the calculation would detract from the beauty of the underlying physical scenario, and by employing the DRG technique, we were able to obtain a controlled representation of the primordial power spectrum for the scalar modes.}

\section{Numerical results and further estimations: A no-go theorem for PBH formation}\label{num}
%%%%%%%%%%%%%%%%%%%%%%FIGURE%%%%%%%%%%%%%%%%%%%%%%%%%%%%%%%%%%%%%
    \begin{figure*}[htb]
    	\centering
    	{
    		\includegraphics[width=16cm,height=11cm] {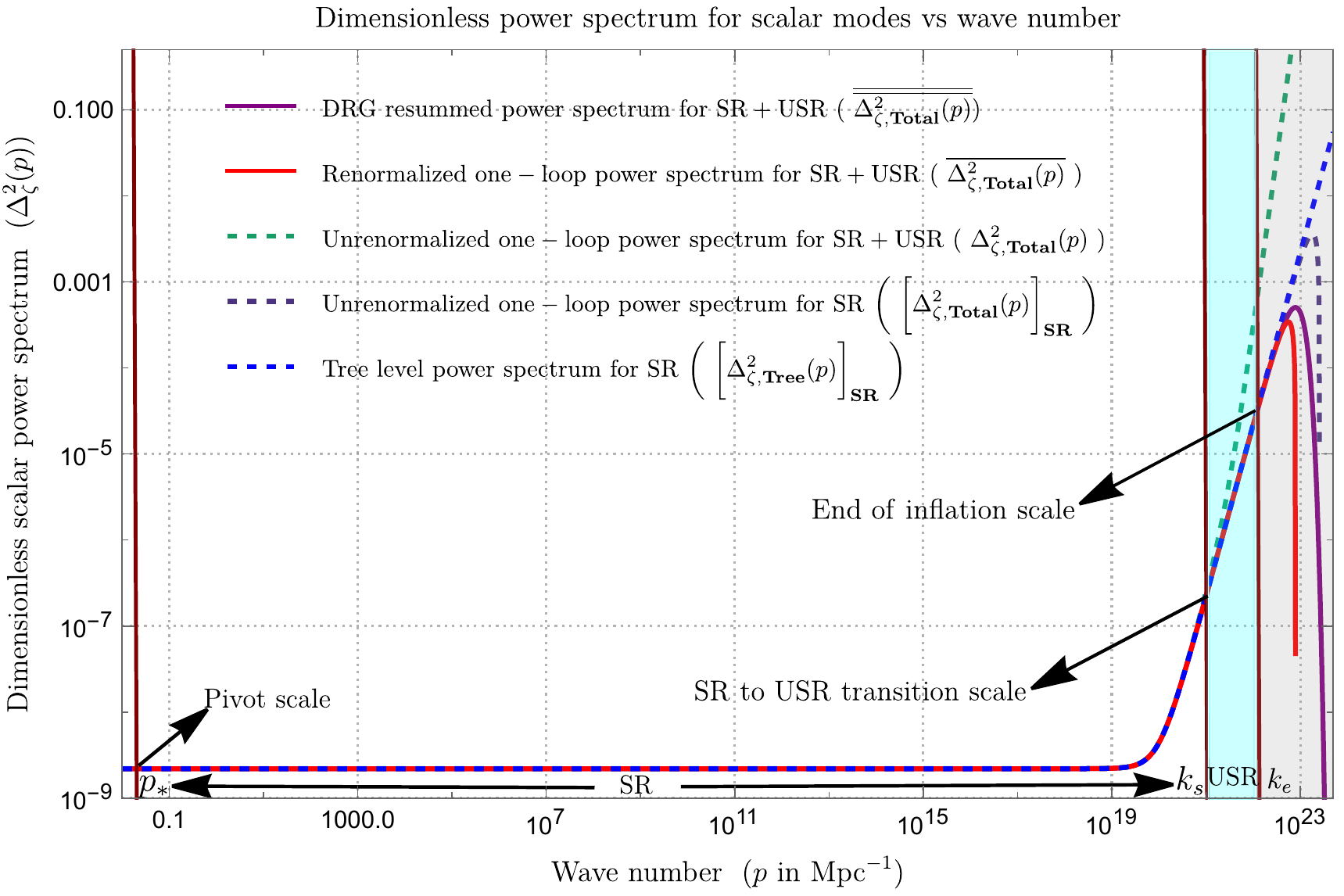}
    	}
    	\caption[Optional caption for list of figures]{Behaviour of the dimensionless power spectrum for scalar modes with respect to the wave number. In this plot we fix the pivot scale at $p_*=0.02\;{\rm Mpc}^{-1}$, transition scale from SR to USR region at $k_s=10^{21}\;{\rm Mpc}^{-1}$ (where we fix the IR cut-off) and the end of inflation at $k_s=10^{22}\;{\rm Mpc}^{-1}$ (where we fix the UV cut-off), the renomalization parameter $c_{\bf SR}=0$ (for SR and SR+USR one-loop corrected and DRG resummed contribution), $\Delta\eta(\tau_e)=1$ and $\Delta\eta(\tau_s)=-6$. In this plot we have found that, $k_{\rm UV}/k_{\rm IR}=k_e/k_s\approx{\cal O}(10)$, which is an extremely useful information for the analysis performed in this paper. } 
    	\label{Spectrum}
    \end{figure*}
    %%%%%%%%%%%%%%%%%%%%%%%%%%%%%%%%%%%%%%%%%%%%%%%%%%%%%%%%%%%%%%%%%%%%%%%%
    %%%%%%%%%%%%%%%%%%%%%%FIGURE%%%%%%%%%%%%%%%%%%%%%%%%%%%%%%%%%%%%%
    \begin{figure*}[htb]
    	\centering
    	{
    		\includegraphics[width=16cm,height=11cm] {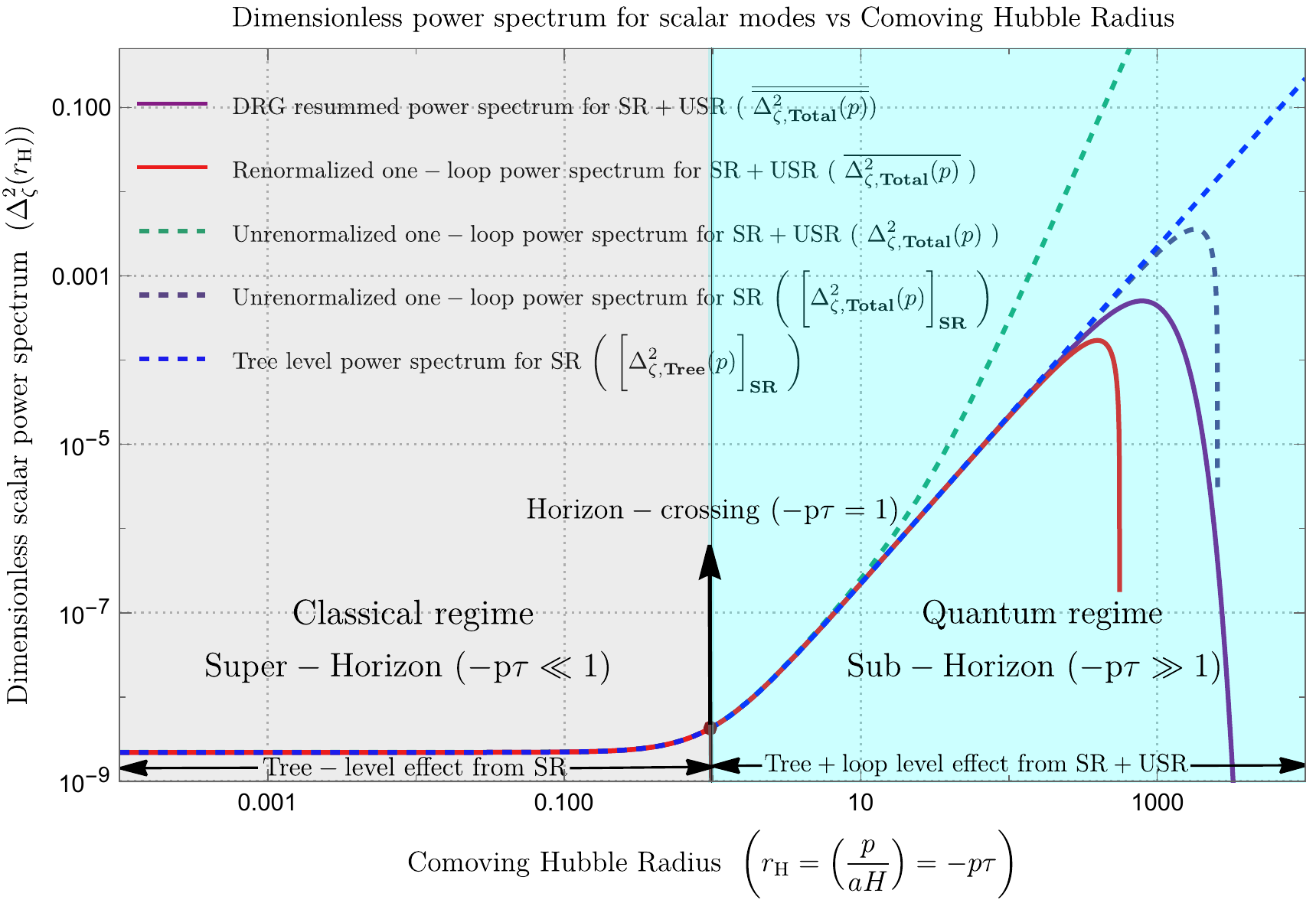}
    	}
    	\caption[Optional caption for list of figures]{Behaviour of the dimensionless power spectrum for scalar modes with respect to the Comoving Hubble Radius. In this plot we fix the pivot scale at $p_*=0.02\;{\rm Mpc}^{-1}$, transition scale from SR to USR region at $k_s=10^{21}\;{\rm Mpc}^{-1}$ (where we fix the IR cut-off) and the end of inflation at $k_s=10^{22}\;{\rm Mpc}^{-1}$ (where we fix the UV cut-off), the renomalization parameter $c_{\bf SR}=0$ (for SR and SR+USR one-loop corrected and DRG resummed contribution), $\Delta\eta(\tau_e)=1$ and $\Delta\eta(\tau_s)=-6$. In this plot we have shown the super-horizon (classical) region, horizon exit point and the sub-horizon (quantum) region explictly, which helps us to understand when the tree level and loop level effects are dominating. } 
    	\label{Spectrum2}
    \end{figure*}
    %%%%%%%%%%%%%%%%%%%%%%%%%%%%%%%%%%%%%%%%%%%%%%%%%%%%%%%%%%%%%%%%%%%%%%%%
    %%%%%%%%%%%%%%%%%%%%%%FIGURE%%%%%%%%%%%%%%%%%%%%%%%%%%%%%%%%%%%%%
    \begin{figure*}[htb]
    	\centering
    	{
    		\includegraphics[width=16cm,height=11cm] {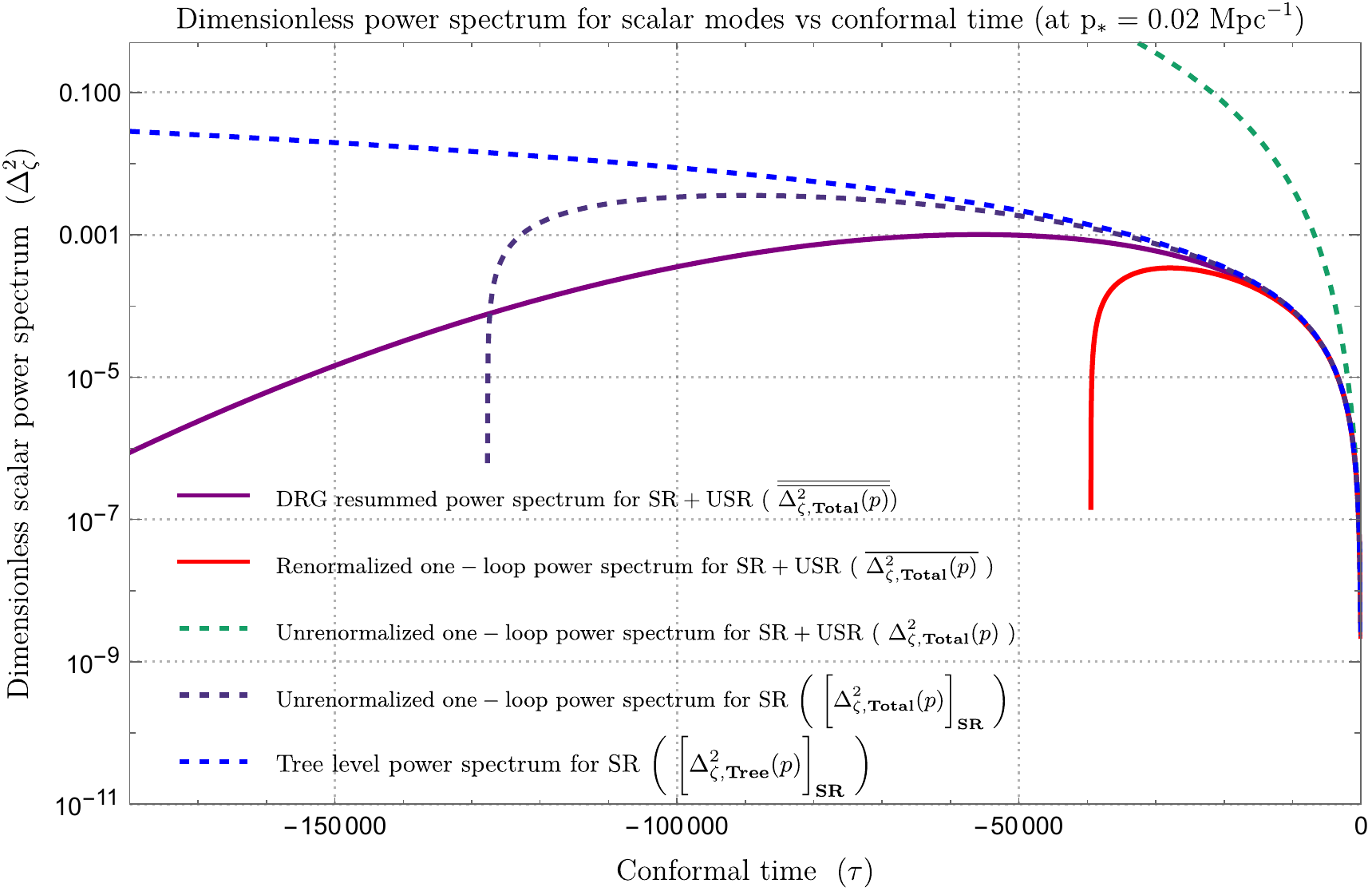}
    	}
    	\caption[Optional caption for list of figures]{Behaviour of the dimensionless power spectrum for scalar modes with respect to conformal time scale. In this plot we fix the pivot scale at $p_*=0.02\;{\rm Mpc}^{-1}$, transition scale from SR to USR region at $k_s=10^{21}\;{\rm Mpc}^{-1}$ (where we fix the IR cut-off) and the end of inflation at $k_s=10^{22}\;{\rm Mpc}^{-1}$ (where we fix the UV cut-off) for SR and SR+USR one-loop corrected and DRG resummed contribution, $\Delta\eta(\tau_e)=1$ and $\Delta\eta(\tau_s)=-6$. In this plot we have shown the sub-horizon behaviour of the spectrum, which implies quantum effects including loop corrections are significant in this region. } 
    	\label{Spectrum4}
    \end{figure*}
    %%%%%%%%%%%%%%%%%%%%%%%%%%%%%%%%%%%%%%%%%%%%%
In figure(\ref{Spectrum}), we have depicted the behaviour of the dimensionless power spectrum for scalar modes with respect to the wave number. We have plotted the individual behaviours of tree level and unrenormalized one-loop contribution from the SR region, unrenormalized and renormalized one-loop corrected contribution from both the SR and USR regions. From the plot, we have found that at the end of inflation, due to the logarithmic one-loop effect, the power spectrum falls off very sharply in the SR region. Otherwise, up to the point of the end of inflation, one can't distinguish the tree level from the unrenormalized one-loop level contribution. This is because the unrenormalized one-loop contribution gives an extremely small correction on top of the tree level contribution in the SR region. Next, we include the effect of the USR region in our analysis, where the PBH formation takes place. The unrenormalized total one-loop corrected power spectrum after including both the contributions from the SR and USR regions exactly follows the same behaviour as the tree level and one-loop corrected power spectrum computed in the SR region up to the scale where we consider the transition from the SR to USR regions. Just after this transition, one-loop contribution coming from the USR dominates over the one-loop contribution coming from the SR region. Consequently, we observe a small but significant deviation in the total unrenormalized one-loop corrected power spectrum. We have used the cut-off regularization technique to compute the one-loop contributions from the momentum integrals in Fourier space. Because of this reason, we have introduced two cut-offs: the IR cut-off $k_{\rm IR}$ and UV cut-off $k_{\rm UV}$. For computational purposes and to implement the PBH formation process within the framework of the single field inflationary paradigm, we have further chosen these cut-off scales as, $k_{\rm IR}=k_s$ and $k_{\rm UV}=k_e$, where $k_s$ and $k_e$ are the corresponding scales where the transition from SR to USR and the end of inflation happened, respectively. We have found that the one-loop effect in the total power spectrum before performing the renormalization is dominated by logarithmic contribution $\ln\left(k_e/k_s\right)$, after the scale $k_s$. Because of this additional contribution we observe a deviation in the unrenormalized one-loop total power spectrum. After implementing the renormalization condition at the pivot scale $p_*=0.02\;{\rm Mpc}^{-1}$ and incorporating the contribution from the counter terms, we have computed the expression for the renormalized one-loop corrected total power spectrum from the scalar modes. From the plot, we have found that, renormalized one-loop spectrum exactly matches the tree level contribution up to the scale where inflation ends. This is the outcome of the fact that we have implemented the renormalization correctly, as all the small logarithmic one-loop corrections are removed at the linear order and dumped to the next sub-leading order. But such sub-leading contributions are numerically extremely small, so that the ultimate corrected contribution becomes extremely small, which gives rise to the final form of the renormalized one-loop corrected total spectrum, which exactly follows the tree level behaviour coming from the SR region contribution. Last but not least, using the DRG resummation method and considering the contribution of all possible diagrams in all loops, we have further plotted the non-perturbative but numerically convergent behaviour of the dimensionless power spectrum for scalar modes with respect to the wave number. In this specific case, we have found that up to the end of inflation, it is completely consistent with all the previously obtained separate contributions except the unrenormalized one-loop result. Due to the resummation over all orders of loop diagrams, one gets the most comprehensible and consistent result with respect to the result obtained from a one-loop renomrmalized power spectrum. To plot the individual behaviours of the spectrum, we have fixed the transition scale from SR to USR region at $k_s=10^{21}\;{\rm Mpc}^{-1}$ (where we fix the IR cut-off) and the end of inflation at $k_s=10^{22}\;{\rm Mpc}^{-1}$ (where we fix the UV cut-off) for SR and SR+USR one-loop corrected and DRG resummed contribution, $\Delta\eta(\tau_e)=1$ and $\Delta\eta(\tau_s)=-6$. In this plot, we have found that, $k_{\rm UV}/k_{\rm IR}=k_e/k_s\approx{\cal O}(10)$, which is the key finding of this calculation. 

\textcolor{black}{Let us now mention the underlying assumptions and few crucial facts that we have followed to analyze the momentum scale dependent behaviour of the primordial power spectrum, which are appended below point-wise:}
\begin{enumerate}
    \item \textcolor{black}{First of all we have assumed that the SR period persists up to the momentum scale $k_s$ and then a sharp transition occur at transition scale $k_s$ to the next USR period which along with inflation ends at the scale $k_e$. In our framework after USR inflation ends and we don't consider any second SR phase after USR period. In refs. \cite{Kristiano:2022maq,Kristiano:2023scm} the authors have included an additional second SR period to sufficiently complete inflation in terms of number of e-foldings. But recently,  in ref \cite{Choudhury:2023hvf} we have explicitly shown that inclusion of an additional second SR phase after USR will not going to suffice the purpose of generating large mass PBHs. We have shown that by adding second SR phase one can able to generate large mass PBHs with having insufficient number of e-foldings, which is obviously not desirable in the present context. In the present paper as we as in the follow up works \cite{Choudhury:2023jlt,Choudhury:2023rks} we have established that to have sufficient inflation in terms of the number of e-foldings only small mass PBHs can be generated after performing the regularization, renormalization and resummation in the total power spectrum in a correct fashion as clearly demonstrated in this paper.}

    \item \textcolor{black}{Next, it is important to note that in this paper in the demonstrated figures we have joined the contributions coming from first SR phase and the USR phase very smoothly, which is the direct outcome of treating the regularization, renormalization and resummation in a correct fashion as mentioned in the previous point. With the help of the detailed computation we have established that the final one-loop corrected spectrum generated after summing over contributions coming from both the SR and USR periods, the problematic quadratic divergence can be completely removed, which is quite consistent from the understanding of Quantum Field Theory of de Sitter space. Only in the final mathematical version of the primordial spectrum IR logarithmic divergent contributions are appearing in very softened form. For this reason in the corresponding representative plots smoothness and softening behaviour is appearing throughout for the regularizede, renormalized and resummed version of the total spectrum.}

    \item \textcolor{black}{In the refs. \cite{Kristiano:2022maq,Kristiano:2023scm} the authors have not performed any renormalization or resummation in the final form of the total one-loop corrected power spectrum. For this reason, the behaviour of the spectrum at the transition point from SR to USR phase and after that in the USR period is a bit different from our result. In refs. \cite{Kristiano:2022maq,Kristiano:2023scm} the authors encounter quadratic UV divergence as well as logarithmic IR divergence, which is clearly reflected in their final plot as could not able to remove and soften from their analysis. In our analysis we have derived a IR softened version of the spectrum that is completely free from quadratic UV divergence. By seeing the overall behaviour of the plots that we have presented in this paper, it is clear that smoothness and softening of the spectrum is maintained in the SR as well as the USR period, which makes our result unique and correct compared to the result obtained in refs. \cite{Kristiano:2022maq,Kristiano:2023scm}. Such unique behaviour is physically consistent with the findings of the Quantum Field Theory de Sitter space, which can able to tackle both short range UV and long range IR modes perfectly in presence large quantum fluctuations generated from one-loop where perturbativity criteria are maintained perfectly by assuring the ratio $k_e/k_s\sim {\cal O}(10)$.}

    \item  \textcolor{black}{Further, it is important to note that, since in the refs. \cite{Kristiano:2022maq,Kristiano:2023scm} the authors have inserted a second SR phase just after USR by considering another sharp transition at the scale $k_e$, in the unrenormalized version of the obtained spectra, small oscillations appear that persist up to the end of inflation, which happened at the later scale $k_{\rm end}$. This oscillations they have included to obtain the required number of e-foldings to have sufficient inflation, as their initial claim is to generate large mass PBHs by fixing the SR to USR transition scale at $k_s\sim 10^{6}{\rm Mpc}^{-1}$. However, from the unrenormalized one-loop amplitude of the spectrum due to the presence of quadratic divergence (which they could not able to remove) they have concluded that PBHs cannot be generated at all. In our work, we have removed quadratic UV divergence, and softened the logarithmic IR divergence by performing adiabatic renormalization followed by DRG resummation, which finally helps us at least to generate small mass PBHs that respect the sufficiency criteria of inflation in terms of the number of e-foldings (for the numerical estimation of PBH mass and other related quantities, look at the end part of this section). From the behaviour of the renormalized and resummed spectrum, it is clearly understood that the USR period has to be very short lived to maintain the perturbativity criteria, which hold perfectly in the present computation. If we include an additional SR phase after USR, that has to also be very short lived. This fact we have clearly analysed and explicitly shown in the ref. \cite{Choudhury:2023hvf} in great detail.}

    \item \textcolor{black}{Finally, it is important to note that during analyzing the representative plots in this paper, we have neglected all other suppressed contributions from the oscillatory terms as well as the inverse power law terms of $(k_s/k_e)(\ll 1)$, as all of them will not be able to significantly change the overall behaviour of the scalar power spectrum in the USR period. For this reason, very small oscillating features do not appear in the final version of the plots. The appearance of all such contributions in the momentum dependent one-loop contribution in the USR phase is explicitly computed in Appendix \ref{App:B} using which one can clearly demonstrate the corresponding suppression. }
\end{enumerate}

In figure(\ref{Spectrum2}), we have depicted the behaviour of the dimensionless power spectrum for scalar modes with respect to the comoving Hubble Radius. In this plot, we have explicitly pointed out the super-horizon region, horizon exit point, and sub-horizon region where classical, semi-classical, and quantum effects dominate. From the behaviour of the plot in the above mentioned three regions, we can clearly see that the quantum loop correction on the tree-level contribution becomes dominant in the sub-horizon regime. On the other hand, starting from the horizon crossing point in the super-horizon region (classical regime), the tree level contribution is important. 

%%%%%%%%%%%%%%%%%%%%%%FIGURE%%%%%%%%%%%%%%%%%%%%%%%%%%%%%%%%%%%%%
    \begin{figure*}[htb]
    	\centering
    	{
    		\includegraphics[width=16cm,height=9cm] {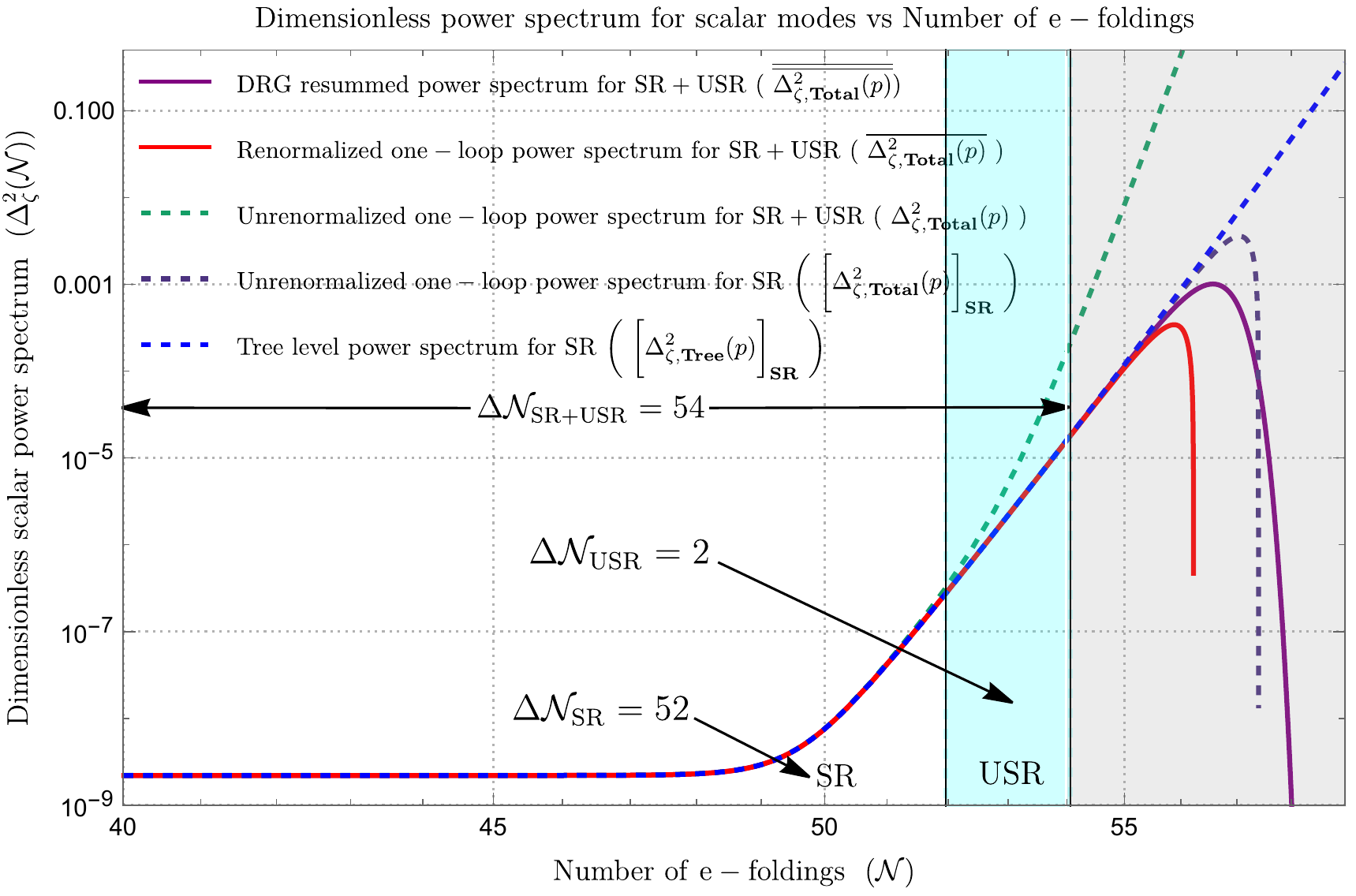}
    	}
    	\caption[Optional caption for list of figures]{Behaviour of the dimensionless power spectrum for scalar modes with respect to the number of e-foldings. In this plot we fix the pivot scale at $p_*=0.02\;{\rm Mpc}^{-1}$, transition scale from SR to USR region at $k_s=10^{21}\;{\rm Mpc}^{-1}$ (where we fix the IR cut-off) and the end of inflation at $k_s=10^{22}\;{\rm Mpc}^{-1}$ (where we fix the UV cut-off), the renomalization parameter $c_{\bf SR}=0$ (for SR and SR+USR one-loop corrected and DRG resummed contribution), $\Delta\eta(\tau_e)=1$ and $\Delta\eta(\tau_s)=-6$. In this plot we have shown the allowed number of e-foldings in the SR period, USR period and found that strictly $2$ e-folds allowed only for PBH formation from the USR period. } 
    	\label{Spectrum3}
    \end{figure*}
    %%%%%%%%%%%%%%%%%%%%%%%%%%%%%%%%%%%%%%%%%%%%%

Next in figure(\ref{Spectrum4}), we have depicted the behaviour of the dimensionless power spectrum for scalar modes with respect to the conformal time scale. In this plot we have pointed the sub-horizon behaviour of the spectrum, which implies quantum effects including loop corrections are significant in this region. We have found that the DRG resummed result gives convergent but sustainable contribution in the power spectrum. This is because of the fact that, in the case of DRG resummed result, we have considered the contribution from all possible allowed diagrams in the loop level and all such infinite possibilities are taken into account during the exponenciation of the final result. On the other hand, in the sub-horizon region we can observe sharp fall in the spectrum for the one-loop corrected SR contribution and for the one-loop corrected SR+USR contribution, which is not there in the resummed result. Nonrenormalizable contribution breaks the perturbative approximation in the conformal time scale after some time, and can't produce the desirable result. Another important point we need to mention that, except the nonrenormalizable contribution, rest of the contributions are consistent with the tree level result obtained for the SR region because all of the loop level originated corrections on the tree level result helps to maintain the perturbative approximation in this region where quantum effects are significant. 

Finally, in figure(\ref{Spectrum3}), we have depicted the behaviour of the dimensionless power spectrum for scalar modes with respect to the number of e-foldings. From this plot we have found that:
\bea \boxed{\boxed{\Delta {\cal N}_{\rm USR}={\cal N}_e-{\cal N}_s=\ln(k_e/k_s)\approx\ln(10)\approx 2}},\eea 
which implies around $2$ e-folds are allowed in the USR period for the PBH formation. On the other hand, we have found from our analysis that the allowed number of e-foldings in the SR+USR period is given by the following expression:
\bea \Delta {\cal N}_{\rm SR+USR}={\cal N}_e-{\cal N}_*=\ln(k_e/k_*)\approx\ln\bigg(\frac{10^{22}\;{\rm Mpc}^{-1}}{0.02\;{\rm Mpc}^{-1}}\bigg)\approx 54,\eea 
which further implies that from SR region following sole contribution is appearing:
\bea \Delta {\cal N}_{\rm SR}=\Delta {\cal N}_{\rm SR+USR}-\Delta {\cal N}_{\rm USR}=\ln(k_s/k_*)\approx\ln\bigg(\frac{10^{21}\;{\rm Mpc}^{-1}}{0.02\;{\rm Mpc}^{-1}}\bigg)\approx 52.\eea 
This is consistent with the required number of e-folds to have observationally consistent single field inflation. 

Further, using the above estimates one can give the following estimation of the field excursion during PBH formation in the USR period using the well known {\it Lyth bound} \cite{Lyth:1996im}:
\bea \boxed{\boxed{\frac{|\Delta \phi|_{\rm USR}}{M_{\rm pl}}=\sqrt{\frac{r_*}{8}}\Delta {\cal N}_{\rm USR}\approx {\cal O}(0.16)<1\quad\quad {\rm where}\quad|\Delta \phi|_{\rm USR}:=|\phi_e-\phi_s|}}.  \eea
Here we have used the value of the tensor to scalar ratio at the pivot scale, $r_*\sim 0.05$, which is the $1\sigma$ upper bound from Planck 2018 data \cite{Planck:2018jri}. Additionally, it is important to note that, $\phi_e$ and $\phi_s$ represent the field value at the end of inflation and SR to USR transition scale respectively. It suggests that the Effective Field Theory (EFT) technique is valid during the USR period when the PBH formation takes place and one can consider {\it sub-Planckian single field inflationary paradigm for PBH formation}. To know more about this bound and related EFT prescription see refs. \cite{Efstathiou:2005tq,Easther:2006qu,Hotchkiss:2011gz,Baumann:2011ws,Choudhury:2013iaa,Choudhury:2013jya,Choudhury:2013woa,Choudhury:2014sua,Choudhury:2014sxa,Choudhury:2014kma,Choudhury:2015pqa,Cai:2021yvq}. We have also found from these estimations that the prolonged USR period is not allowed for PBH formation. Within this short span one can further give the estimation of PBH mass, using the following expression:
\bea \frac{M_{\rm PBH}}{M_{\odot}}=1.13\times 10^{15}\bigg(\frac{\gamma}{0.2}\bigg)\bigg(\frac{g_*}{106.75}\bigg)^{-1/6}\bigg(\frac{k_s}{p_*}\bigg)^{-2}.\eea
Here $k_s=10^{21}\;{\rm Mpc}^{-1}$ represents not only the SR to USR transition scale, but also the PBH formation scale. Here $M_{\odot}\sim 2\times 10^{30}{\rm kg}$ is the solar mass. Look at the ref. \cite{Ballesteros:2017fsr} for more details. Hence the PBH mass at the pivot scale $p_*=0.02\;{\rm Mpc}^{-1}$ can be estimated as:
\bea \boxed{\boxed{\frac{M_{\rm PBH}}{M_{\odot}}=4.52\times {10}^{-31}\bigg(\frac{\gamma}{0.2}\bigg)\bigg(\frac{g_*}{106.75}\bigg)^{-1/6}}}.\eea 
Finally, using the above result we can give the following estimation of the evaporation time scale of PBH:
\bea \boxed{\boxed{t^{\rm evap}_{\rm PBH}={10}^{64}\bigg(\frac{M_{\rm PBH}}{M_{\odot}}\bigg)^{3}{\rm years}\approx 2.02\times10^{-20}\bigg(\frac{\gamma}{0.2}\bigg)\bigg(\frac{g_*}{106.75}\bigg)^{-1/6}{\rm sec}}}.\eea 
Here $\gamma\sim 0.2$ is known as the efficiency factor for collapse and $g_*$ is the relativistic degrees of freedom which is, $g_*\sim 106.75$ for Standard Model and $g_*\sim 226$ for SUSY d.o.f (see ref.\cite{Hawking:1974rv} for details).  The above numerical estimations not only suggest that the span for PBH formation in terms of the number of e-foldings is small, but inevitably the estimated PBH mass is extremely small, and the corresponding evaporation time scale is also very small and highly constrained.

This calculation indicates that if one takes into account the one loop quantum effects or the contribution from all loop order DRG ressumed result of the power spectrum of scalar modes, then the USR phase will be very short lived ($\Delta{\cal N}_{\rm USR}\approx 2$), the implication of this in case of PBH and the phenomenology associated with this is quite huge. If one takes this method of the SR period is followed by a USR period and then completion of inflation as the procedure to produce PBHs, then our calculation points towards a {\it no-go theorem} on PBH mass with $M_{\rm PBH}\sim 10^{2}{\rm gm}$. But this  can have rich phenomenology in the early universe, such as baryogenesis\cite{Hamada:2016jnq,Blinnikov:2016bxu,Hasegawa:2018yuy,Morrison:2018xla,Kawasaki:2019iis,Wu:2021mwy,Gehrman:2022imk}. Another major motivation for believing that PBHs can be the answer to the identity of the dark matter remains unresolved, at least following this production mechanism of PBHs. Thus, in a nutshell, we propose a {\it no-go theorem} beyond $10^{2}{\rm gm}$ PBHs to maintain the purterbative sanctity of cosmological perturbation theory in the sub-horizon region, where the quantum loop effects are significant.

\section{Conclusion}
\label{conclu}
%%%
In this paper, we study the formation of PBHs in single-field models of inflation. In particular, we investigate one-loop corrections to the renormalized primordial power spectrum in the framework where PBHs are produced during the transition from SR to USR followed by the end of inflation.  
Using a general single-field inflationary paradigm, we considered the cosmological perturbation in the unitary gauge, which gives rise to the second order perturbed action for the scalar perturbation, which we used to compute the explicit expression for the scalar modes in the SR period by solving the Mukhanov Sasaki equation and applying the Bunch Davies quantum initial condition. We then study the behaviour of the solution at the sub and super-horizon scales and at the horizon crossing. The quantum features become dominant in the sub-horizon region, so this region should contribute significantly during loop computations. The super-horizon result, on the other hand, becomes fully classical where the scalar modes are frozen. Last but not least, the obtained result can be treated semi-classically at the horizon-crossing point. Next, we obtained the explicit general solution of the scalar modes during PBH formation by exploiting the continuity of the scalar modes and their canonically conjugate momenta at the sharp transition point from the SR to the USR region. The behaviour of these derived modes has also been investigated at the sub-horizon and super-horizon scales, as well as at the horizon crossing. Furthermore, we computed the one-loop quantum effects from the SR and USR regions using the well-known "{\it in-in" formalism}. \textcolor{black}{We have explicitly demonstrated that on the sub-horizon scale, due to the contribution from the short-range UV modes, both the quadratic UV divergent and IR dominated logarithmic divergent contributions survive in the one-loop correction in both the SR and USR regions. We also found from our computation that in the USR period, many more oscillating and power law suppressed contributions are appearing, which is adding a very small amplitude and negligible fluctuation to the overall one-loop correction}. We have also shown that at the horizon crossing point and super-horizon region, all other quantum effects that appear in the sub-horizon region are diluted by choosing proper counter terms. \textcolor{black}{To implement this, we have further used the well known adiabatic renormalization scheme, using which we have computed the expression for the counter terms both in the SR and USR phases in terms of an arbitrary renormalization scale  $\mu$.  Further, by choosing such a scale appropriately in the present computation, we have explicitly demonstrated that the quadratic UV divergence can be completely removed from the final result.  Also, the additional IR dominated logarithmic divergent contribution can be softened very easily, and it is further possible to shift such perturbative effects into the next order of perturbation theory by implementing power spectrum renormalization. With the help of these established facts, we have computed the expression for the renormalized one-loop corrected power spectrum for scalar modes by implementing the renormalization condition at the pivot scale where CMB observation takes place.} We have also found that the renormalized one-loop corrected power spectrum for scalar modes becomes free from all quantum effects at the pivot scale, which is consistent with the findings from cosmological observations. Additionally, we have found that, away from the pivot scale, the logarithmically divergent contribution appears in the second and higher order in the final result after including the effect of the determined {\it counter term} from this computation. 
 We have found that the spectral tilt, its running, and running of the running of the renormalized power spectrum are free from all divergent quantum effects at the pivot scale. This is the immediate outcome of having no quantum effect dependence on the renormalized power spectrum at the pivot scale. Away from the pivot scale, the said dependence is not important during the time of estimation because CMB observations can probe the information at the pivot scale only. 
Further, using the DRG resummation method, we have explicitly computed the expression for the convergent form of the power spectrum for the scalar modes. This resummation was performed to account for the contributions from the quantum IR logarithmic divergent effects in the power spectrum. We found that after performing the summation over the total convergent series made up of these loop effects, we get a final result that takes care of all possible allowed diagrams in the perturbative expansion. This finite resummed result is consistent with the one-loop results obtained from the SR period without renormalization, SR and USR periods with renormalization. The representative plots \ref{Spectrum},\ref{Spectrum2},\ref{Spectrum4} and \ref{Spectrum3} clearly show that, in some ways, we obtained better results after performing the resummation using the DRG method at late time scales. It actually helps us to dilute the logarithmically divergent contributions in a considerable manner and give rise to a finite re-summed contribution that is consistent with observational constraints.
From our analysis, we have found that the span of PBH formation in terms of the number of e-folds is extremely small, which is $2$ e-folds. In addition, we discovered that the PBH has a mass of $M_{\rm PBH}\sim 4.52\times {10}^{-31}M_{\odot}$, which is very small for SM and SUSY particles. This is going to directly affect the estimation of the evaporation time scale of PBH, which we have found $t_{\rm PBH}\sim 2.02\times10^{-20}{\rm sec}$, which is again very small. This implies that the re-summed finite quantum loop effects may be significant for this small mass of PBH, which evaporated at a very early time scale and formed for a very short period of time. If one takes this method of the SR period being followed by a USR period and then completion of inflation as the procedure to produce PBH's, then our calculation points towards the possibility of having a very small mass PBH ($M_{\rm PBH}\sim 10^{2}{\rm gm}$). In this paper we propose a {\it no-go theorem} beyond $10^{2}{\rm gm}$ PBH's (formation of large mass PBHs is not allowed by quantum loop effects) to maintain the purterbative sanctity of cosmological perturbation theory in the sub-horizon region, where the quantum loop effects are important. Finally, based on the proposed no-go theorem, we concluded that PBH formation cannot be ruled out for the {\it single field inflationary paradigm} due to having one loop quantum correction in the power spectrum for small mass PBHs. For large mass PBHs, the one-loop and DRG resummation prescriptions will fail due to a large wave number difference between the SR to USR transition scale and the end of the inflation scale, resulting in a large number of e-foldings ${\Delta \cal N}_{\rm USR}\gg 2$, which goes against the necessary requirement of having $\Delta {\cal N}_{\rm USR}= 2$ strictly within the current framework. The obtained quantum loop corrected results in this paper can easily refute the strong claim {\it Ruling Out Primordial Black Hole Formation From Single-Field Inflation} made in ref \cite{Kristiano:2022maq} with detailed proof and justification, at least for the small mass PBH generation. Thus we have shown that one-loop corrections do not rule out the PBHs formation but astonishingly enough, as a by-product of our calculations, we found a no-go to the amount of USR region, which puts a severe bound on the PBHs mass, and wipes out the hope to claim PBHs to be the dark matter at least through the framework where the SR to USR transition occurs followed by the end of inflation.

The fruitful immediate future prospects of our work are as follows: Due to an additional temperature-dependent parameter, the warm inflationary paradigm naturally describes the PBH formation at the tree level (see refs.\cite{Arya:2019wck,Bastero-Gil:2021fac,Correa:2022ngq} for details). It would be interesting to study the quantum one-loop effects and renormalization of the primordial power spectrum at a finite temperature in the framework of warm inflation. The next promising framework is the multi-field inflationary paradigm, particularly hybrid inflation, which can naturally describe the phenomenon of PBH formation at the tree level (see \cite{Brown:2017osf,Palma:2020ejf,Geller:2022nkr,Braglia:2022phb,Kawai:2022emp}); it would be interesting to extend our computations to multi-field inflation. Since the model independent framework with  SR to USR transition followed by the end of inflation only allows the generation of small mass PBHs (no-go for large mass PBHs) due to one loop effects, it is important to explore whether the same/different result will be obtained from the model-dependent prescription, where the inflationary effective potential has a dip or bump introduced by hand \cite{Mishra:2019pzq,ZhengRuiFeng:2021zoz}.

%%%%%%%%%%%%%%%%%%%%

%%%%%%%%%%%%%%%%%%%%

	\subsection*{Acknowledgements}
The authors would like to thank Grant J. Mathews and Katherine Freese for bringing our attention to this problem initially. The authors would also like to thank Romesh Kaul, Yogesh and Mohit Kumar Sharma for fruitful discussions. Additionally, the authors of this paper would like to sincerely thank 
Antonio Riotto, Vincent Vennin, Adam Brown, Maxim Khlopov, Amjad Ashoorioon, Gerasimos Rigopoulos, Ying-li, Shu-Lin Cheng, Da-Shin Lee, Kin-Wang Ng, Jinsu Kim, Theodoros Papanikolaou, Xinpeng Wang, H. V. Ragavendra, Alexandros Karam for fruitful communication. %\textcolor{red}{The authors would like to thank the Editor and the anonymous referee for constructive and fruitful suggestions which allowed us to significantly improve the content and the presentation of the paper.} 
SC would like to thank the work-friendly environment of The Thanu Padmanabhan Centre For Cosmology and Science Popularization (CCSP), SGT University, Gurugram, Delhi-NCR, for providing tremendous support in research and offering the Assistant Professor (Senior Grade) position. SC would like to thank The North American Nanohertz Observatory for Gravitational Waves (NANOGrav) collaboration and the National Academy of Sciences (NASI), Prayagraj, India, for being elected as an associate member and the member of the academy respectively. SC would also like to thank all the members of our newly formed virtual international non-profit consortium Quantum Aspects of the Space-Time \& Matter (QASTM) for elaborative discussions. The work of MRG is supported by DST,
Government of India under the Grant Agreement number
IF18-PH-228 (DST INSPIRE Faculty Award). The work of MRG and MS is supported by Science and Engineering Research Board (SERB), DST, Government of India under the Grant Agreement number CRG/2022/004120 (Core Research Grant). MS is also partially supported by the Ministry of Education and Science of the Republic of Kazakhstan, Grant
No. 0118RK00935 and CAS President's International Fellowship Initiative (PIFI).
%\newpage
\appendix
%\section{Appendix}

\section{Details of the one loop computation of two point function using in-in formalism in USR phase}\label{App:1}

\textcolor{black}{We now explicitly compute the contribution from the quantum one-loop correction to the power spectrum of the scalar mode during PBH formation from the last term of the third order expanded action as stated in equation(\ref{third}).  For this purpose we use the well known {\it in-in formalism},  which is actually motivated from the {\it Schwinger-Keldysh path integral formalism}.  Within the framework of {\it in-in formalism},  the correlation function of any quantum operator $\widehat{\cal W}(\tau)$ at the fixed conformal time scale $\tau$ can be expressed as \cite{Kristiano:2022maq}:}
\bea \textcolor{black}{\langle\widehat{\cal W}(\tau)\rangle:=\left\langle\bigg[\overline{T}\exp\bigg(i\int^{\tau}_{-\infty}d\tau^{'}\;H_{\rm int}(\tau^{'})\bigg)\bigg]\;\;\widehat{\cal W}(\tau)\;\;\bigg[{T}\exp\bigg(-i\int^{\tau}_{-\infty}d\tau^{'}\;H_{\rm int}(\tau^{'})\bigg)\bigg]\right\rangle_{\tau\rightarrow 0}},\quad\quad \eea
\textcolor{black}{where $\overline{T}$ and $T$ represent the anti-time and time ordering operation in the present context respectively. In the present context we are interested in the quantum operator,  $\widehat{\cal W}(\tau\rightarrow 0)=\hat{\zeta}_{\bf p}\hat{\zeta}_{-{\bf p}}$. The interaction Hamiltonian appearing in the above expression can be computed by the following {\it Legendre transformed} expression:}
\bea \textcolor{black}{H_{\rm int}(\tau)=-\int d^3x\; {\cal L}_{\rm int}(\tau)}.\eea
\textcolor{black}{Since we are interested in only on the contribution from last term in the equation(\ref{third}), which physically represents the leading cubic self-interaction. The corresponding interaction Hamiltonian can be expressed as:}
\bea \textcolor{black}{H_{\rm int}(\tau)=-\frac{M^2_{\rm pl}}{2}\int d^3x\;  a^2\epsilon\eta^{'}\zeta^{'}\zeta^{2}}.\eea
\textcolor{black}{Consequently,  we have the following expression for the correlation function of any quantum operator $\widehat{\cal W}(\tau)$ considering the contribution up to the one-loop level:}
\bea  &&\textcolor{black}{\langle\widehat{\cal W}(\tau)\rangle= \underbrace{\langle\widehat{\cal W}(\tau)\rangle_{(0,0)}}_{\bf Tree\;level}+\underbrace{\langle\widehat{\cal W}(\tau)\rangle_{(0,1)}+\langle\widehat{\cal W}(\tau)\rangle^{\dagger}_{(0,1)}+\langle\widehat{\cal W}(\tau)\rangle_{(0,2)}+\langle\widehat{\cal W}(\tau)\rangle^{\dagger}_{(0,2)}+\langle\widehat{\cal W}(\tau)\rangle_{(1,1)}}_{\bf One-loop\;level}},\eea
\textcolor{black}{where the first term represents the tree level VEV with respect to the Bunch Davies quantum vacuum state.  The rest of the one-loop contributions are given by the following expressions:}
\bea &&\textcolor{black}{\langle\widehat{\cal W}(\tau)\rangle_{(0,1)}=\int^{\tau}_{-\infty}d\tau_1\;\langle \widehat{\cal W}(\tau)H_{\rm int}(\tau_1)\rangle=0,}\\
 &&\textcolor{black}{\langle\widehat{\cal W}(\tau)\rangle^{\dagger}_{(0,1)}=\int^{\tau}_{-\infty}d\tau_1\;\langle \widehat{\cal W}(\tau)H_{\rm int}(\tau_1)\rangle^{\dagger}=0,}\\
 &&\textcolor{black}{\langle\widehat{\cal W}(\tau)\rangle_{(0,2)}=\int^{\tau}_{-\infty}d\tau_1\;\int^{\tau}_{-\infty}d\tau_2\;\langle \widehat{\cal W}(\tau)H_{\rm int}(\tau_1)H_{\rm int}(\tau_2)\rangle\neq 0,}\\
 &&\textcolor{black}{\langle\widehat{\cal W}(\tau)\rangle^{\dagger}_{(0,2)}=\int^{\tau}_{-\infty}d\tau_1\;\int^{\tau}_{-\infty}d\tau_2\;\langle \widehat{\cal W}(\tau)H_{\rm int}(\tau_1)H_{\rm int}(\tau_2)\rangle^{\dagger}\neq 0,}\\
  &&\textcolor{black}{\langle\widehat{\cal W}(\tau)\rangle^{\dagger}_{(1,1)}=\int^{\tau}_{-\infty}d\tau_1\;\int^{\tau}_{-\infty}d\tau_2\;\langle H_{\rm int}(\tau_1)\widehat{\cal W}(\tau)H_{\rm int}(\tau_2)\rangle^{\dagger}\neq 0.}\eea

  \textcolor{black}{After substituting the specific form of the cubic self-interaction as appearing in the Hamiltonian, we have the following non-vanishing contributions coming up in the present computation:}
  \bea  \textcolor{black}{\langle\hat{\zeta}_{\bf p}\hat{\zeta}_{-{\bf p}}\rangle_{(0,2)}}&=& \textcolor{black}{-\frac{M^4_{\rm pl}}{4}\int^{0}_{-\infty}d\tau_1\;a^2(\tau_1)\epsilon(\tau_1)\eta^{'}(\tau_1)\;\int^{0}_{-\infty}d\tau_2\;a^2(\tau_2)\epsilon(\tau_2)\eta^{'}(\tau_2)}\nonumber\\
  &&\textcolor{black}{\times\int \frac{d^{3}{\bf k}_1}{(2\pi)^3} \int \frac{d^{3}{\bf k}_2}{(2\pi)^3} \int \frac{d^{3}{\bf k}_3}{(2\pi)^3} \int \frac{d^{3}{\bf k}_4}{(2\pi)^3} \int \frac{d^{3}{\bf k}_5}{(2\pi)^3} \int \frac{d^{3}{\bf k}_6}{(2\pi)^3}}\nonumber\\
  &&\textcolor{black}{\times \delta^3\bigg({\bf k}_1+{\bf k}_2+{\bf k}_3\bigg) \delta^3\bigg({\bf k}_4+{\bf k}_5+{\bf k}_6\bigg)}\nonumber\\
  &&\textcolor{black}{\times \langle \hat{\zeta}_{\bf p}\hat{\zeta}_{-{\bf p}}\hat{\zeta}^{'}_{{\bf k}_1}(\tau_1)\hat{\zeta}_{{\bf k}_2}(\tau_1)\hat{\zeta}_{{\bf k}_3}(\tau_1)\hat{\zeta}^{'}_{{\bf k}_4}(\tau_2)\hat{\zeta}_{{\bf k}_5}(\tau_2)\hat{\zeta}_{{\bf k}_6}(\tau_2)\rangle,}\eea\bea
  \textcolor{black}{\langle\hat{\zeta}_{\bf p}\hat{\zeta}_{-{\bf p}}\rangle^{\dagger}_{(0,2)}}&=& \textcolor{black}{-\frac{M^4_{\rm pl}}{4}\int^{0}_{-\infty}d\tau_1\;a^2(\tau_1)\epsilon(\tau_1)\eta^{'}(\tau_1)\;\int^{0}_{-\infty}d\tau_2\;a^2(\tau_2)\epsilon(\tau_2)\eta^{'}(\tau_2)}\nonumber\\
  &&\textcolor{black}{\times\int \frac{d^{3}{\bf k}_1}{(2\pi)^3} \int \frac{d^{3}{\bf k}_2}{(2\pi)^3} \int \frac{d^{3}{\bf k}_3}{(2\pi)^3} \int \frac{d^{3}{\bf k}_4}{(2\pi)^3} \int \frac{d^{3}{\bf k}_5}{(2\pi)^3} \int \frac{d^{3}{\bf k}_6}{(2\pi)^3}}\nonumber\\
  &&\textcolor{black}{\times \delta^3\bigg({\bf k}_1+{\bf k}_2+{\bf k}_3\bigg) \delta^3\bigg({\bf k}_4+{\bf k}_5+{\bf k}_6\bigg)}\nonumber\\
  &&\textcolor{black}{\times \langle \hat{\zeta}_{\bf p}\hat{\zeta}_{-{\bf p}}\hat{\zeta}^{'}_{{\bf k}_1}(\tau_1)\hat{\zeta}_{{\bf k}_2}(\tau_1)\hat{\zeta}_{{\bf k}_3}(\tau_1)\hat{\zeta}^{'}_{{\bf k}_4}(\tau_2)\hat{\zeta}_{{\bf k}_5}(\tau_2)\hat{\zeta}_{{\bf k}_6}(\tau_2)\rangle^{\dagger},}\\ 
 \textcolor{black}{\langle\hat{\zeta}_{\bf p}\hat{\zeta}_{-{\bf p}}\rangle_{(1,1)}}&=& \textcolor{black}{\frac{M^4_{\rm pl}}{4}\int^{0}_{-\infty}d\tau_1\;a^2(\tau_1)\epsilon(\tau_1)\eta^{'}(\tau_1)\;\int^{0}_{-\infty}d\tau_2\;a^2(\tau_2)\epsilon(\tau_2)\eta^{'}(\tau_2)}\nonumber\\
  &&\textcolor{black}{\times\int \frac{d^{3}{\bf k}_1}{(2\pi)^3} \int \frac{d^{3}{\bf k}_2}{(2\pi)^3} \int \frac{d^{3}{\bf k}_3}{(2\pi)^3} \int \frac{d^{3}{\bf k}_4}{(2\pi)^3} \int \frac{d^{3}{\bf k}_5}{(2\pi)^3} \int \frac{d^{3}{\bf k}_6}{(2\pi)^3}}\nonumber\\
  &&\textcolor{black}{\times \delta^3\bigg({\bf k}_1+{\bf k}_2+{\bf k}_3\bigg) \delta^3\bigg({\bf k}_4+{\bf k}_5+{\bf k}_6\bigg)}\nonumber\\
  &&\textcolor{black}{\times \langle \hat{\zeta}^{'}_{{\bf k}_1}(\tau_1)\hat{\zeta}_{{\bf k}_2}(\tau_1)\hat{\zeta}_{{\bf k}_3}(\tau_1)\hat{\zeta}_{\bf p}\hat{\zeta}_{-{\bf p}}\hat{\zeta}^{'}_{{\bf k}_4}(\tau_2)\hat{\zeta}_{{\bf k}_5}(\tau_2)\hat{\zeta}_{{\bf k}_6}(\tau_2)\rangle.} \eea
  \textcolor{black}{It is important to note that,  in the SR and USR regions,  the second slow roll parameter $\eta$ behaves as a constant for which one can approximately consider $\eta^{'}(\tau)\approx 0$.  However,  this approximation does not hold good at the conformal time scales at $\tau=\tau_s$ and $\tau=\tau_e$.  So instead of having the contribution from the full time scale $-\infty<\tau<0$ the only significant contributions will appear at $\tau=\tau_s$ and $\tau=\tau_e$.  Consequently,  the associated conformal time integral part as appearing in the above mentioned correlation functions can be evaluated by considering the following approximated expression  \cite{Kristiano:2022maq}:}
  \bea \textcolor{black}{\int^{0}_{-\infty}d\tau\;\eta^{'}(\tau)\; {\cal F}(\tau)
 }&\approx&\textcolor{black}{\bigg(\Delta \eta(\tau_e)\; {\cal F}(\tau_e)-\Delta \eta(\tau_s)\; {\cal F}(\tau_s)\bigg)-\underbrace{\int^{0}_{-\infty}d\tau\;\eta(\tau)\; {\cal F}^{'}(\tau)}_{\approx 0}}\nonumber\\&\approx& \textcolor{black}{\bigg(\Delta \eta(\tau_e)\; {\cal F}(\tau_e)-\Delta \eta(\tau_s)\; {\cal F}(\tau_s)\bigg),}
 \eea
 \textcolor{black}{where ${\cal F}(\tau)$ represents the conformal time dependent contributions in each of the integrals appearing in the above mentioned correlations which, in principle, are continuous function.  The scalar modes are functions of a fixed time $\tau_s$, where the SR to USR sharp transition occurs, in the sub-horizon region and at the horizon crossing, where quantum effects are dominant. On the other hand,  the super horizon scale modes are function of the fixed time scale $\tau_e$ when the inflation ends. But it is strictly not allowed to consider the effects of the superhorizon scalar modes in this computation as it becomes classical, and such contributions will not matter for the quantum one-loop corrected part of the two-point correlation function and its associated scalar power spectrum.  Another important fact is that,  the first slow roll parameter $\epsilon$ is constant in SR and USR regions including all time scales.  As a consequence one can immediately consider $\epsilon^{'}(\tau)\approx 0$,   which further implies ${\cal F}^{'}(\tau)\approx 0$. Consequently, one can immediately ignore the contribution of the last integral in the above mentioned expression which is appearing as an outcome of integration by parts.}
 
\textcolor{black}{As a result,  after applying the all possible Wick contraction the one-loop contribution to the two-point correlation function of the scalar perturbation can be further simplified as:}
 \bea \textcolor{black}{\langle\langle\hat{\zeta}_{\bf p}\hat{\zeta}_{-{\bf p}}\rangle\rangle_{\bf One-loop}}&=&\textcolor{black}{\langle\langle\hat{\zeta}_{\bf p}\hat{\zeta}_{-{\bf p}}\rangle\rangle_{(1,1)}+2{\rm Re}\bigg[\langle\langle\hat{\zeta}_{\bf p}\hat{\zeta}_{-{\bf p}}\rangle\rangle_{(0,2)}\bigg]}\nonumber\\
 &\approx&\textcolor{black}{\frac{M^4_{\rm pl}}{4}a^4(\tau_e)\epsilon^2(\tau_e)\left(\Delta\eta(\tau_e)\right)^2}\nonumber\\
 &&\textcolor{black}{\times\int \frac{d^{3}{\bf k}}{(2\pi)^3}\bigg[4\zeta_{{\bf p}}\zeta^{*}_{{\bf p}}\zeta^{'}_{{\bf p}}\zeta^{*'}_{{\bf p}}\zeta_{{\bf k}}\zeta^{*}_{{\bf k}}\zeta_{{\bf k}-{\bf p}}\zeta^{*}_{{\bf k}-{\bf p}}+8\zeta_{{\bf p}}\zeta^{*}_{{\bf p}}\zeta^{*'}_{{\bf p}}\zeta_{{\bf p}}\zeta^{'}_{{\bf k}}\zeta^{*}_{{\bf k}}\zeta_{{\bf k}-{\bf p}}\zeta^{*}_{{\bf k}-{\bf p}}}\nonumber\\
 &&\textcolor{black}{+8\zeta_{{\bf p}}\zeta^{*}_{{\bf p}}\zeta^{'}_{{\bf p}}\zeta^{*}_{{\bf p}}\zeta^{*'}_{{\bf k}}\zeta_{{\bf k}}\zeta_{{\bf k}-{\bf p}}\zeta^{*}_{{\bf k}-{\bf p}}-{\rm Re}\bigg(4\zeta_{{\bf p}}\zeta_{{\bf p}}\zeta^{*'}_{{\bf p}}\zeta^{*'}_{{\bf p}}\zeta_{{\bf k}}\zeta^{*}_{{\bf k}}\zeta_{{\bf k}-{\bf p}}\zeta^{*}_{{\bf k}-{\bf p}}}\nonumber\\
 &&\textcolor{black}{+8\zeta_{{\bf p}}\zeta_{{\bf p}}\zeta^{'*}_{{\bf p}}\zeta^{*}_{{\bf p}}\zeta^{'}_{{\bf k}}\zeta^{*}_{{\bf k}}\zeta_{{\bf k}-{\bf p}}\zeta^{*}_{{\bf k}-{\bf p}}+8\zeta_{{\bf p}}\zeta_{{\bf p}}\zeta^{*'}_{{\bf p}}\zeta^{*}_{{\bf p}}\zeta^{*'}_{{\bf k}}\zeta_{{\bf k}}\zeta_{{\bf k}-{\bf p}}\zeta^{*}_{{\bf k}-{\bf p}}\bigg)\bigg]_{\tau=\tau_e}}\nonumber\eea\bea
 &&\textcolor{black}{-\frac{M^4_{\rm pl}}{4}a^4(\tau_s)\epsilon^2(\tau_s)\left(\Delta\eta(\tau_s)\right)^2}\nonumber\\
 &&\textcolor{black}{\times\int \frac{d^{3}{\bf k}}{(2\pi)^3}\bigg[4\zeta_{{\bf p}}\zeta^{*}_{{\bf p}}\zeta^{'}_{{\bf p}}\zeta^{*'}_{{\bf p}}\zeta_{{\bf k}}\zeta^{*}_{{\bf k}}\zeta_{{\bf k}-{\bf p}}\zeta^{*}_{{\bf k}-{\bf p}}+8\zeta_{{\bf p}}\zeta^{*}_{{\bf p}}\zeta^{*'}_{{\bf p}}\zeta_{{\bf p}}\zeta^{'}_{{\bf k}}\zeta^{*}_{{\bf k}}\zeta_{{\bf k}-{\bf p}}\zeta^{*}_{{\bf k}-{\bf p}}}\nonumber\\
 &&\textcolor{black}{+8\zeta_{{\bf p}}\zeta^{*}_{{\bf p}}\zeta^{'}_{{\bf p}}\zeta^{*}_{{\bf p}}\zeta^{*'}_{{\bf k}}\zeta_{{\bf k}}\zeta_{{\bf k}-{\bf p}}\zeta^{*}_{{\bf k}-{\bf p}}-{\rm Re}\bigg(4\zeta_{{\bf p}}\zeta_{{\bf p}}\zeta^{*'}_{{\bf p}}\zeta^{*'}_{{\bf p}}\zeta_{{\bf k}}\zeta^{*}_{{\bf k}}\zeta_{{\bf k}-{\bf p}}\zeta^{*}_{{\bf k}-{\bf p}}}\nonumber\\
 &&\textcolor{black}{+8\zeta_{{\bf p}}\zeta_{{\bf p}}\zeta^{'*}_{{\bf p}}\zeta^{*}_{{\bf p}}\zeta^{'}_{{\bf k}}\zeta^{*}_{{\bf k}}\zeta_{{\bf k}-{\bf p}}\zeta^{*}_{{\bf k}-{\bf p}}+8\zeta_{{\bf p}}\zeta_{{\bf p}}\zeta^{*'}_{{\bf p}}\zeta^{*}_{{\bf p}}\zeta^{*'}_{{\bf k}}\zeta_{{\bf k}}\zeta_{{\bf k}-{\bf p}}\zeta^{*}_{{\bf k}-{\bf p}}\bigg)\bigg]_{\tau=\tau_s}}\nonumber\\
 &=&\textcolor{black}{\frac{M^4_{\rm pl}}{4}a^4(\tau_e)\epsilon^2(\tau_e)\left(\Delta\eta(\tau_e)\right)^2\times 16\int \frac{d^{3}{\bf k}}{(2\pi)^3}\bigg[|\zeta_{\bf p}|^{2}|\zeta_{{\bf k}-{\bf p}}|^{2}{\rm Im}\bigg(\zeta^{'}_{{\bf p}}\zeta^{*}_{{\bf p}}\bigg){\rm Im}\bigg(\zeta^{'}_{{\bf k}}\zeta^{*}_{{\bf k}}\bigg)\bigg]_{\tau=\tau_e}}\nonumber\\
 &&\textcolor{black}{-\frac{M^4_{\rm pl}}{4}a^4(\tau_s)\epsilon^2(\tau_s)\left(\Delta\eta(\tau_s)\right)^2\times 16\int \frac{d^{3}{\bf k}}{(2\pi)^3}\bigg[|\zeta_{\bf p}|^{2}|\zeta_{{\bf k}-{\bf p}}|^{2}{\rm Im}\bigg(\zeta^{'}_{{\bf p}}\zeta^{*}_{{\bf p}}\bigg){\rm Im}\bigg(\zeta^{'}_{{\bf k}}\zeta^{*}_{{\bf k}}\bigg)\bigg]_{\tau=\tau_s}.}
 \nonumber\\
 &&\eea
 \textcolor{black}{Here it is important to note that in general we have the following expressions:}
 \bea &&\label{im1}\textcolor{black}{\bigg[{\rm Im}\bigg(\zeta^{'}_{{\bf k}}\zeta^{*}_{{\bf k}}\bigg)\bigg]=-\frac{1}{3k^3_e}\left(\frac{k}{k_s}\right)^3\left(\frac{k_e}{k_s}\right)^{3}\left(\frac{H^{2}}{4\pi^{2}M^{2}_{\rm pl}\epsilon}\right),}\\
&&\label{im2}\textcolor{black}{\bigg[{\rm Im}\bigg(\zeta^{'}_{{\bf p}}\zeta^{*}_{{\bf p}}\bigg)\bigg]=-\frac{1}{3p^3_e}\left(\frac{p}{k_s}\right)^3\left(\frac{k_e}{k_s}\right)^{3}\left(\frac{H^{2}}{4\pi^{2}M^{2}_{\rm pl}\epsilon}\right).} \eea
 \textcolor{black}{using which at the conformal time scale $\tau=\tau_e$ and $\tau=\tau_s$ we have the following simplified expressions:}
 \bea &&\label{ima}\textcolor{black}{\bigg[{\rm Im}\bigg(\zeta^{'}_{{\bf k}}\zeta^{*}_{{\bf k}}\bigg)\bigg]_{\tau=\tau_e}=\bigg[{\rm Im}\bigg(\zeta^{'}_{{\bf p}}\zeta^{*}_{{\bf p}}\bigg)\bigg]_{\tau=\tau_e}=-\frac{1}{3k^3_e}\left(\frac{k_e}{k_s}\right)^{6}\left(\frac{H^{2}}{4\pi^{2}M^{2}_{\rm pl}\epsilon}\right)},\\
 &&\label{imb}\textcolor{black}{\bigg[{\rm Im}\bigg(\zeta^{'}_{{\bf k}}\zeta^{*}_{{\bf k}}\bigg)\bigg]_{\tau=\tau_s}=\bigg[{\rm Im}\bigg(\zeta^{'}_{{\bf p}}\zeta^{*}_{{\bf p}}\bigg)\bigg]_{\tau=\tau_s}=-\frac{1}{3k^3_s}\left(\frac{H^{2}}{4\pi^{2}M^{2}_{\rm pl}\epsilon}\right).}\quad\quad\quad\eea
 \textcolor{black}{This is only possible as in eqn (\ref{ima}) and eqn (\ref{imb}), for the imaginary parts of the integrands, the contributions are coming from the two points,   which are the wave numbers $k_s$ and $k_e$ associated with the conformal time scale $\tau_s$ and $\tau_e$ respectively. }

 \section{First step on renormalization}
\label{App:R}

\textcolor{black}{Translating the computation into a more comprehensible language will allow us to link it to the traditional renormalization techniques employed in quantum field theory with more ease. In this part, instead of utilizing several methods of this kind, we will just focus on the process of eliminating divergences at the level of the unrenormalized/bare action through the introduction of counter-terms. Ultimately, this will provide the renormalized form of the action, where, upon successful completion of the technique, any potentially harmful divergences—specifically, quadratic UV divergence—may be entirely eliminated and logarithmic IR divergences smoothed.}

\textcolor{black}{Let us start with the following bare action:}
\bea
\label{action3}
         \textcolor{black}{S_{\zeta,{\bf B}}^{(3)}} &=& \textcolor{black}{M^{2}_{p}\int d\tau\;d^3x\;\bigg [\left({\bf g}_1\right)_{\bf B}\; \zeta^{\prime} {^2}_{\bf B} \zeta_{\bf B} + \left({\bf g}_2\right)_{\bf B}\;(\partial_i \zeta_{\bf B})^2 \zeta_{\bf B}  -  \left({\bf g}_3\right)_{\bf B}\;\zeta^{\prime}_{\bf B} (\partial_i \zeta_{\bf B}) \bigg (\partial_i \partial ^{-2}\bigg(\epsilon \zeta^{\prime}_{\bf B}\bigg)\bigg)} \nonumber \\ 
        && \quad \quad \quad \quad\textcolor{black}{-   \left({\bf g}_4\right)_{\bf B}\;\zeta_{\bf B} \bigg(\partial_i \partial_j \partial^{-2}\bigg (\epsilon \zeta^{\prime}_{\bf B}\bigg)\bigg)^2 + \underbrace{ \left({\bf g}_5\right)_{\bf B}\zeta^{\prime}_{\bf B} \zeta^2_{\bf B}}_{\textbf{Dominant term in USR}}+.....\bigg]},\quad\quad\quad
   \eea 
   \textcolor{black}{where the bare coupling parameters $\left({\bf g}_i\right)_{\bf B}\forall i=1,2,\cdots,5$ are defined as:}
   \bea  &&\textcolor{black}{\left({\bf g}_1\right)_{\bf B}=\bigg(\epsilon ^2 - \frac{\epsilon ^3}{2}\bigg )a^2,~~~~~\left({\bf g}_2\right)_{\bf B}=\epsilon^2a^2}, \nonumber\\
    &&\textcolor{black}{\left({\bf g}_3\right)_{\bf B}=\epsilon a^2,~~\left({\bf g}_4\right)_{\bf B}=\frac{\epsilon}{2}a^2,~~\left({\bf g}_5\right)_{\bf B}=\frac{\epsilon}{2}\eta^{'}a^2}.\eea
    \textcolor{black}{The renormalized form of the third order action for the comoving scalar curvature perturbation can be obtained by using the rescaling ansatz of the gauge invariant modes, which is very helpful in figuring out the relationship between the renormalized, unrenormalized/bare, and counter-term contributions. The renormalized version of the third-order action is given by the following expression:}
\bea \textcolor{black}{\zeta_{\bf R}=\zeta_{\bf B}-\zeta_{\bf C}=\sqrt{{\bf Z}^{\rm IR}}\times \zeta_{\bf B}\quad\quad\quad{\rm where}\quad\quad\quad {\bf Z}^{\rm IR}:=\left(1+\delta_{{\bf Z}^{\rm IR}}\right)}.\eea
\textcolor{black}{Renormalized contributions are denoted by the subscripts {\bf R}, {\bf B}, and {\bf C}, respectively. Firstly, it should be noted that the amount ${\bf Z}^{\rm IR}$, or more accurately $\delta_{{\bf Z}^{\rm IR}}$, is often called the counter-term. To estimate this amount directly, we must use the renormalization condition.}

\textcolor{black}{The expression for the third-order unrenormalized/bare action must then be translated into terms of the renormalized version using the previously mentioned rescaled renormalized form of the curvature perturbation. This may be simply accomplished by carrying out the following tasks:}
\begin{itemize}
    \item[$\bigstar$] \textcolor{black}{The renormalized coupling parameters of the third-order perturbed action may be expressed in terms of the contributions of the bare and counter-terms as follows, using the previously given ansatz:}
\bea \textcolor{black}{\left({\bf g}_i\right)_{\bf R}}&=&\textcolor{black}{\left({\bf g}_i\right)_{\bf B}-\left({\bf g}_i\right)_{\bf C}={\bf Z}_{{\bf g}_i}\times \left({\bf g}_i\right)_{\bf B}\quad\quad\quad{\rm where}\quad\quad\quad {\bf Z}_{{\bf g}_i}:=\left(1+\delta_{{\bf Z}_{{\bf g}_i}}\right)\forall~i=1,2,\cdots,5,}\eea

    \item[$\bigstar$]\textcolor{black}{In this case, the couplings and the independent operator contributions may be expressed as follows:}
\bea &&\textcolor{black}{\left({\bf g}_1\right)_{\bf R} \zeta^{\prime} {^2}_{\bf R} \zeta_{\bf R}
     = \left(1+\delta_{{\bf Z}_{{\bf g}_1}}+\frac{3}{2}\delta_{{\bf Z}^{\rm IR}}+\cdots\right)\times \left({\bf g}_1\right)_{\bf B}\zeta^{\prime} {^2}_{\bf B} \zeta_{\bf B}},\nonumber\\
&&\textcolor{black}{\left({\bf g}_2\right)_{\bf R} (\partial_i \zeta_{\bf R})^2 \zeta_{\bf R}
     = \left(1+\delta_{{\bf Z}_{{\bf g}_2}}+\frac{3}{2}\delta_{{\bf Z}^{\rm IR}}+\cdots\right)\times \left({\bf g}_2\right)_{\bf B}(\partial_i \zeta_{\bf B})^2 \zeta_{\bf B}},\nonumber\\
    && \textcolor{black}{\left({\bf g}_3\right)_{\bf R} \zeta^{\prime}_{\bf R} (\partial_i \zeta_{\bf R}) \bigg (\partial_i \partial ^{-2}\bigg(\epsilon \zeta^{\prime}_{\bf R}\bigg)\bigg)= \left(1+\delta_{{\bf Z}_{{\bf g}_3}}+\frac{3}{2}\delta_{{\bf Z}^{\rm IR}}+\cdots\right)\times \left({\bf g}_3\right)_{\bf B}(\partial_i \zeta_{\bf B}) \bigg (\partial_i \partial ^{-2}\bigg(\epsilon \zeta^{\prime}_{\bf B}\bigg)\bigg)},\nonumber\\
&&\textcolor{black}{\left({\bf g}_4\right)_{\bf R} \zeta_{\bf R} \bigg (\partial_i\partial_j \partial ^{-2}\bigg(\epsilon \zeta^{\prime}_{\bf R}\bigg)\bigg)^2= \left(1+\delta_{{\bf Z}_{{\bf g}_4}}+\frac{3}{2}\delta_{{\bf Z}^{\rm IR}}+\cdots\right)\times \left({\bf g}_4\right)_{\bf B}\zeta_{\bf B}\bigg (\partial_i\partial_j \partial ^{-2}\bigg(\epsilon \zeta^{\prime}_{\bf B}\bigg)\bigg)^2},\nonumber\\
     &&\textcolor{black}{\left({\bf g}_5\right)_{\bf R} \zeta^{\prime}_{\bf R} \zeta^2_{\bf R}
     = \left(1+\delta_{{\bf Z}_{{\bf g}_5}}+\frac{3}{2}\delta_{{\bf Z}^{\rm IR}}+\cdots\right)\times \left({\bf g}_5\right)_{\bf B}\zeta^{\prime}_{\bf B} \zeta^2_{\bf B}}.
   \eea
   \textcolor{black}{In this instance, we have used the subsequent universal scaling relationship:}
 \bea &&\textcolor{black}{{\bf Z}_{{\bf g}_i}\left({\bf Z}^{\rm IR}\right)^{\frac{3}{2}}\approx \left(1+\delta_{{\bf Z}_{{\bf g}_i}}+\frac{3}{2}\delta_{{\bf Z}^{\rm IR}}+\cdots\right)\quad\quad\forall \quad i=1,2,\cdots,5}.\eea
\textcolor{black}{In this case, the dotted contributions $\cdots$ represent the higher-order terms in the associated power-series expansion. All higher-order minor effects have been disregarded, and we have limited our study to first-order variables. Consequently, this suggests that we have completed the last calculations needed to ascertain the exact contributions of the counter-terms within the linear domain of the parallel expansion.}
    \end{itemize}
\textcolor{black}{Thus, for the gauge invariant comoving curvature perturbation, the third-ordered renormalized version of the perturbed actions may be obtained by completing all the previously specified steps:}
\bea
\label{ractionx}
         \textcolor{black}{S_{\zeta,{\bf R}}^{(3)}} &=&\textcolor{black}{M^{2}_{\rm pl}\int d\tau\;d^3x\;\bigg [\left({\bf g}_1\right)_{\bf R}\; \zeta^{\prime} {^2}_{\bf R} \zeta_{\bf R} + \left({\bf g}_2\right)_{\bf R}\;(\partial_i \zeta_{\bf R})^2 \zeta_{\bf R}  -  \left({\bf g}_3\right)_{\bf R}\;\zeta^{\prime}_{\bf R} (\partial_i \zeta_{\bf R}) \bigg (\partial_i \partial ^{-2}\bigg(\epsilon \zeta^{\prime}_{\bf R}\bigg)\bigg) }\nonumber \\ 
        && \quad \quad \quad \quad \quad \quad \quad \textcolor{black}{  -  \left({\bf g}_4\right)_{\bf R}\;\zeta_{\bf R} \bigg(\partial_i \partial_j \partial^{-2}\bigg (\epsilon \zeta^{\prime}_{\bf R}\bigg)\bigg)^2 + \underbrace{ \left({\bf g}_5\right)_{\bf R}\zeta^{\prime}_{\bf R} \zeta^2_{\bf R}}_{\textbf{Dominant term in USR}}+.....\bigg]}\nonumber\\
        &=&\textcolor{black}{M^{2}_{\rm pl}\int d\tau\;d^3x\;\bigg [\left(1+\delta_{{\bf Z}_{{\bf g}_1}}+\frac{3}{2}\delta_{{\bf Z}^{\rm IR}}+\cdots\right)\left({\bf g}_1\right)_{\bf B}\; \zeta^{\prime} {^2}_{\bf B} \zeta_{\bf B} +\left(1+\delta_{{\bf Z}_{{\bf g}_2}}+\frac{3}{2}\delta_{{\bf Z}^{\rm IR}}+\cdots\right)\left({\bf g}_2\right)_{\bf B}\;(\partial_i \zeta_{\bf B})^2 \zeta_{\bf B}}\nonumber \\ 
        && \quad \quad \quad \quad \quad \quad \quad   \textcolor{black}{ -  \left(1+\delta_{{\bf Z}_{{\bf g}_3}}+\frac{3}{2}\delta_{{\bf Z}^{\rm IR}}+\cdots\right)\left({\bf g}_3\right)_{\bf B}\;\zeta^{\prime}_{\bf B} (\partial_i \zeta_{\bf B}) \bigg (\partial_i \partial ^{-2}\bigg(\epsilon \zeta^{\prime}_{\bf B}\bigg)\bigg)} \nonumber \\
        && \quad \quad \quad \quad \quad \quad \quad \textcolor{black}{ -  \left(1+\delta_{{\bf Z}_{{\bf g}_4}}+\frac{3}{2}\delta_{{\bf Z}^{\rm IR}}+\cdots\right)\left({\bf g}_4\right)_{\bf B}\;\zeta_{\bf B} \bigg(\partial_i \partial_j \partial^{-2}\bigg (\epsilon \zeta^{\prime}_{\bf B}\bigg)\bigg)^2}\nonumber \\ 
        && \quad \quad \quad \quad \quad \quad \quad \textcolor{black}{ + \underbrace{\left(1+\delta_{{\bf Z}_{{\bf g}_5}}+\frac{3}{2}\delta_{{\bf Z}^{\rm IR}}+\cdots\right) \left({\bf g}_5\right)_{\bf B}\zeta^{\prime}_{\bf B} \zeta^2_{\bf B}}_{\textbf{Dominant term in USR}}+.....\bigg]}\nonumber\\
        &=&\textcolor{black}{S_{\zeta,{\bf B}}^{(3)}-S_{\zeta,{\bf C}}^{(3)}},
   \eea 
   \textcolor{black}{where $S_{\zeta,{\bf B}}^{(3)}$, the bare component of the third-order perturbed action, is specified directly in the equation (\ref{action3}). The third-order action may be further explained by the following phrase once the counter-terms have been introduced:}
\bea
\label{conaction}
        \textcolor{black}{ S_{\zeta,{\bf C}}^{(3)}} &=& \textcolor{black}{M^{2}_{\rm pl}\int d\tau\;d^3x\;\bigg [\left({\bf g}_1\right)_{\bf C}\; \zeta^{\prime} {^2}_{\bf C} \zeta_{\bf C} + \left({\bf g}_2\right)_{\bf C}\;(\partial_i \zeta_{\bf C})^2 \zeta_{\bf C}  -  \left({\bf g}_3\right)_{\bf C}\;\zeta^{\prime}_{\bf C} (\partial_i \zeta_{\bf C}) \bigg (\partial_i \partial ^{-2}\bigg(\epsilon \zeta^{\prime}_{\bf R}\bigg)\bigg)} \nonumber \\ 
        && \quad \quad \quad \quad \quad \quad \quad  \textcolor{black}{ - \left({\bf g}_4\right)_{\bf C}\;\zeta_{\bf C} \bigg(\partial_i \partial_j \partial^{-2}\bigg (\epsilon \zeta^{\prime}_{\bf C}\bigg)\bigg)^2 + \underbrace{ \left({\bf g}_5\right)_{\bf C}\zeta^{\prime}_{\bf C} \zeta^2_{\bf C}}_{\textbf{Dominant term}}+.....\bigg]}\nonumber\\
        &=&\textcolor{black}{\Bigg[S_{\zeta,{\bf C}}^{(3)}\Bigg]_{\bf UV}+\Bigg[S_{\zeta,{\bf C}}^{(3)}\Bigg]_{\bf IR}}.
   \eea 
  \textcolor{black}{ In the one-loop corrected primordial power spectrum, the counter-term contribution of the third-order action that is capable of eliminating the quadratic UV divergence contribution entirely is represented by the following expression:}
\bea
\label{UVaction3}
        \textcolor{black}{ \Bigg[S_{\zeta,{\bf C}}^{(3)}\Bigg]_{\bf UV} }
        &=&\textcolor{black}{-M^{2}_{\rm pl}\int d\tau\;d^3x\;\bigg [\delta_{{\bf Z}_{{\bf g}_1}}\left({\bf g}_1\right)_{\bf B}\; \zeta^{\prime} {^2}_{\bf B} \zeta_{\bf B} +\delta_{{\bf Z}_{{\bf g}_2}}\left({\bf g}_2\right)_{\bf B}\;(\partial_i \zeta_{\bf B})^2 \zeta_{\bf B}   -  \delta_{{\bf Z}_{{\bf g}_3}}\left({\bf g}_3\right)_{\bf B}\;\zeta^{\prime}_{\bf B} (\partial_i \zeta_{\bf B}) \bigg (\partial_i \partial ^{-2}\bigg(\epsilon \zeta^{\prime}_{\bf B}\bigg)\bigg)} \nonumber \\ 
        && \quad \quad \quad \quad \quad \quad \quad \textcolor{black}{ -   \delta_{{\bf Z}_{{\bf g}_4}}\left({\bf g}_4\right)_{\bf B}\;\zeta_{\bf B} \bigg(\partial_i \partial_j \partial^{-2}\bigg (\epsilon \zeta^{\prime}_{\bf B}\bigg)\bigg)^2 + \underbrace{\delta_{{\bf Z}_{{\bf g}_5}} \left({\bf g}_5\right)_{\bf B}\zeta^{\prime}_{\bf B} \zeta^2_{\bf B}}_{\textbf{Dominant term}}+.....\bigg]},\quad\quad\quad\eea
       \textcolor{black}{ Likewise, the following expression characterizes the counter-term contribution of the third-order action that might soften the behavior of the logarithmic IR divergence contribution in the one-loop adjusted primordial power spectrum:}
 \bea\label{IRaction3}
        \textcolor{black}{ \Bigg[S_{\zeta,{\bf C}}^{(3)}\Bigg]_{\bf IR} }
        &=&\textcolor{black}{-\frac{3}{2}M^{2}_{\rm pl}\int d\tau\;d^3x\;\delta_{{\bf Z}^{\rm IR}}\times\bigg [\left({\bf g}_1\right)_{\bf B}\; \zeta^{\prime} {^2}_{\bf B} \zeta_{\bf B} +\left({\bf g}_2\right)_{\bf B}\;(\partial_i \zeta_{\bf B})^2 \zeta_{\bf B}   -  \left({\bf g}_3\right)_{\bf B}\;\zeta^{\prime}_{\bf B} (\partial_i \zeta_{\bf B}) \bigg (\partial_i \partial ^{-2}\bigg(\epsilon \zeta^{\prime}_{\bf B}\bigg)\bigg)} \nonumber \\ 
        && \quad \quad \quad \quad \quad \quad \quad\quad \quad \quad \quad \quad \textcolor{black}{ -  \left({\bf g}_4\right)_{\bf B}\;\zeta_{\bf B} \bigg(\partial_i \partial_j \partial^{-2}\bigg (\epsilon \zeta^{\prime}_{\bf B}\bigg)\bigg)^2+ \underbrace{\left({\bf g}_5\right)_{\bf B}\zeta^{\prime}_{\bf B} \zeta^2_{\bf B}}_{\textbf{Dominant term}}+.....\bigg]}.\quad\quad\quad
   \eea
   \textcolor{black}{Utilizing the above-mentioned facts, for scalar modes, the complete regularized and renormalized one-loop corrected power spectrum may thus be written as follows:}
   \bea \textcolor{black}{ \overline{\Delta^{2}_{\zeta,{\bf Total}}(p)}}&=&\textcolor{black}{{\bf Z}^{\rm IR}\Delta^{2}_{\zeta, {\bf Total}}(p)={\bf Z}^{\rm UV}{\bf Z}^{\rm IR}\bigg[\Delta^{2}_{\zeta, {\bf Tree}}(p)\bigg]_{\bf SR}},\eea
   \textcolor{black}{where the definitions of IR and UV counter-terms are:}
\bea && \textcolor{black}{{\bf Z}^{\rm IR}=\left(1+\delta_{{\bf Z}^{\rm IR}}\right),~~~~~~~~~{\rm and}~~~~~~~~~ {\bf Z}^{\rm UV}=\left(1+\delta_{{\bf Z}^{\rm UV}}\right)}.\eea
\textcolor{black}{Specifically, the UV counter can be further expressed in terms counter-terms of the coupling parameters as appearing in the action as stated in equation (\ref{UVaction3}) for the SR and USR phases are given by the following expressions:}
\bea &&\textcolor{black}{\underline{\bf For~SR:}~~~~~~~~~~~~~~~~~~~~~~~~\delta_{{\bf Z}^{\rm UV}}:=\left(\delta_{{\bf Z}_{{\bf g}_1}}+\delta_{{\bf Z}_{{\bf g}_2}}+\delta_{{\bf Z}_{{\bf g}_3}}+\delta_{{\bf Z}_{{\bf g}_4}}\right)~~~~~{\rm as}~~~~~\delta_{{\bf Z}_{{\bf g}_5}}\approx 0},\\
&&\textcolor{black}{\underline{\bf For~USR:}~~~~~~~~~~~~~~~~~~~~~~~~\delta_{{\bf Z}^{\rm UV}}:=\delta_{{\bf Z}_{{\bf g}_5}}~~~~~{\rm as}~~~~~\left(\delta_{{\bf Z}_{{\bf g}_1}}+\delta_{{\bf Z}_{{\bf g}_2}}+\delta_{{\bf Z}_{{\bf g}_3}}+\delta_{{\bf Z}_{{\bf g}_4}}\right)\approx 0}.\eea
\textcolor{black}{Next, we use the renormalization condition to get the expression for the counter-terms as they show up in the derived result previously described. The current problem may be understood clearly using Renormalization Group (RG) flow, which will also successfully correct the structure of any counter-terms arising in the calculation that are UV and IR sensitive. The renormalized spectrum is represented in Fourier space, and this is made feasible by the flow equation and matching beta functions provided for the appropriate 1PI one-loop corrected renormalized two-point amplitude. In this case, the Callan–Symanzik equation for this cosmological arrangement may be expressed as follows:}
\bea \textcolor{black}{\frac{d}{d\ln \mu}\Bigg\{\bigg[\Delta^{2}_{\zeta,\textbf{Tree}}(k)\bigg]_{\textbf{SR}}\Bigg\}=\frac{d}{d\ln \mu}\Bigg\{\frac{\left[\overline{\Delta_{\zeta,\textbf{Total}}^{2}(k)}\right]}{{\bf Z}^{\rm IR}{\bf Z}^{\rm UV}}\Bigg\}=0}.\eea
\textcolor{black}{It is now crucial to keep in mind that, in this case, the appropriate total differential operator may be further minimized using the following formula:}
\bea \textcolor{black}{\frac{d}{d\ln \mu}=\bigg(\frac{\partial}{\partial\ln \mu}+\beta_{{\bf g}_1}\frac{\partial}{\partial {\bf g}_1}+\beta_{{\bf g}_2}\frac{\partial}{\partial {\bf g}_2}+\beta_{{\bf g}_3}\frac{\partial}{\partial {\bf g}_3}+\beta_{{\bf g}_4}\frac{\partial}{\partial {\bf g}_4}+\beta_{{\bf g}_5}\frac{\partial}{\partial {\bf g}_5}-\gamma_{\bf IR}-\gamma_{\bf UV}\Bigg)},\eea
\textcolor{black}{where we define:}
\bea \textcolor{black}{\gamma_{\rm IR}:=\left(\frac{\partial\ln {\bf Z}^{\rm IR}}{\partial\ln \mu}\right),~~~~~~~
 \gamma_{\bf UV}:=\left(\frac{\partial\ln {\bf Z}^{\rm UV}}{\partial\ln \mu}\right)}.\quad\eea
 \textcolor{black}{This further reduces to the following differential equation:}
\bea \textcolor{black}{\bigg(\frac{\partial}{\partial\ln \mu}+\beta_{{\bf g}_1}\frac{\partial}{\partial {\bf g}_1}+\beta_{{\bf g}_2}\frac{\partial}{\partial {\bf g}_2}+\beta_{{\bf g}_3}\frac{\partial}{\partial {\bf g}_3}+\beta_{{\bf g}_4}\frac{\partial}{\partial {\bf g}_4}+\beta_{{\bf g}_5}\frac{\partial}{\partial {\bf g}_5}-\gamma_{\bf IR}-\gamma_{\bf UV}\Bigg)\overline{\Delta_{\zeta,\textbf{Total}}^{2}(p)}=0}.\eea
\textcolor{black}{where the following describes the beta functions:}
\bea \textcolor{black}{\beta_{{\bf g}_1}}&=&\textcolor{black}{\left(\frac{\partial {\bf g}_1}{\partial\ln \mu}\right)= 2\epsilon a^2\Bigg[(\epsilon-\eta)\bigg(2\epsilon-\frac{3\epsilon^2}{2}\bigg)+\bigg(\epsilon-\frac{\epsilon^2}{2}\bigg)\Bigg]\left(1+\epsilon\right)},\\
\textcolor{black}{\beta_{{\bf g}_2}}
&=&\textcolor{black}{\left(\frac{\partial {\bf g}_2}{\partial\ln \mu}\right)=2\epsilon a^2\left[2\epsilon(\epsilon-\eta)+\epsilon\right]\left(1+\epsilon\right)},\quad\quad\quad\\
\textcolor{black}{\beta_{{\bf g}_3}}
&=&\textcolor{black}{\left(\frac{\partial {\bf g}_3}{\partial\ln \mu}\right)=4\epsilon a^2\left(\epsilon-\eta\right)\left(1+\epsilon\right)},\\
\textcolor{black}{\beta_{{\bf g}_4}}
&=&\textcolor{black}{\left(\frac{\partial {\bf g}_4}{\partial\ln \mu}\right)=\epsilon a^2 \left(\epsilon-\eta+1\right)\left(1+\epsilon\right)},\\
\textcolor{black}{\beta_{{\bf g}_5}}
&=&\textcolor{black}{\left(\frac{\partial {\bf g}_5}{\partial\ln \mu}\right)=a^2\epsilon\eta^{'}\Bigg[\left(\epsilon-\eta\right)+\frac{1}{2}\frac{d}{d\ln \mu}\ln \eta^{'}\Bigg]\left(1+\epsilon\right)}.
\eea
\textcolor{black}{It is possible to compute the flow equations below by identifying the IR and UV counter-terms in the current environment at the renormalization scale:}
\begin{itemize}
\item[$\blacksquare$] \textcolor{black}{The renormalized form of the spectral tilt for the scalar modes is described by the following first flow equation:}
 \bea 
\textcolor{black}{\left[\overline{n_{\zeta,\textbf{Total}}(p)-1}\right]}&=&\textcolor{black}{\frac{d}{d\ln p}\bigg(\ln \left[\overline{\Delta_{\zeta,\textbf{Total}}^{2}(p)}\right]\bigg)}\nonumber\\
&=&\textcolor{black}{{\bf Z}^{\bf IR}\times\Bigg[{\bf Z}^{\bf UV}\bigg(\bigg[n_{\zeta,\textbf{Tree}}(p)\bigg]_{{\bf SR}}-1\bigg)+\bigg(\frac{d{\bf Z}^{\bf UV}}{d\ln p}\bigg)\bigg(\ln \bigg[\Delta^{2}_{\zeta,\textbf{Tree}}(p)\bigg]_{\textbf{SR}}\bigg)\Bigg]}.\eea 

\item[$\blacksquare$] \textcolor{black}{The renormalized form of the spectral tilt running for the scalar modes is described by the following second flow equation:}
\bea 
\textcolor{black}{\left[\overline{\alpha_{\zeta,\textbf{Total}}(p)}\right]} &=&\textcolor{black}{\frac{d}{d\ln p}\bigg(\left[\overline{n_{\zeta,\textbf{Total}}(p)}\right]\bigg)}\nonumber\\
&=&\textcolor{black}{{\bf Z}^{\bf IR}\times\Bigg[{\bf Z}^{\bf UV}\bigg(\bigg[\alpha_{\zeta,\textbf{Tree}}(p)\bigg]_{{\bf SR}}\bigg)+2\bigg(\frac{d{\bf Z}^{\bf UV}}{d\ln p}\bigg)\bigg(\bigg[n_{\zeta,\textbf{Tree}}(p)\bigg]_{{\bf SR}}-1\bigg)}\nonumber\\
&&\quad\quad\quad\quad\quad\quad\quad\quad\quad\quad\quad\quad\quad\quad\quad\quad\textcolor{black}{+\bigg(\frac{d^2{\bf Z}^{\bf UV}}{d\ln p^2}\bigg)\bigg(\ln \bigg[\Delta^{2}_{\zeta,\textbf{Tree}}(p)\bigg]_{\textbf{SR}}\bigg)\Bigg]}.\eea

\item[$\blacksquare$] \textcolor{black}{The renormalized form of the running of the spectral tilt running for the scalar modes is described by the third flow equation, which is as follows:}
\bea 
\textcolor{black}{\left[\overline{\beta_{\zeta,\textbf{Total}}(k)}\right]}&=&\textcolor{black}{\frac{d}{d\ln p}\bigg(\left[\overline{\alpha_{\zeta,\textbf{Total}}(p)}\right]\bigg)}\nonumber\\
&=&\textcolor{black}{{\bf Z}^{\bf IR}\times\Bigg[{\bf Z}^{\bf UV}\bigg(\bigg[\beta_{\zeta,\textbf{Tree}}(p)\bigg]_{{\bf SR}}\bigg)+2\bigg(\frac{d{\bf Z}^{\bf UV}}{d\ln p}\bigg)\bigg(\bigg[\alpha_{\zeta,\textbf{Tree}}(p)\bigg]_{{\bf SR}}\bigg)}\nonumber\\
&&\quad\quad\quad\quad\quad\textcolor{black}{+3\bigg(\frac{d^2{\bf Z}^{\bf UV}}{d\ln p^2}\bigg)\bigg(\bigg[n_{\zeta,\textbf{Tree}}(p)\bigg]_{{\bf SR}}-1\bigg)+\bigg(\frac{d^3{\bf Z}^{\bf UV}}{d\ln p^3}\bigg)\bigg(\ln \bigg[\Delta^{2}_{\zeta,\textbf{Tree}}(p)\bigg]_{\textbf{SR}}\bigg)\Bigg]}.\quad\quad\eea

\end{itemize}
\textcolor{black}{The flow equations that come before it demonstrate that the two-point amplitude of the power spectrum of primordial scalar modes exhibits a scale-dependent feature when the IR and UV counter-term effects are seen in each individual expression. When we apply the renormalization criterion, the structure of the counter-terms that are both UV and IR-sensitive will be fixed. We will now proceed with this utilizing the established facts at the CMB pivot scale $p_*$, which prevents us from taking into account the following important constraints, which are explained in terms of renormalization conditions:}
\begin{itemize}
    \item[$\checkmark$] \textcolor{black}{\underline{\bf Condition I:} The first renormalization requirement is that the tree-level contribution recorded during the first SR phase must precisely match the two-point amplitude of the scalar power spectrum following renormalization at the CMB pivot scale $k_*$. Technically speaking, this claim may be expressed as follows:}
\bea \label{recon1z}\textcolor{black}{\left[\overline{\Delta_{\zeta,\textbf{Total}}^{2}(p_*)}\right]} &=& \textcolor{black}{\bigg[\Delta^{2}_{\zeta,\textbf{Tree}}(p_*)\bigg]_{\textbf{SR}}}.\eea

    \item[$\checkmark$] \textcolor{black}{\underline{\bf Condition II:} The second renormalization criteria is that the logarithmic derivative of the amplitude of the scalar power spectrum with respect to the momentum scale following renormalization must exactly match the tree-level contribution estimated in the first SR phase at the CMB pivot scale $p_*$. In a technical sense, this claim may be stated as follows:}
 \bea 
\textcolor{black}{\left[\overline{n_{\zeta,\textbf{Total}}(p_*)-1}\right]}&=&\textcolor{black}{\Bigg[\frac{d}{d\ln p}\bigg(\ln \left[\overline{\Delta_{\zeta,\textbf{Total}}^{2}(p)}\right]\bigg)\Bigg]_{p=p_*}=\bigg(\bigg[n_{\zeta,\textbf{Tree}}(p_*)\bigg]_{{\bf SR}}-1\bigg)}.\eea

    \item[$\checkmark$] \textcolor{black}{\underline{\bf Condition III:} The third renormalization requirement, which is the second logarithmic derivative of the two-point amplitude of the scalar power spectrum with respect to the momentum scale, must be fully satisfied by the tree-level contribution computed during the first slow-roll phase. Alternatively, this claim can be expressed technically as: }
\bea 
\textcolor{black}{\left[\overline{\alpha_{\zeta,\textbf{Total}}(p_*)}\right]}&=&\textcolor{black}{\Bigg[\frac{d}{d\ln p}\bigg(\left[\overline{n_{\zeta,\textbf{Total}}(p)}\right]\bigg)\Bigg]_{p=p_*}=\bigg[\alpha_{\zeta,\textbf{Tree}}(p_*)\bigg]_{{\bf SR}}}.\eea

\item[$\checkmark$] \textcolor{black}{\underline{\bf Condition IV:} According to the fourth renormalization criterion, the tree-level contribution computed in the first slow-roll phase at the CMB pivot scale $k_*$ must exactly equal the third logarithmic derivative of the two-point amplitude of the scalar power spectrum with respect to the momentum scale. Stated otherwise, the technical form of this claim is as follows:}
\bea 
\textcolor{black}{\left[\overline{\beta_{\zeta,\textbf{Total}}(p_*)}\right]}&=&\textcolor{black}{\Bigg[\frac{d}{d\ln p}\bigg(\left[\overline{\alpha_{\zeta,\textbf{Total}}(p)}\right]\bigg)\Bigg]_{p=p_*}=\bigg(\bigg[\beta_{\zeta,\textbf{Tree}}(p_*)\bigg]_{{\bf SR}}\bigg)}.\eea

\end{itemize}

\textcolor{black}{Further limits on the parameters of the UV and IR-sensitive counter-terms are obtained as a direct consequence of the four renormalization requirements previously mentioned. A detailed summary of these limitations is provided below:}
\begin{itemize}[label=\ding{212}]

\item \textcolor{black}{\underline{\textbf{Consequence I:}} The following constraint condition expresses the immediate result of the first renormalization condition:}
\bea \textcolor{black}{{\bf Z}^{\bf IR}(p_*)=  \frac{\small[\overline{\Delta_{\zeta,\textbf{Total}}^{2}(p_{*})}\small]}{\small[\Delta_{\zeta,\textbf{Total}}^{2}(p_{*})\small]} = \frac{\small[  \Delta_{\zeta,\textbf{Tree}}^{2}(p_{*})\small]_{\textbf{SR}}}{\small[\Delta_{\zeta,\textbf{Total}}^{2}(p_{*})\small]}=\frac{\small[  \Delta_{\zeta,\textbf{Tree}}^{2}(p_{*})\small]_{\textbf{SR}}}{{\bf Z}^{\bf UV}(p_*)\small[  \Delta_{\zeta,\textbf{Tree}}^{2}(p_{*})\small]_{\textbf{SR}}}\quad\Longrightarrow {\bf Z}^{\bf IR}(p_*){\bf Z}^{\bf UV}(p_*)=1}.\eea

\item \textcolor{black}{\underline{\textbf{Consequence II:}} The subsequent constraint condition expresses the immediate result of the second renormalization requirement:}
\bea \textcolor{black}{{\bf Z}^{\bf IR}(p_*){\bf Z}^{\bf UV}(p_*)=1\quad {\rm and}\quad \bigg(\frac{d{\bf Z}^{\rm UV}}{d\ln p}\bigg)_{p=p_*}=0}.\eea

\item \textcolor{black}{\underline{\textbf{Consequence III:}} The third renormalization condition's immediate result is conveyed by the following constraint condition:}
 \bea \textcolor{black}{\quad{\bf Z}^{\bf IR}(p_*){\bf Z}^{\bf UV}(p_*)=1,\quad \bigg(\frac{d{\bf Z}^{\bf UV}}{d\ln p}\bigg)_{p=p_*}=0\quad{\rm and}\quad\bigg(\frac{d^2{\bf Z}^{\bf UV}}{d\ln p^2}\bigg)_{p=p_*}=0}.\eea

\item \textcolor{black}{\underline{\textbf{Consequence IV:}} The following constraint condition expresses the immediate result of the fourth renormalization condition:}
 \bea \textcolor{black}{{\bf Z}^{\bf IR}(p_*){\bf Z}^{\bf UV}(p_*)=1,\bigg(\frac{d{\bf Z}^{\bf UV}}{d\ln p}\bigg)_{p=p_*}=0,\bigg(\frac{d^2{\bf Z}^{\bf UV}}{d\ln p^2}\bigg)_{p=p_*}=0\quad{\rm and}\quad \bigg(\frac{d^3{\bf Z}^{\bf UV}}{d\ln p^3}\bigg)_{p=p_*}=0}.\quad\quad\quad\eea
\end{itemize}
\textcolor{black}{Our thorough analysis of the issue revealed that, in order to meet the previously acquired sets of constraint criteria, we must very carefully define the UV counter-term ${\bf Z}^{\bf UV}$. Only if we appropriately eliminate the quadratic divergence component can this be accomplished. Upon completion, the IR counter-term ${\bf Z}^{\bf IR}$ will have its form automatically fixed. Notwithstanding the aforementioned limits at the CMB pivot scale, determining the precise form of the UV counter-term ${\bf Z}^{\bf UV}$ is incredibly challenging at this level. The primary challenge at the technical level arises from the necessity for counter-terms in the SR and USR phases to independently eliminate the quadratic divergences' contribution. As simple as it may appear, completing the computation we have done so far is rather challenging from a technological standpoint. The significance of the next three sections lies in the way the adiabatic renormalization approach allows us to fully eliminate the quadratic UV divergences' contributions from each of the SR and USR phases separately. 
Following completion of this, we may use the condition ${\bf Z}^{\bf IR}(p_*){\bf Z}^{\bf UV}(p_*)=1$ to quickly ascertain the explicit form of the IR counter-term. Together with the quadratic divergence-free result for ${\bf Z}^{\bf UV}$ that comes from the adiabatic renormalization scheme, we have used this condition in the context of power spectrum renormalization.}

\textcolor{black}{We firmly believe that providing a clear explanation of the relationship between adding a counter-term at the level of action—a standard approach within the framework of Quantum Field Theory—and the renormalization schemes for adiabatic/wave function, and power spectrum will aid in understanding the applicability of the results derived in this paper. To prevent any further misunderstandings regarding the renormalization schemes employed or the interconnection among various tools and techniques used in this paper, let us discuss it in detail in the following subsections.}
\begin{itemize}
\item \textcolor{black}{In the current computation, the counter-term contribution of the third order perturbed action—which we have indicated in our computation by $\Bigg[S_{\zeta,{\bf C}}^{(3)}\Bigg]_{\bf UV}$ in the equation (\ref{UVaction3})—is directly linked to the complete removal of the harmful quadratic UV divergence. It is feasible to clearly demonstrate as an immediate result of the computation with this particular section that the sum of the counter-terms $\delta_{{\cal Z}_{{\bf g}_i}}\forall i=1,2,\cdots,5$ for the five operators that were previously described may be represented in terms of a cumulative factor. We determine this factor to be the counter-term contribution that, when applied to the wave function/adiabatic renormalization scheme—a topic we will address in the following portion of this paper—would entirely eliminate the quadratic UV divergence.}

\item \textcolor{black}{However, in our computation, we have denoted the third-order perturbed action as $\Bigg[S_{\zeta,{\bf C}}^{(3)}\Bigg]_{\bf IR}$ in the equation (\ref{IRaction3}), which is directly linked to the coarse-graining and smoothing of the logarithmic IR divergence's behaviour. It is feasible to clearly demonstrate that the single counter term $\delta_{{\bf Z}^{\bf IR}}$ represents the instantaneous result of the calculation using this particular portion. This factor will smooth the logarithmic IR divergence's behaviour by moving it to the higher order during the computation of the 1PI one-loop corrected two-point amplitude. This higher order corresponds to the higher even loop diagrams that arise in the perturbative expansion. Later in this work, we shall address this issue in depth in the context of the power spectrum renormalization technique.}

\item \textcolor{black}{Further, we justify more deeply the applicability of two renormalization schemes within one
theoretical framework that we are studying in this paper. At the level of third-order perturbed action representing the counter-term contributions we found that the UV and IR sensitive terms are completely decoupled, as presented in equation (\ref{UVaction3}) and equation (\ref{IRaction3}). On the other hand, in the expression for the renormalized power spectrum the UV and IR counter terms appear as a product. Just by applying the previously mentioned renormalization conditions, one cannot fix the structure of both of these counter-terms simultaneously. For this reason, the adiabatic renormalization scheme, which responds to the quadratic divergences successfully, is used to determine the UV counter-term. Implementation of this scheme helps to remove quadratic divergences fully from this computation. This fixed structure of the UV counter-term is inserted in the expression for the power spectrum and further using the renormalization conditions one can further fix the structure of the IR counter-term, which helps to smoothen the logarithmic divergences. The latter we identify as the power spectrum renormalization scheme. It seems like we have used two completely separate schemes for the renormalization of a single theory, however, from the discussions presented in this section and its underlying connections with the renormalization in Quantum Field Theory of curved space-time it is now quite clear that these two schemes of renormalizations, i.e. adiabatic and power-spectrum are related to each other. Without using the adiabatic scheme, one cannot further use the power-spectrum scheme, as IR counter-terms cannot be fixed without fixing UV counter-terms. In the next two subsections where the details of the adiabatic and wave function renormalization schemes are presented, we further establish these claims in more detail. } 

 \section{Computation of cut-off regulated one-loop momentum integrals}\label{App:B}

\textcolor{black}{Our main goal in this appendix is to explicitly assess the entire contributions of the momentum integrals that were used to measure the one-loop contributions from the USR and SR regions, respectively, to the primordial scalar power spectrum. }

\subsection{One-loop momentum integrals in USR period}

\textcolor{black}{Let's first express the contribution as an integral that depends on momentum and appears in the USR region:}
\bea  \label{gk1} &&\textcolor{black}{{\bf E}(\tau):=\int^{k_e}_{k_s}\frac{dk}{k}\;\left|{\cal S}_{\bf k}(\tau)\right|^{2}},\eea
\textcolor{black}{where we define a new momentum and conformal time dependent function ${\cal S}_{\bf k}(\tau)$, which is defined by the following expression:}
		\bea \label{hhgx} \textcolor{black}{{\cal S}_{\bf k}(\tau)=\bigg[\alpha^{\rm USR}_{\bf k}\left(1+ik\tau\right)\; e^{-ik\tau}-\beta^{\rm USR}_{\bf k}\left(1-ik\tau\right)\; e^{ik\tau}\bigg].}\eea
  \textcolor{black}{Here the Bogoliubov coefficients $\alpha^{\rm USR}_{\bf k}$ and $\beta^{\rm USR}_{\bf k}$ in the USR region is given by:}
  \bea && \textcolor{black}{\alpha^{\rm USR}_{\bf k}=1-\frac{3}{2ik^{3}\tau^{3}_s}\left(1+k^{2}\tau^{2}_s\right),}\\
&& \textcolor{black}{\beta^{\rm USR}_{\bf k}=-\frac{3}{2ik^{3}\tau^{3}_s}\left(1+ik\tau_s\right)^{2}\; e^{-2ik\tau_s}.}\eea
 \textcolor{black}{After substituting the specific mathematical form of the momentum and conformal time dependent function ${\cal S}_{\bf k}(\tau)$ in equation (\ref{gk1}), we found:}
\bea \label{gk2} && \textcolor{black}{{\bf E}(\tau)= {\bf E}_1(\tau)+{\bf E}_2(\tau)+{\bf E}_3(\tau)+{\bf E}_4(\tau)=\sum^{4}_{i=1}{\bf E}_i(\tau),}\eea
 \textcolor{black}{where the four contributions ${\bf E}_i(\tau)\forall i=1,2,3,4$, can be written as:}
\bea  \textcolor{black}{{\bf E}_1(\tau)}&=& \textcolor{black}{\int^{k_e}_{k_s}\frac{dk}{k}\;\bigg(1+\frac{9}{4}\frac{\left(1+k^2\tau^2_s\right)^2}{k^6\tau^6_s}\bigg) \left(1+k^2\tau^2\right),}\\
 \textcolor{black}{{\bf E}_2(\tau)}&=& \textcolor{black}{\int^{k_e}_{k_s}\frac{dk}{k}\;\bigg(\frac{9}{4}\frac{\left(1+k^2\tau^2_s\right)^2}{k^6\tau^6_s}\bigg) \left(1+k^2\tau^2\right),}\\
 \textcolor{black}{{\bf E}_3(\tau)}&=& \textcolor{black}{-\int^{k_e}_{k_s}\frac{dk}{k}\;\bigg(1+\frac{3}{2i}\frac{\left(1+k^2\tau^2_s\right)}{k^3\tau^3_s}\bigg)\bigg(-\frac{3}{2i} \frac{\left(1+ik\tau_s\right)^2}{k^3\tau^3_s}\bigg)\left(1-ik\tau\right)^2\; e^{2ik(\tau-\tau_s)},}\\
 \textcolor{black}{{\bf E}_4(\tau)}&=& \textcolor{black}{-\int^{k_e}_{k_s}\frac{dk}{k}\;\bigg(1-\frac{3}{2i}\frac{\left(1+k^2\tau^2_s\right)}{k^3\tau^3_s}\bigg)\bigg(\frac{3}{2i} \frac{\left(1-ik\tau_s\right)^2}{k^3\tau^3_s}\bigg)\left(1+ik\tau\right)^2\; e^{-2ik(\tau-\tau_s)}.}\eea
 \textcolor{black}{The outcomes of the aforementioned integrals are as follows after a few algebraic operations:}
\bea \label{c1} \textcolor{black}{{\bf E}_1(\tau)}&=& \textcolor{black}{\bigg[\frac{1}{2}\left(k^2_e-k^2_s\right)\tau^2+\left(1+\frac{9}{4}\left(\frac{\tau}{\tau_s}\right)^2\right) \ln\left(\frac{k_e}{k_s}\right)-\frac{9}{8}\frac{1}{\tau^4_s}\bigg(1+\left(\frac{\tau}{\tau_s}\right)^2\bigg)\bigg(\frac{1}{k^4_e}-\frac{1}{k^4_s}\bigg)}\nonumber\\
&&\quad\quad\quad\quad\quad\quad\quad\quad\quad\quad\quad\quad \textcolor{black}{ -\frac{9}{8}\frac{1}{\tau^2_s}\bigg(1+2\left(\frac{\tau}{\tau_s}\right)^2\bigg)\bigg(\frac{1}{k^2_e}-\frac{1}{k^2_s}\bigg)-\frac{3}{8}\frac{1}{\tau^6_s}\bigg(\frac{1}{k^6_e}-\frac{1}{k^6_s}\bigg)\bigg],}\\
\label{c2} \textcolor{black}{{\bf E}_2(\tau)}&=& \textcolor{black}{\bigg[\frac{9}{4}\left(\frac{\tau}{\tau_s}\right)^2 \ln\left(\frac{k_e}{k_s}\right)-\frac{9}{8}\frac{1}{\tau^4_s}\bigg(1+\frac{1}{2}\left(\frac{\tau}{\tau_s}\right)^2\bigg)\bigg(\frac{1}{k^4_e}-\frac{1}{k^4_s}\bigg)}\nonumber\\
&&\quad\quad\quad\quad\quad\quad\quad\quad\quad\quad\quad\quad  \textcolor{black}{-\frac{9}{8}\frac{1}{\tau^2_s}\bigg(1+2\left(\frac{\tau}{\tau_s}\right)^2\bigg)\bigg(\frac{1}{k^2_e}-\frac{1}{k^2_s}\bigg)-\frac{3}{8}\frac{1}{\tau^6_s}\bigg(\frac{1}{k^6_e}-\frac{1}{k^6_s}\bigg)\bigg],}\\
\label{c3} \textcolor{black}{{\bf E}_3(\tau)}&=& \textcolor{black}{-\frac{1}{16 \tau _s^6}\bigg[\bigg(36 \tau^2 \tau _s^4+8 \tau _s^6-8 \tau^6 \bigg)\bigg(\text{Ei}\left(2 i k_e \left(\tau-\tau _s\right)\right)-\text{Ei}\left(2 i k_s \left(\tau-\tau _s\right)\right)\bigg)}\nonumber\\
&& \textcolor{black}{+ \bigg(\frac{12 \tau^2 \tau _s^5}{\tau-\tau _s}\bigg(e^{2 i k_e \left(\tau-\tau _s\right)}-e^{2 i k_s \left(\tau-\tau _s\right)}\bigg)}\nonumber\\
&& \textcolor{black}{-4 i \left(\tau^4 \tau _s+\tau^3 \tau _s^2+\tau^2 \tau _s^3+7 \tau \tau _s^4+\tau^5+\tau _s^5\right)\bigg(\frac{e^{2 i k_e \left(\tau-\tau _s\right)}}{k_e}-\frac{e^{2 i k_s \left(\tau-\tau _s\right)}}{k_s}\bigg)}\nonumber\\
&& \textcolor{black}{-2 \left(2 \tau^3 \tau _s+3 \tau^2 \tau _s^2+28 \tau \tau _s^3+\tau^4-7 \tau _s^4\right)\bigg(\frac{e^{2 i k_e \left(\tau-\tau _s\right)}}{k^2_e}-\frac{e^{2 i k_s \left(\tau-\tau _s\right)}}{k^2_s}\bigg)}\nonumber\\
&& \textcolor{black}{+2 i \left(3 \tau^2 \tau _s+6 \tau \tau _s^2+\tau^3-14 \tau _s^3\right)\bigg(\frac{e^{2 i k_e \left(\tau-\tau _s\right)}}{k^3_e}-\frac{e^{2 i k_s \left(\tau-\tau _s\right)}}{k^3_s}\bigg)}\nonumber\\
&& \textcolor{black}{+3 \left(-8 \tau \tau _s+\tau^2-2 \tau _s^2\right)\bigg(\frac{e^{2 i k_e \left(\tau-\tau _s\right)}}{k^4_e}-\frac{e^{2 i k_s \left(\tau-\tau _s\right)}}{k^4_s}\bigg)}\nonumber\\
&& \textcolor{black}{+12 i \left(\tau-\tau _s\right)\bigg(\frac{e^{2 i k_e \left(\tau-\tau _s\right)}}{k^5_e}-\frac{e^{2 i k_s \left(\tau-\tau _s\right)}}{k^5_s}\bigg)-6\bigg(\frac{e^{2 i k_e \left(\tau-\tau _s\right)}}{k^6_e}-\frac{e^{2 i k_s \left(\tau-\tau _s\right)}}{k^6_s}\bigg)\bigg)\bigg],}\eea\bea
\label{c4} \textcolor{black}{{\bf E}_4(\tau)}&=& \textcolor{black}{-\frac{1}{16 \tau _s^6}\bigg[\bigg(36 \tau^2 \tau _s^4+8 \tau _s^6-8 \tau^6 \bigg)\bigg(\text{Ei}\left(-2 i k_e \left(\tau-\tau _s\right)\right)-\text{Ei}\left(-2 i k_s \left(\tau-\tau _s\right)\right)\bigg)}\nonumber\\
&& \textcolor{black}{+ \bigg(\frac{12 \tau^2 \tau _s^5}{\tau-\tau _s}\bigg(e^{-2 i k_e \left(\tau-\tau _s\right)}-e^{-2 i k_s \left(\tau-\tau _s\right)}\bigg)}\nonumber\\
&& \textcolor{black}{+4 i \left(\tau ^4 \tau _s+\tau ^3 \tau _s^2+\tau ^2 \tau _s^3+7 \tau  \tau _s^4+\tau _s^5+\tau ^5\right)\bigg(\frac{e^{-2 i k_e \left(\tau-\tau _s\right)}}{k_e}-\frac{e^{-2 i k_s \left(\tau-\tau _s\right)}}{k_s}\bigg)}\nonumber\\
&& \textcolor{black}{-2 \left(2 \tau ^3 \tau _s+3 \tau ^2 \tau _s^2+28 \tau  \tau _s^3-7 \tau _s^4+\tau ^4\right)\bigg(\frac{e^{-2 i k_e \left(\tau-\tau _s\right)}}{k^2_e}-\frac{e^{-2 i k_s \left(\tau-\tau _s\right)}}{k^2_s}\bigg)}\nonumber\\
&& \textcolor{black}{-2 i \left(3 \tau ^2 \tau _s+6 \tau  \tau _s^2-14 \tau _s^3+\tau ^3\right)\bigg(\frac{e^{-2 i k_e \left(\tau-\tau _s\right)}}{k^3_e}-\frac{e^{-2 i k_s \left(\tau-\tau _s\right)}}{k^3_s}\bigg)}\nonumber\\
&& \textcolor{black}{+3 \left(-8 \tau  \tau _s-2 \tau _s^2+\tau ^2\right)\bigg(\frac{e^{-2 i k_e \left(\tau-\tau _s\right)}}{k^4_e}-\frac{e^{-2 i k_s \left(\tau-\tau _s\right)}}{k^4_s}\bigg)}\nonumber\\
&& \textcolor{black}{-12 i \left(\tau -\tau _s\right)\bigg(\frac{e^{-2 i k_e \left(\tau-\tau _s\right)}}{k^5_e}-\frac{e^{-2 i k_s \left(\tau-\tau _s\right)}}{k^5_s}\bigg)-6\bigg(\frac{e^{-2 i k_e \left(\tau-\tau _s\right)}}{k^6_e}-\frac{e^{-2 i k_s \left(\tau-\tau _s\right)}}{k^6_s}\bigg)\bigg)\bigg].}\eea
 \textcolor{black}{Let's put the contributions from equations (\ref{c1}), (\ref{c2}),(\ref{c3}) and (\ref{c4})) together now to better understand how the findings behave in the USR regime, which will result in the following results that have been simplified:}
\bea \label{c11} \textcolor{black}{{\bf E}(\tau)}&=& \textcolor{black}{{\bf E}_1(\tau)+{\bf E}_2(\tau)+{\bf E}_3(\tau)+{\bf E}_4(\tau)}\nonumber\\
&=& \textcolor{black}{\Bigg\{\bigg[\frac{1}{2}\left(k^2_e-k^2_s\right)\tau^2+\left(1+\frac{9}{2}\left(\frac{\tau}{\tau_s}\right)^2\right) \ln\left(\frac{k_e}{k_s}\right)-\frac{9}{4}\frac{1}{\tau^4_s}\bigg(1+\frac{3}{4}\left(\frac{\tau}{\tau_s}\right)^2\bigg)\bigg(\frac{1}{k^4_e}-\frac{1}{k^4_s}\bigg)}\nonumber\\
&&\quad\quad\quad\quad\quad\quad\quad\quad\quad\quad\quad\quad  \textcolor{black}{-\frac{9}{4}\frac{1}{\tau^2_s}\bigg(1+2\left(\frac{\tau}{\tau_s}\right)^2\bigg)\bigg(\frac{1}{k^2_e}-\frac{1}{k^2_s}\bigg)-\frac{3}{4}\frac{1}{\tau^6_s}\bigg(\frac{1}{k^6_e}-\frac{1}{k^6_s}\bigg)\bigg]}\nonumber\\
&&\quad\quad\quad\quad \textcolor{black}{-\frac{1}{16 \tau _s^6}\bigg[\bigg(36 \tau^2 \tau _s^4+8 \tau _s^6-8 \tau^6 \bigg)}\nonumber\\
&&\quad\quad\quad\quad\quad\quad\quad\quad \textcolor{black}{\times\bigg(\text{Ei}\left(2 i k_e \left(\tau-\tau _s\right)\right)+\text{Ei}\left(-2 i k_e \left(\tau-\tau _s\right)\right)-\text{Ei}\left(2 i k_s \left(\tau-\tau _s\right)\right)-\text{Ei}\left(-2 i k_s \left(\tau-\tau _s\right)\right)\bigg)}\nonumber\\
&&\quad\quad\quad\quad \textcolor{black}{+ \bigg(\frac{24 \tau^2 \tau _s^5}{\tau-\tau _s}\bigg(\cos \left(k_e \left(\tau-\tau _s\right)\right)-\cos \left(k_s \left(\tau-\tau _s\right)\right)\bigg)}\nonumber\\
&&\quad\quad\quad\quad \textcolor{black}{+8 \left(\tau ^4 \tau _s+\tau ^3 \tau _s^2+\tau ^2 \tau _s^3+7 \tau  \tau _s^4+\tau _s^5+\tau ^5\right)\bigg(\frac{\sin \left(k_e \left(\tau-\tau _s\right)\right)}{k_e}-\frac{\sin \left(k_s \left(\tau-\tau _s\right)\right)}{k_s}\bigg)}\nonumber\\
&&\quad\quad\quad\quad \textcolor{black}{-4 \left(2 \tau ^3 \tau _s+3 \tau ^2 \tau _s^2+28 \tau  \tau _s^3-7 \tau _s^4+\tau ^4\right)\bigg(\frac{\cos \left(k_e \left(\tau-\tau _s\right)\right)}{k^2_e}-\frac{\cos \left(k_s \left(\tau-\tau _s\right)\right)}{k^2_s}\bigg)}\\
&&\quad\quad\quad\quad \textcolor{black}{-4 \left(3 \tau ^2 \tau _s+6 \tau  \tau _s^2-14 \tau _s^3+\tau ^3\right)\bigg(\frac{\sin \left(k_e \left(\tau-\tau _s\right)\right)}{k^3_e}-\frac{\sin \left(k_s \left(\tau-\tau _s\right)\right)}{k^3_s}\bigg)}\\
&&\quad\quad\quad\quad \textcolor{black}{+6 \left(-8 \tau  \tau _s-2 \tau _s^2+\tau ^2\right)\bigg(\frac{\cos \left(k_e \left(\tau-\tau _s\right)\right)}{k^4_e}-\frac{\cos \left(k_s \left(\tau-\tau _s\right)\right)}{k^4_s}\bigg)}\nonumber\\
&&\quad\quad\quad\quad \textcolor{black}{-24 \left(\tau -\tau _s\right)\bigg(\frac{\sin \left(k_e \left(\tau-\tau _s\right)\right)}{k^5_e}-\frac{\sin \left(k_s \left(\tau-\tau _s\right)\right)}{k^5_s}\bigg)}\nonumber\\
&&\quad\quad\quad\quad \textcolor{black}{-6\bigg(\frac{\cos \left(k_e \left(\tau-\tau _s\right)\right)}{k^6_e}-\frac{\cos \left(k_s \left(\tau-\tau _s\right)\right)}{k^6_s}\bigg)\bigg)\bigg]\Bigg\}.}\eea
\textcolor{black}{Here the symbol "Ei'' signifies the exponential integral function, which is defined by the following expression:}
\bea \textcolor{black}{{\rm Ei}(x):=-\int^{\infty}_{-x}\frac{e^{-t}}{t}~dt=\int^{x}_{-\infty}\frac{e^{t}}{t}~dt}.\eea
 \textcolor{black}{Then considering both the contribution from the short range as well as the long range quantum fluctuations we found the following simplified result:}
\bea  \textcolor{black}{\boxed{{\bf E}(\tau_e)={\bf E}(\tau_s)\approx \ln\left(\frac{k_e}{k_s}\right)+\frac{1}{2}\bigg(\left(\frac{k_e}{k_s}\right)^2-1\bigg)=\ln\left(\frac{k_{\rm UV}}{k_{\rm IR}}\right)+\frac{1}{2}\bigg(\left(\frac{k_{\rm UV}}{k_{\rm IR}}\right)^2-1\bigg)}.}\eea
\textcolor{black}{To obtain this simplified result we have neglected all other suppressed contributions from the oscillatory terms as well as inverse power law terms of $(k_s/k_e)(\ll 1)$, as all of them will not be able to significant change the overall behaviour of the scalar power spectrum in USR period. The obtained results suggests that in the one-loop result IR sensitive logarithmic divergent and quadratic UV divergent contributions are dominating over all the other terms appearing in this computation. With the detailed analysis we have shown in the text portion of the paper that the quadratic UV divergence can be completely removed by implementing the well known adiabatic renormalization scheme.  Additionally, by implementing the DRG resummation technique we have shown that the IR sensitivity can be further softened and shifted to the next order of perturbation theory. It's also crucial to note that in order to extract the finite contributions from the one-loop momentum integrals in the USR period, we have constrained the momentum integration within a window, $k_s<k<k_e$, by introducing two well known physical cut-offs: the IR cut-off $k_{\rm IR}=k_s$ and the UV cut-off $k_{\rm UV}=k_e$ which are the corresponding cut-off scales of UV and IR divergences inserted through the implementation of cut-of regularization before implementing the adiabatic renormalization and DRG resummation. }

\subsection{One-loop momentum integral in SR period}
\textcolor{black}{Next, we'll discuss the integral that appears in the computation of the one-loop correction to the primordial power spectrum of scalar modes in the SR area. Let's assess the ensuing integral:}
\bea\label{intSR} \textcolor{black}{{\bf D}(\tau):=\int^{k_e}_{p_*}\frac{dk}{k}\;\left|{\cal M}_{\bf k}(\tau)\right|^{2},}\eea
\textcolor{black}{where we define a new momentum and conformal time dependent function ${\cal M}_{\bf k}(\tau)$ in presence of Bunch Davies quantum initial condition, which is defined as:}
		\bea \textcolor{black}{{\cal M}_{\bf k}(\tau)=\left(1+ik\tau\right)\; e^{-ik\tau}.}\eea
\textcolor{black}{After substituting the explicit mathematical form of the above function ${\cal M}_{\bf k}(\tau)$ in the equation (\ref{intSR}), we get the following simplified result:}
\bea\textcolor{black}{{\bf D}(\tau)=\int^{k_e}_{p_*}\frac{dk}{k}\;\left(1+k^2\tau^2\right)=\bigg[\ln\left(\frac{k_e}{p_*}\right)+\frac{1}{2}\left(k^2_e-p^2_*\right)\tau^2\bigg].}\eea
\textcolor{black}{Then considering both the contribution from the short range as well as the long range quantum fluctuations we found the following simplified result:}
\bea \textcolor{black}{\boxed{{\bf D}(\tau_e)=\ln\left(\frac{k_e}{p_*}\right)+\frac{1}{2}\bigg(\left(\frac{k_{e}}{p_*}\right)^2-1\bigg)=\ln\left(\frac{k_{\rm UV}}{p_*}\right)+\frac{1}{2}\bigg(\left(\frac{k_{\rm UV}}{p_*}\right)^2-1\bigg)}.}\eea
\textcolor{black}{Here in SR phase $p_*$ represents the pivot scale which is expected to be $p_*\ll k_s$ in the present framework under consideration. In the SR region the final one-loop result will be controlled by the above mentioned IR logarithmic divergent contribution and a quadratic UV divergent contribution. We have shown in this paper by applying adiabatic renormalization scheme UV divergence can be completely by choosing the underlying renormalization scale properly. Also we have shown that IR divergence can be softened and shifted to next order in the perturbation theory.}

\end{itemize}

%\clearpage

\newpage
%\phantomsection
%\addcontentsline{toc}{section}{References}
\bibliographystyle{utphys}
\bibliography{reference3}

\end{document}